National Academy of Sciences of Ukraine
Institute for Nuclear Research, Kyiv, Ukraine

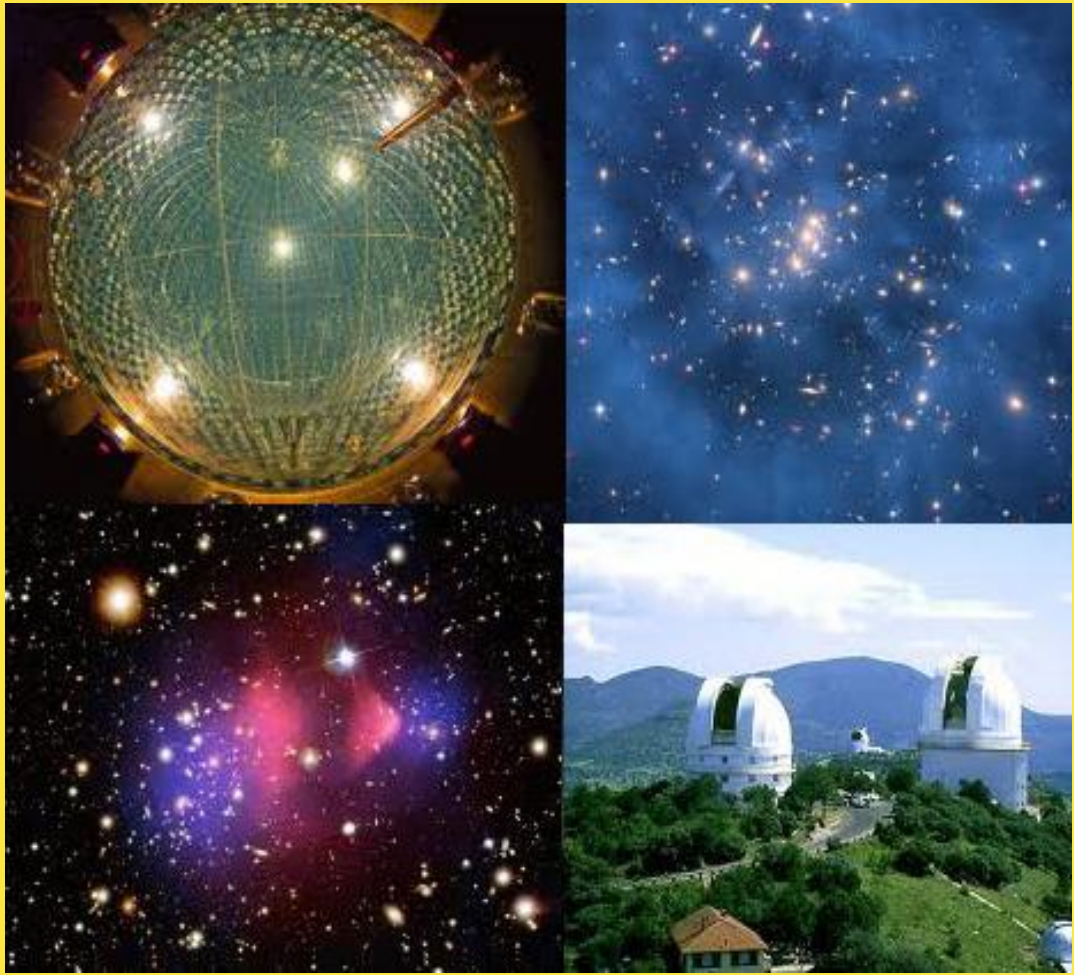

Proceedings of the Workshop on results of the Project Kosmomikrofizyka-2 (Astroparticle Physics) of the National Academy of Sciences of Ukraine

# Astrophysical and cosmological problems of invisible mass and dark energy in the Universe

November 21 – 22, 2012
Institute for Nuclear Research, Kyiv, Ukraine

National Academy of Sciences of Ukraine

Institute for Nuclear Research, Kyiv, Ukraine

Proceedings of the Workshop

on results of the Project Kosmomikrofizyka-2 (Astroparticle Physics) of the National Academy of Sciences of Ukraine

# Astrophysical and cosmological problems of invisible mass and dark energy in the Universe

November 21-22, 2012

Institute for Nuclear Research, Kyiv, Ukraine

Kyiv 2013

# Contents

# Foreword

The Workshop on results of the Project Kosmomikrofizyka-2 (Astroparticle Physics) of the National Academy of Sciences (NAS) of Ukraine "Astrophysical and cosmological problems of invisible mass and dark energy in the Universe" was held on November 21-22, 2012 in the Institute for Nuclear Research, Kyiv, Ukraine (http://lpd.kinr.kiev.ua/kmf12). This Project was carried out during three years (2010 – 2012) by scientists from various universities and institutes of the National Academy of Sciences of Ukraine; it was a logical continuation of the previous scientific program of the NAS of Ukraine "Researches of structure and composition of the Universe, hidden mass and dark energy (Kosmomikrofizyka)" in 2007 – 2009. These programs were devoted to theoretical and experimental investigations in astronomy, astrophysics, cosmology, physics of atomic nuclei and particle physics, which are related with the problems of dark matter and dark energy in the Universe.

Scientists from the following institutions of Ukraine took part in the Workshop:
1. Main Astronomical Observatory of the NAS of Ukraine;
2. Institute of Astronomy of V.N. Karazin Kharkiv National University;
3. Ya.S. Pidstryhach Institute for Applied Problems of Mechanics and Mathematics of the NAS of Ukraine;
4. Bogolyubov Institute for Theoretical Physics of the NAS of Ukraine;
5. Institute for Nuclear Research of the NAS of Ukraine;
6. Taras Shevchenko National University of Kyiv;
7. Ivan Franko National University of Lviv;
8. Odessa National Polytechnic University;
9. Ilya Mechnikov National University of Odessa;
10. Institute of Radio Astronomy of the NAS of Ukraine;
11. Taras Shevchenko National Pedagogical University of Chernigiv.

We hope that the materials of the Workshop will be interesting and useful for a reader fascinated by one of the biggest puzzles of the Universe which is evidently the challenge for the science of the 21$^{st}$ century.

V.M. Shulga, Head of the Program, Academician of the NAS of Ukraine

# List of participants

| | |
|---|---|
| Agienko Kateryna | Main Astronomical Observatory of the NAS of Ukraine |
| Berdina Lyudmyla | Institute of Radio Astronomy of the NAS of Ukraine |
| Bogdan Artem | Taras Shevchenko National University of Kyiv |
| Vavilova Iryna | Main Astronomical Observatory of the NAS of Ukraine |
| Vasylenko Olga | Taras Shevchenko National University of Kyiv |
| Hnatyk Bohdan | Taras Shevchenko National University of Kyiv |
| Danevich Fedor | Institute for Nuclear Research of the NAS of Ukraine |
| Eingorn Maksym | Odessa National Polytechnic University; Ilya Mechnikov National University of Odessa |
| Zhdanov Valery | Taras Shevchenko National University of Kyiv |
| Zhuk Oleksandr | Ilya Mechnikov National University of Odessa |
| Ivashchenko Ganna | Taras Shevchenko National University of Kyiv |
| Kobychev Vladislav | Institute for Nuclear Research of the NAS of Ukraine |
| Kobzar Oleg | Taras Shevchenko National Pedagogical University of Chernigiv |
| Marchenko Volodymyr | Taras Shevchenko National Pedagogical University of Chernigiv |
| Mokina Valentyna | Institute for Nuclear Research of the NAS of Ukraine |
| Novosyadlyj Bohdan | Ivan Franko National University of Lviv |
| Pavlenko Yakiv | Main Astronomical Observatory of the NAS of Ukraine |
| Pelykh Volodymyr | Ya.S. Pidstryhach Institute for Applied Problems of Mechanics and Mathematics of the NAS of Ukraine |
| Polischuk Oksana | Institute for Nuclear Research of the NAS of Ukraine |
| Sliusar Vitaly | Taras Shevchenko National University of Kyiv |
| Sushchov Oleksandr | Taras Shevchenko National Pedagogical University of Chernigiv |
| Torbaniuk Olena | Taras Shevchenko National University of Kyiv |
| Fedorov Petro | Institute of Astronomy of V.N. Karazin Kharkiv National University |
| Fedorova Olena | Taras Shevchenko National University of Kyiv |
| Tsvetkova Viktoriya | Institute of Radio Astronomy of the NAS of Ukraine |
| Shtanov Yuri | Bogolyubov Institute for Theoretical Physics of the NAS of Ukraine |
| Shulga Valery | Institute of Radio Astronomy of the NAS of Ukraine |
| Chernyak Dmytro | Institute for Nuclear Research of the NAS of Ukraine |

# Program

**21 November 2012, Wednesday**

| 14:00-14:20 | Registration |
|---|---|
| 14:20-14:40 | Opening of the Workshop. Greetings from the Director of the Institute for Nuclear Research of the NAS of Ukraine **I.M. Vyshnevskyi**, members of the Presidium of the NAS of Ukraine **Ya.S. Yatskiv** and **V.M. Loktev**, opening address of the Head of the Program, Academician of the NAS of Ukraine **V.M. Shulga** |

**1st session. Reports: achievements, shortcomings, criticism and approving estimations**

| 14:40-15:00 | **F.A. Danevich,** Studies of neutrino properties and search for dark matter |
|---|---|
| 15:00-15:20 | **V.O. Pelykh,** Review of investigations in IAPMM in framework of the Program:<br>• group classification of models of axion electrodynamics<br>• discovery and study of ultrarelativistic spin-gravitation interaction<br>• problems of gravitational energy and spinor fields |
| 15:20-15:40 | **Yu.V. Shtanov,** Main scientific results obtained in framework of the program "Cosmological and astrophysical tests of fundamental physics" |
| 15:40-16:00 | **V.S. Tsvetkova, V.M. Shulga,** The phenomenon of gravitational lensing as applied to the search of dark matter in the Universe: summary of results in the framework of the Program |
| 16:00-16:40 | **Coffee break, presentation and discussion of posters** |
| 16:40-17:00 | **V.I. Zhdanov,** Dark matter, dark energy and gravitation lenses: results and perspectives |
| 17:00-17:20 | **A.I. Zhuk,** Problem aspects of many-dimensional models |
| 17:20-17:35 | **M. Eingorn,** Dynamics of the Universe inside homogenious cell |
| 17:45-19:00 | **Banquet, informal discussions** |

**22 November 2012, Thursday**

| 10:00-10:20 | **I.B. Vavilova,** Properties of baryonic and non-baryonic matter in galaxies and clusters of galaxies. Report for the 2010-2012 period. |
|---|---|

**2nd session. First steps toward the Program of the NAS of Ukraine on cosmic researches**

| 10:20-10:40 | **B.S. Novosyadlyj,** Status and perspectives of determination of nature of dark energy in the earth and space experiments |
|---|---|
| 10:40-11:00 | **P.N. Fedorov, V.M. Shulga,** Catalog XPM as an instrument to study kinematics of galaxies |
| 11:00-11:20 | **Coffee break** |
| 11:20-11:40 | **B.I. Hnatyk,** Cosmic ray astrophysics: Cosmological aspect |
| 11:40-12:00 | **V. Kobychev,** Proposal of satellite and underground experiments in framework of the Program of space researches of the NAS of Ukraine |
| 12:00-13:00 | **Lunch, summary and conclusion of the Workshop** |

**Posters**

1. O.V. Fedorova, V.I. Zhdanov. Atypical behaviour of X-ray spectrum of active galaxy nucleus NGC 4945.
2. V.M. Sliusar, V.I. Zhdanov. Statistical simulation of the density inhomogeneity spectrum in one-dimensional hydrodynamic problem.
3. O.M. Aleksandrov, C.M. Koval, V.I. Zhdanov. Equation of gravitation lense: critical solutions and enhancement near cusp point.
4. O. Vasylenko, G. Ivaschenko. Clustering of quasars on the basis of catalogue SDSS NBCKDE.
5. O. Torbaniuk, G. Ivaschenko. The mean transmission of the Lyα-forest at $2<z<4$ from a sample of 33 high-resolution quasar spectra.
6. L.A. Berdina. Effect of the microlens-substar location on brightness of a remote source.
7. O. Polischuk. Search for Li-7 solar axions.
8. V. Mokina. Optimization of cryogenic scitillating detectors to search for dark matter and double beta decay.
9. D. Chernyak. Cryogenic scintillating bolometers with crystals $ZnMoO_4$ for high-sensitive investigations of neutrino properties.
10. S.H. Fakhr. Fine-tuning problem in five-dimensional brane world models.
11. A. Burgazli. Quintessence and phantom field inhomogeneities at late stages of Universe evolution.
12. A. Kudinova. Motion of astrophysical objects in the open Universe.
13. O. Kobzar, O. Sushchov, V. Marchenko, B. Hnatyk. Search for sources of ultra high energy cosmic rays with taking into account influence of galactic and extragalactic magnetic fields.
14. V. Marchenko, M. Ostrovsky, D. Harris, L. Stavazh, A. Bogdan, B. Hnatyk. The structure of X-ray jet of quasar 3C 273.

# Метод розщеплення для дослідження системи рівнянь Максвелла і Сена-Віттена

В.О. Пелих [1], Ю.В. Тайстра [1], Б.С. Новосядлий [2]

[1] *Інститут прикладних проблем механіки і математики ім. Я. Підстригача НАН України,
вул. Наукова, 3-б, Львів, 79060, Україна*
[2] *Астрономічна обсерваторія Львівського національного університету імені Івана Франка
вул. Кирила і Мефодія, 8, Львів, 79005, Україна*

Запропоновано метод отримання розщеплених рівнянь другого порядку електромагнітного поля при калібровці Лоренца в термінах скалярів Ньюмена-Пенроуза та рівнянь Сена-Віттена, доведено калібрувальний характер поля Сена-Віттена у просторах визначеного типу.

## 1. ВСТУП

Вивчення впливу гравітаційного поля на усі інші фізичні поля, навіть за припущення про їх класичний характер та про відсутність зворотного впливу на гравітаційне, як і опис поширення гравітаційних збурень на викривленому гравітацією фоні, наштовхуються на значні труднощі принципового характеру.

Основною із них є зв'язана структура рівнянь, яка виникає із переходом від рівнянь полів у плоскому просторі до рівнянь у рімановому просторі. Вона стосується усіх полів (за винятком скалярного), не усувається жодними калібрувальними умовами і перешкоджає не лише отриманню точних розв'язків рівнянь, а й чисельному їх інтегруванню та, навіть, дослідженню якісних властивостей, наприклад, коливальності чи існування нулів. Розв'язання проблеми можливе або на шляху обмеження локальної загальності (алгебрично спеціальні простори), чи властивостей "в цілому" (клас гомології) ріманового простору, або на шляху вибору часткових точних розв'язків рівнянь Ейнштейна.

Для рівнянь Максвелла Когеном та Кегелесом [1] на випадок викривленого простору узагальнено підхід, відомий під назвою формалізму потенціалів Герца, до побудови їх розв'язків на основі використання додаткових симетрій цих рівнянь та узагальнено двокомпонентний підхід Дебая (потенціали Герца-Бромвіча-Дебая-Вітеккера-Пенроуза). Авторами встановлено, що формалізм Герца може бути розширений до всіх викривлених просторів, а формалізм Дебая може бути розширений до простору, де безсувні ізотропні конгруенції геодезійних можуть мати дотичними головні кратні напрямки тензора Вейля; таким є, зокрема простір типу *D*. В цих просторах застосування потенціалів Герца та підходу Дебая зводить систему рівнянь Максвелла до відокремлених лінійних хвильових рівнянь для комплексного бівектор-потенціалу поля.

У роботі Біні [2] хвильова система рівнянь для векторного поля, доповнена калібрувальною умовою Лоренца, розщеплюється на окремі рівняння і точно розв'язується у просторі де Сіттера. Альтернативні до методу Когена і Кегелеса підходи до вивчення збурень, які також ґрунтуються на відокремленні рівнянь (розщеплені системи рівнянь) розвинули Хшановскі [3], Стюарт [4] і Валд [5], однак всі вони також припускають, що простори є алгебрично-спеціальними. У низці публікацій розв'язки та (або) властивості рівнянь Дірака і Вейля, Раріта-Швінгера, гравітаційна самодія отримуються у полях Керра або й Шварцшільда завдяки тому, що відповідні системи рівнянь розщеплюються у цих просторах на окремі рівняння. Завдяки цьому ж у єдиному до цього часу випадку – у конформно-плоскому просторі – отримано розв'язки рівняння Сена-Віттена та описано властивості його вузлових множин.

Існує значна кількість причин, які зумовлюють необхідність інтегрування чи дослідження властивостей рівнянь фізичних полів у просторах, відмінних від згаданих вище, чи у просторах більшої загальності. Для випадку електромагнітного поля це викликано, наприклад, актуальною потребою опису його поширення у системі двох чорних дір. Для випадку поля Сена-Віттена припущення про те, що простір є конформно-плоским, виключає можливість застосування отриманих результатів навіть для одного із найбільш використовуваних в астрофізиці полів – для поля Керра.

## 2. СИСТЕМА РІВНЯНЬ МАКСВЕЛЛА ДРУГОГО ПОРЯДКУ У ФОРМАЛІЗМІ СПІНОВИХ КОЕФІЦІЄНТІВ

Оскільки усі виявлені до цього часу умови розщеплення є достатніми, становить інтерес відкриття інших можливостей отримання розв'язків рівнянь Максвелла шляхом розщеплення системи рівнянь на основі інших підходів. Нами [6] запропоновано метод, у якому система рівнянь Максвелла другого порядку у поданні через спінор-потенціал з умовою Лоренца у тетраді Ньюмена-Пенроуза стає послідовно розщеплюваною.

Загальноковаріантну систему рівнянь Максвелла у вакуумі за калібрувальної умови Лоренца

$$\nabla_b(\nabla^b A^a) - R^a{}_c A^c = 0,$$

де $R^a{}_c$ – тензор Річчі, $A_0, A_1, A_2, A_3$ – компоненти «спінор-потенціалу» електромагнітного поля, $A^a = A_0 l^a + A_1 n^a + A_2 m^a + A_3 \overline{m}^a$, подаємо через вектор-потенціал електромагнітного поля $A^a$ в ізотропній тетраді Ньюмена-Пенроуза (НП) $l^a = o^A o^{A'}$, $n^a = i^A i^{A'}$, $m^a = o^A i^{A'}$, $\overline{m}^a = i^A o^{A'}$.

Оператор коваріантної похідної подається у вигляді:

$$\nabla_a = g_a{}^b \nabla_b = (l_a n^b + n_a l^b - m_a \overline{m}^b - \overline{m}_a m^b)\nabla_b = n_a D + l_a \Delta - \overline{m}_a \delta - m_a \overline{\delta},$$

де $D = l^a \nabla_a$, $\Delta = n^a \nabla_a$, $\delta = m^a \nabla_a$, $\overline{\delta} = \overline{m}^a \nabla_a$ – оператори похідних за напрямками ізотропної тетради.

Використовуючи представлення тензора Річчі в НП-тетраді, вираз для спінорного девіатор тензора Річчі через НП-скаляри та властивості спінових коефіцієнтів, отримаємо наступний вигляд системи рівнянь Максвелла:

$$D\Delta A \begin{pmatrix} D\Delta A_0 \\ D\Delta A_1 \\ D\Delta A_2 \\ D\Delta A_3 \end{pmatrix} + \delta\overline{\delta}A \begin{pmatrix} \delta\overline{\delta}A_0 \\ \delta\overline{\delta}A_1 \\ \delta\overline{\delta}A_2 \\ \delta\overline{\delta}A_3 \end{pmatrix} + DA \begin{pmatrix} DA_0 \\ DA_1 \\ DA_2 \\ DA_3 \end{pmatrix} + \Delta A \begin{pmatrix} \Delta A_0 \\ \Delta A_1 \\ \Delta A_2 \\ \Delta A_3 \end{pmatrix} + \delta A \begin{pmatrix} \delta A_0 \\ \delta A_1 \\ \delta A_2 \\ \delta A_3 \end{pmatrix} + \overline{\delta}A \begin{pmatrix} \overline{\delta}A_0 \\ \overline{\delta}A_1 \\ \overline{\delta}A_2 \\ \overline{\delta}A_3 \end{pmatrix} + A \begin{pmatrix} A_0 \\ A_1 \\ A_2 \\ A_3 \end{pmatrix} = \begin{pmatrix} 0 \\ 0 \\ 0 \\ 0 \end{pmatrix},$$

де $D\Delta A$, $\delta\overline{\delta}A$, $DA$, $\Delta A$, $\delta A$, $\overline{\delta}A$, $A$ - матриці коефіцієнтів при відповідних похідних, $A_{ij} = \varphi_{ij}(\alpha, \beta, \gamma, \varepsilon, \kappa, \tau, \sigma, \rho, \pi, \nu, \mu, \lambda)$ – деякі білінійні форми.

Розщеплюваність системи можна забезпечити кількома принципово відмінними умовами на скаляри Ньюмена-Пенроуза, з-посеред яких ми тут зупинимось на двох.

Зокрема, для послідовної розщеплюваності системи достатньо, щоб виконувались наступні додаткові умови для спінових коефіцієнтів:

$$\kappa = \sigma = \rho = \tau = 0. \tag{1}$$

При цьому послідовно відокремлюються друге, четверте, третє, перше рівняння системи. При виконанні умов (1) виконуються наступні співвідношення для компонент спінорів кривизни: $\Psi_0 = \Psi_1 = \Phi_{00} = \Phi_{01} = \Phi_{02} = \Phi_{10} = 0$. Ці умови виокремлюють простір типу *D*.

Якщо ж обмежитись умовою ізотропності електромагнітного поля, то дві компоненти спінора поля $\varphi_{AB} = \varphi_0 o_A o_B + \varphi_1 o_A \iota_B + \varphi_1 \iota_A o_B + \varphi_2 \iota_A \iota_B$ обертаються в нуль: $\varphi_1 = 0$, $\varphi_2 = 0$. Тоді система рівнянь Максвелла розщеплюється при виконанні таких умов на спінові коефіцієнти: $\kappa = \sigma = \nu = \lambda = \pi = \tau = 0$, $\Phi_{22} = \mu\bar{\mu}$, компоненти спінорів кривизни набувають значень $\Psi_0 = \Psi_4 = \Phi_{20} = \Phi_{02} = 0$, і простір вже не є простором типу *D*. Описано усі випадки, коли умови, що забезпечують розщеплення, не вимагають належності простору до типу *D*.

### 3. СПІНОРНЕ ПОЛЕ СЕНА-ВІТТЕНА ЯК КАЛІБРУВАЛЬНЕ І ЯК КООРДИНАТНА УМОВА У НЕКОНФОРМНО-ПЛОСКОМУ ПРОСТОРІ

Розвинутий кількома авторами тензорний (тетрадний) метод доведення теореми про додатну визначеність гравітаційної енергії асимптотично плоского простору із запровадженням шляхом задавання спеціальних додаткових умов (special gauge, SG) спеціального ортонормованого репера (special orthonormal frame, SOF) був більш відповідним до духу класичної ЗТВ у порівнянні із спінорним методом Віттена, у якому виникала проблема інтерпретації фізичного змісту спінорного поля. Однак тензорний метод, за визнанням самих авторів, містив істотні прогалини, серед яких найважливішою була відсутність доведення самого існування SOF.

Тензорний метод отримав своє повне завершення після доведення існування у наших публікаціях [7-10] існування перетворення лоренцової бази до такої, у якій виконується SG. Перешкодою для існування такого перетворення було існування нулів (вузлових многовидів) розв'язків рівняння Сена-Віттена, яке є проекцією на тривимірний простір рівняння Дірака. Спростовуючи висловлене творцями методу припущення про обумовленість існування нулів топологією тривимірного простору та самовизнання внаслідок цього неповноцінності методу у порівнянні із спінорним, у наших роботах доведено, що існування вузлових многовидів залежить від коефіцієнтів рівняння та граничних умов, і встановлено умови їх відсутності. Це дозволило довести існування у асимптотично плоскому просторі широкого класу гіперповерхонь, у всіх точках яких існує $2 \leftrightarrow 1$ відповідність між спінором Сена-Віттена і деяким ортонормованим 3-репером, який узагальнює SOF, і співпадає з ним у випадку максимальних гіперповерхонь; ми назвали його ортонормованим репером Сена-Віттена. Цим, з одного боку, ми обґрунтували тензорний метод Нестера в асимптотично плоскому просторі, а, з іншого, довели виключно геометричний характер спінорного поля Сена-Віттена у цьому просторі. Ключовим елементом усіх доведень теореми про додатність повної енергії гравітаційного поля є умова енергодомінантності, але ми довели, що порушення темною енергією умови енергодомінантності не порушує додатної визначеності повної гравітаційної енергії.

Актуальним завданням є розширення класу просторів, у яких виконується SG і існує SOF. Це важливо з точки зору реалізації концепції Пенроуза квазілокалізації енергії, зокрема, у космологічних просторах, також з точки зору numerical relativity – числових методів у загальній теорії відносності (серед іншого, моделювання злиття чорних дір), а також з точки можливості трактування у будь-якому просторі поля Сена-Віттена як калібрувального у калібрувальній супергравітації.

Дж. Нестер разом із співавторами [10], відмовившись від заперечень проти нашого висновку про обумовленість нулів рівняння Сена-Віттена властивостями самого рівняння і граничних умов та використавши цю нашу ідею, описали граничні умови, які забезпечують відсутність нулів і існування SG на максимальній глобально конформно-плоскій просторово-подібній гіперповерхні. Однак такі умови на гіперповерхню є надзвичайно обтяжливими, оскільки вони не виходять поза межі найпростіших множин початкових даних (плоских, Нордстрема-Рейснера, Брілла-Ліндквіста), не охоплюючи важливий випадок даних Керра (автори помилково тверджять протилежне, див., наприклад, [11-13]).

У пошуку підходів до розширення класу просторів, у яких рівняння Сена-Віттена не має вузлових ліній, ми реалізуємо концепцію розщеплення, встановлюючи необхідні і достатні умови розщеплення.

Рівняння Сена-Віттена (SWE) в обмеженій замкнутій області $\Omega$ на просторовоподібній гіперповерхні $\Sigma_t$ із 1-формою нормалі $dt$ у чотиривимірному многовиді $M = \Sigma \times R$ у формалізмі просторових спінорів Соммерса-Сена має вигляд

$$D^B{}_C \beta^C = 0, \qquad (2)$$

де дія оператора $D_{AB}$ на спінорні поля визначається так:

$$D_{AB}\lambda_C = \nabla_{AB}\lambda_C + \frac{\sqrt{2}}{2} K_{ABC}{}^D \lambda_D,$$
$$\nabla_{AB}\lambda_C = (\sqrt{2}) t_{(A}{}^{A'} \nabla_{B)A'} \lambda_C$$

Оператор $D_{AB}$ відомий під назвою оператора (зв'язності) Сена, $K_{ABCD}$ – спінорний тензор зовнішньої кривини $\Sigma_t$.

Із класичної теореми для системи рівнянь першого порядку випливає, що вузлові многовиди рівняння (2) розмірності 2 є відсутніми, однак можуть існувати вузлові лінії і точки. Для їх дослідження ми розглядаємо диференціальний наслідок рівняння Сена-Віттена

$$D_{AB} D^B{}_C \beta^C = 0$$

і доводимо відсутність вузлових точок рівняння Сена-Віттена в обмеженій замкнутій області $\Omega$ на просторовоподібній гіперповерхні $\Sigma_t$ при виконанні наступних умов:

a) $\frac{1}{2}K^2 + \frac{1}{8}K_{пр}K^{пр} \geq -\mu$ або
$\frac{1}{2}K^2 + \frac{1}{8}K_{пр}K^{пр} \leq -\mu$, $\frac{1}{2}(K^2 + \frac{1}{8}K_{пр}K^{пр} + \mu)^2 - \frac{1}{16}[(\partial_1 K)^2 + (\partial_2 K)^2 + (\partial_3 K)^2] \geq 0$.

b) спінор Річчі гіперповерхні в області $\Omega$ задовольняє умову $R_{(A/C/B)D}=0$;

c) $\operatorname{Re}\lambda^0_\infty$ або $\operatorname{Im}\lambda^0_\infty$ в жодній точці межі області $\Omega$ не обертаються в нуль.

Виконання умови b) забезпечує локальне існування паралельного у зв'язності Сена спінорного поля, а існування останнього – розщеплення системи чотирьох рівнянь відносно дійсної і уявної частини компонент спінорного поля на незалежні рівняння. Паралельні та кілінгові спінорні поля є предметом інтенсивних досліджень фізиків і геометрів у зв'язку із їх значенням у струнних і бранних теоріях, однак вперше запроваджено поле, паралельне у зв'язності Сена. Завдяки цьому умовою існування поля не є достатньо тривіальна з фізичної точки зору умова Річчі-площинності.

За вказаних умов в області $\Omega$ існує відповідність між спінорним полем Сена-Віттена і полем спеціального ортонормованого репера (гіпотеза енергодомінантності не зобов'язана виконуватись) [14]. Умови а)–с) не виокремлюють певний алгебраїчно спеціальний простір; зокрема, вони не вимагають, щоб простір належав до типу $O$, до якого належить конформно-плоский простір. Отриманий результат є підтвердженням калібрувального характеру поля Сена-Віттена за вказаних умов і створює основу для продовження розвитку теорії квазілокалізації гравітаційної енергії у космологічних моделях.

# Decoupling method for investigation of system of Maxwell and Sen-Witten equations

**V.O. Pelykh[1], Y.V. Taistra[1], B.S. Novosyadlyj[2]**

[1]*Ya.S. Pidstryhach Institute for Applied Problems of Mechanics and Mathematics
Naukova str., 3-b, Lviv, 79060, Ukraine*
[2]*Astronomical Observatory of Ivan Franko National University of Lviv
Kyryla i Methodia str., 8, Lviv, 79005, Ukraine*

Method for obtaining decoupled second order equations for electromagnetic field and Sen-Witten equation is proposed, gauge character of Sen-Witten field on spaces of defined type is proven (in Ukrainian).

# Явление гравитационного линзирования в применении к поиску темной материи во Вселенной

**В.С. Цветкова[1], В.М. Шульга[1], А.А. Минаков[1], В.Г. Вакулик[1,2]**

[1]*Радиоастрономический институт НАН Украины, 61002 Харьков, Украина*
[2]*НИИ Астрономии Харьковского национального университета им. В.Н. Каразина, 61077 Харьков, Украина*

Излагаются основные результаты, полученные объединенной харьковской группой (РИ НАНУ и НИИ Астрономии) в рамках программы НАН Украины «Астрофизические и космологические проблемы скрытой массы и темной энергии Вселенной (Космомикрофизика-2)». Кратко изложены направления теоретических исследований различных аспектов гравитационного линзирования и приводятся результаты наблюдений конкретных гравитационно линзовых систем, измерения временных запаздываний между колебаниями собственного блеска квазара, оценки значения постоянной Хаббла, содержания темной материи в линзирующих галактиках.

## 1. ВВЕДЕНИЕ

Наиболее естественный путь изучения темной материи – поиск проявлений ее гравитационного взаимодействия с обычными объектами Вселенной. Наблюдения объектов, порожденных гравитационным линзированием, позволяют обнаруживать присутствие темной материи и изучать ее обилие и пространственное распределение на различных масштабах Вселенной – от отдельных компактных объектов до скоплений галактик.

Различают три типа гравитационного линзирования: ближнее, или галактическое, сильное и слабое. Основное внимание в работе уделялось теоретическим исследованиям явления сильного гравитационного линзирования и наблюдениям квазаров, изображения которых деформированы гравитационными полями близлежащих галактик – «гравитационно линзированных квазаров» (ГЛК).

Сильное линзирование наблюдается в случаях, когда гравитационное поле галактики (или группы галактик), расположенной вблизи луча зрения «квазар-наблюдатель», формирует либо двойное, либо кратное, либо дугообразное (кольцевое) изображение удаленного квазара. В настоящее время обнаружено порядка двух сотен гравитационно линзированных квазаров (ГЛК) и ежегодно открывается несколько новых.

ГЛК – переменные объекты: изменяется блеск макроизображений, их показатели цвета и спектры. Причины изменений связаны с собственной переменностью квазаров и с событиями микролинзирования, которые вызываются движениями компактных тел, составляющих население линзирующей галактики. Анализ переменных явлений в ГЛК – источник информации для решения таких задач, как 1) получение оценок характерных масс компактных объектов, вызывающих события микролинзирования; 2) изучение пространственной структуры квазаров; 3) получение оценок постоянной Хаббла и/или уточнения распределения гравитационного потенциала в галактике-линзе; 4) обнаружение темной материи в линзирующей галактике и оценка ее относительного содержания, и др.

При выполнении работ в рамках научной Программы Национальной академии наук Украины и Космического агентства Украины «Исследования структуры и состава Вселенной, скрытая масса и темная энергия» выполнялись теоретические исследования различных аспектов гравитационного линзирования и наблюдения ряда гравитационно линзированных квазаров на 1.5-м телескопе АЗТ-22 Высокогорной обсерватории на горе Майданак (Узбекистан).

## 2. ТЕОРЕТИЧЕСКИЕ ИССЛЕДОВАНИЯ И РАЗРАБОТКА СМЕЖНЫХ ПРОБЛЕМ

Основной вклад в развитие теоретических исследований феномена гравитационного линзирования был сделан ныне покойными В.Г. Вакуликом и А.А. Минаковым, руководителем работ в РИ НАНУ до 2012 года. В 2010 году была издана их монография «Статистический анализ гравитационного микролинзирования» [1], в которой подведены итоги их многолетней совместной работы. Основная цель монографии – исследование эффекта микролинзирования в самом общем случае, в том числе при критических значениях параметров распределения массы.

В монографии показано, в частности, что анализ микролинзирования можно достаточно просто выполнить на основе известной аналогии с задачами электродинамики сплошных сред. Исследованы структуры критических и каустических поверхностей, обусловленные взаимодействием полей макролинзы-галактики и микролинз-звезд. Введено понятие эффективной апертуры линзы, с помощью которого показана связь между структурой наблюдаемого изображения протяженного источника и коэффициентом усиления его блеска. Для ряда задач впервые получены аналитические решения.

Предсказано и детально исследовано возникновение неизвестного ранее эффекта, связанного с действием поля тяготения массивного небесного тела на излучение удаленного переменного во времени источника и аналогичного действию пространственно-временного фильтра, пропускающего медленные вариации блеска источника и сглаживающего быстрые.

Разработан новый обобщенный метод фазового экрана и с его помощью рассмотрено действие сложной линзы (макролинза-галактика и микролинза) [2]. Впервые получено аналитическое решение для коэффициента усиления при произвольном расположении микролинзы вдоль трассы «источник – макролинза – наблюдатель».

В 2011 году работы А.А. Минакова и В.Г. Вакулика по развитию статистических методов исследования эффектов микролинзирования была удостоена премии НАН Украины имени академика С.Я. Брауде.

## 3. СИЛЬНОЕ ЛИНЗИРОВАНИЕ: ТРИ ПОДХОДА К ПОИСКУ ТЕМНОЙ МАТЕРИИ

Изучение изображений удаленных квазаров, сформированных действием гравитационного поля близлежащей галактики, расположенной вблизи луча зрения квазар-наблюдатель, предоставляет несколько методов обнаружения в линзирующей галактике, или где-либо вдоль трассы распространения излучения, темной материи или ее субструктур. Ниже кратко рассматриваются три наиболее часто используемых метода. Исходным материалом для них являются кривые блеска компонентов ГЛК, полученные в результате длительного мониторинга.

*Метод временных запаздываний.* Первоначально метод был предложен [3] в качестве способа получения оценок величины постоянной Хаббла, независимых от каких-либо вспомогательных индикаторов расстояний. Принципиально идея метода проста. Действительно, из-за различия геометрических длин путей для лучей, соответствующих различным макроизображениям, колебания собственного блеска квазара наблюдаются в них с некоторыми запаздываниями, $\tau_i$ и $\tau_j$. Из наблюдений мы можем получить только относительные запаздывания, $\Delta t_{ij} = \tau_i - \tau_j$, так как величины абсолютных запаздываний неизвестны. Несложные выкладки позволяют получить следующее выражение для зависимости величины постоянной Хаббла $H_0$ от измеренных значений временных запаздываний и параметров системы:

$$H_0 = \frac{1+z_L}{\Delta t_{ij}} \frac{\tilde{D}_L \tilde{D}_S}{\tilde{D}_{LS}} \cdot \left\{ \frac{1}{2} \left[ (\theta_i - \beta)^2 - (\theta_j - \beta)^2 \right] - \left[ \psi(\theta_i) - \psi(\theta_j) \right] \right\}.$$

Здесь $\beta$ и $\theta_i$, $\theta_j$ – угловые координаты источника и двух его изображений, соответственно, $z_L$ – красное смещение линзы, $\psi$ – распределение гравитационного потенциала линзы, $D_L$, $D_S$ и $D_{LS}$ – приведенные расстояния угловых диаметров до линзы, квазара и расстояние между ними (нормированы на радиус Хаббла $r_H=c/H_0$).

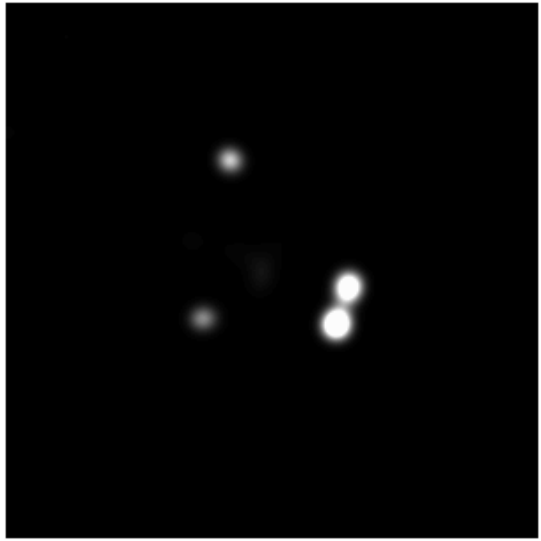

Рис. 1. PG 1115+080 - ГЛК типа «источник вблизи складки каустики». Изображения получены на 1.5-м телескопе Высокогорной обсерватории на горе Майданак. Максимальное расстояние между компонентами ~2".5, между компонентами тесной пары ~0."5.

При измеренной разности времен запаздывания $\Delta t_{ij}$ величина $H_0$ может быть получена: 1) в рамках принятой космологической модели (содержится во втором сомножителе); 2) по наблюдаемой геометрии системы (измеренные $\beta$ и $\theta_i$, $\theta_j$); 3) при известных красных смещениях линзы и источника; 4) с использованием физически обоснованной модели распределения массы в линзирующей галактике (функция $\psi$).

Задачу можно инвертировать: будем искать модель линзирующей галактики, которая бы удовлетворяла, для измеренных запаздываний, наиболее надежному значению $H_0$, полученному другими методами (например, по Цефеидам [4]). Это – путь к изучению распределения общей массы в галактике-линзе – как светящейся, так и темной.

*Метод аномалий блеска.* Об аномалиях относительного блеска говорят в случаях, когда наблюдаемые отношения блеска макроизображений не воспроизводятся никакими регулярными (гладкими) моделями галактики-линзы. В 1998 году было высказано предположение [5], что аномалии блеска могут вызываться наличием субструктур темной материи в линзе или где-либо на пути распространения излучения

Для квадрупольно линзированного квазара типа «источник вблизи складки каустики» (рис. 1) теория предсказывает практически одинаковую яркость для компонентов тесной пары макроизображений. В работе [6] .исследовано асимптотическое поведение отношения блеска компонентов такой тесной пары, которое позволяет учесть вклад геометрического фактора (степени близости источника к каустике) в наблюдаемое отношение блеска и оценить таким образом составляющую, обусловленную возможным наличием неоднородности темной материи (субструктуры).

*Распределения вероятностей усиления (метод гистограмм).* Основная идея метода изложена в работе 2002 года [7]. Представленные в ней результаты моделирования событий микролинзирования при различных соотношениях между массой диффузно распределенной (темной) материи и массой в компактных объектах показывают, что распределения вероятностей усиления при микролинзировании чувствительны к относительной доле темной (диффузно распределенной) материи в общей поверхностной плотности массы линзы.

## 4. РЕЗУЛЬТАТЫ НАБЛЮДЕНИЙ ИЗБРАННЫХ ГЛК

В процессе выполнения программы «Космомикрофизика-2» были использованы все три перечисленных подхода к проблеме поиска темной материи во Вселенной. Анализировались данные наблюдений, выполненных на 1.5-м телескопе АЗТ-22 Высокогорной обсерватории на горе Майданак (Узбекистан). Это единственный телескоп и единственный астропункт на территории бывшего Союза, где возможны наблюдения ГЛК. Сложность наблюдений этих объектов обусловлена их чрезвычайной компактностью, низкой поверхностной яркостью и высокими требованиями к фотометрической точности исходного материала.

*PG 1115+08; временные запаздывания и аномалии блеска.* Система удобна как для измерения временных запаздываний, так и для анализа аномалии отношения блеска компонентов тесной пары макроизображений (рис. 1). Впервые временные запаздывания были измерены в 1997 году [8]. Других регулярных наблюдений этого объекта с тех пор не было.

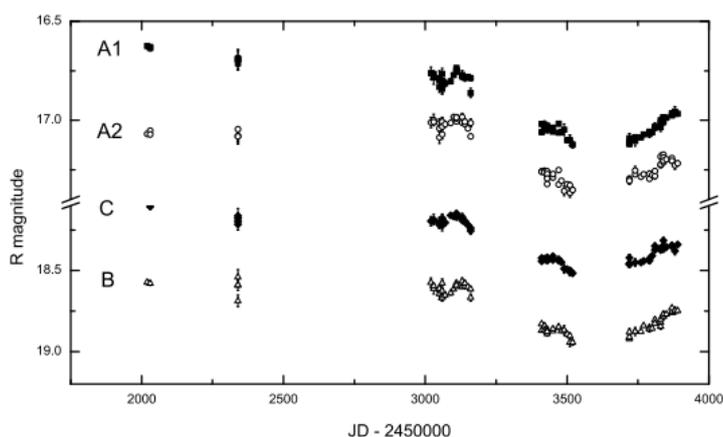

Рис. 2. Кривые блеска ГЛК PG 1115+080, охватывающие период времени с 2001 по 2006 год.

Результаты наших наблюдений [9], использованных при определении временных запаздываний, охватывают, по сравнению с исходными данными работы [10], больший период времени (три последних сезона на рис. 2), при заметно большей амплитуде изменений блеска, что чрезвычайно важно для обеспечения надежности измерений временных запаздываний. В таблице 1 приведено сравнение наших результатов с результатами [8] для пары BC, для которой временные запаздывания измеряются наиболее уверенно.

Таблица 1. Временные запаздывания для пары изображений BC и соответствующие значения постоянной Хаббла, вычисленные с использованием для галактики модели изотермической сферы с внешним shear (ISES).

| Schechter et al. [10] | $\Delta t_{BC} \approx 25d$ | **ISES** model | $H_0 = 44$ km s$^{-1}$Mpc$^{-1}$ |
| Our data [11] | $\Delta t_{BC} \approx 16.4d$ | **ISES** model | $H_0 = 63$ km s$^{-1}$Mpc$^{-1}$ |

Выполненный нами анализ позволяет утверждать, что оценки запаздываний в [10], по-видимому, ошибочны. Наши оценки [9] получены на более качественном наблюдательном материале, и к тому же, как видно из таблицы, обеспечивают более реалистичные значения постоянной Хаббла $H_0$ при использовании изотермической модели распределения массы в линзирующей галактике, которая считается наиболее вероятной и предполагает наличие гало темной материи.

Данные наблюдений за 2001-2006 гг., приведенные на рис. 2, были использованы также для анализа аномалий блеска, отмечавшихся ранее другими авторами. Наш анализ [10] показывает, что наблюдаемое отклонение от единицы отношения блеска в тесной паре макроизображений невелико и объясняется событиями микролинзирования. Это не исключает возможного присутствия субструктуры темной материи, но накладывает ограничение на верхний предел ее массы: $m_{sub} < 10^4 M_\odot - 10^5 M_\odot$.

Помимо ГЛК PG 1115+080, временные запаздывания были измерены также в системах Q2237+0305, H1413+117, SDSS 0909+532, UM 673, причем, для систем Q2237+0305, SDSS 0909+532 и UM 673 – впервые.

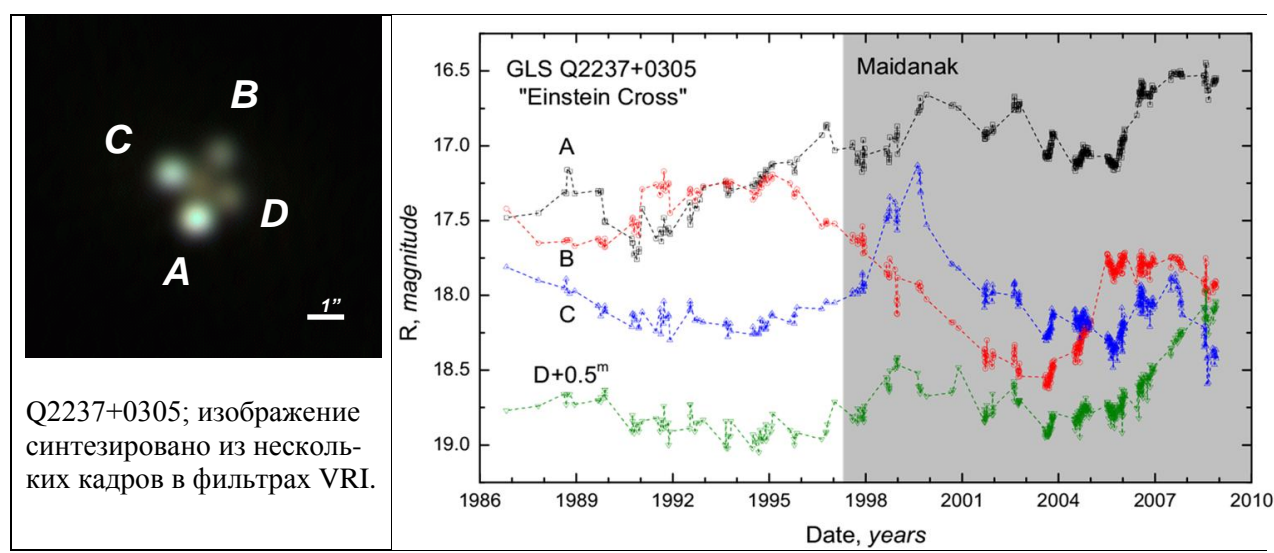

Рис. 3. Слева – изображение одного из наиболее эффектных проявлений гравитационного линзирования, синтезированное из ПЗС-кадров, полученных на 1.5-м телескопе на г. Майданак. Справа – сводные кривые блеска в полосе R. Более темным фоном отмечены наши наблюдения.

***Q2237+0305 (Крест Эйнштейна); оценка содержания темной материи.*** Из всех известных ГЛК линзирующая галактика системы Q2237+0305 – самая близкая к наблюдателю. Компактность системы и близость галактики делают этот объект уникальным для изучения событий микролинзирования. На кривых блеска (рис. 3) переменность квазара проявляется в виде синхронных изменений блеска, микролинзирования же вызывают нескоррелированные вариации блеска.

Для оценки содержания темной (диффузно распределенной) материи в линзирующей галактике ГЛК Q2237+0305 были построены, с использованием всех имеющихся данных фотометрии этой системы (рис. 3), распределения плотностей вероятностей усиления блеска вследствие событий микролинзирования. Здесь следует отметить, что при интерпретации данных наблюдений эффектов гравитационного линзирования корректная формулировка и решение обратной задачи часто оказываются невозможными, и поэтому оценка параметров источников излучения и распределения полной массы линзирующих объектов выполняется решением прямых модельных задач.

Поэтому далее наблюдаемые гистограммы сравнивались с распределениями вероятностей усиления, полученными по результатам компьютерного моделирования, выполненного при различных соотношениях между поверхностной плотностью массы галактики в непрерывно распределенной (темной) материи и в компактных объектах.

Проведенный анализ позволяет утверждать, что доля темной материи в общей поверхностной плотности массы в системе Q2237+0305 не превышает 50% на расстоянии от центра галактики ~0.8 Кпс, где наблюдаются линзированные изображения квазара [11].

## ЛИТЕРАТУРА

# The phenomenon of gravitational lensing as applied to the search of dark matter in the Universe

V.S. Tsvetkova[1], V.M. Shulga[1], A.A. Minakov[1], V.G. Vakulik[1,2]

[1]*Institute of Radio Astronomy of Nat. Ac. Sci. of Ukraine, 61002 Kharkiv, Ukraine*
[2]*Institute of Astronomy of V.N. Karazin Kharkiv National University, 61077 Kharkiv, Ukraine*

Principal results obtained by a joint Kharkiv group (IRA NASU and IA of KhNU) within the framework of the NASU program "Astrophysical and cosmological problems of the hidden mass and dark energy in the Universe (Cosmomicrophysics-2)" are presented. Theoretical studies of various aspects of gravitational lensing are briefly described, and some results of observations of gravitational lens systems are presented, such as measurements of the time delays between the intrinsic quasar brightness variations, estimates of the Hubble constant value and the dark matter content in lensing galaxies (in Russian).

# Скалярно-польові моделі темної енергії: властивості та спостережувані обмеження

**Б.С. Новосядлий[1], О.М. Сергієнко[1], Ю.А. Кулініч[1], В.О. Пелих[2]**

[1]*Астрономічна обсерваторія Львівського національного університету імені Івана Франка, вул. Кирила і Мефодія, 8, Львів, 79005, Україна*

[2]*Інститут прикладних проблем механіки і математики ім. Я. Підстригача НАН України, вул. Наукова, 3-б, Львів, 79060, Україна*

Досліджено властивості квінтесенційних та фантомних скалярних полів з баротропним рівнянням стану, їх вплив на динаміку розширення та формування великомасштабної структури Всесвіту. Визначено найбільш оптимальні параметри космологічних моделей та моделей темної енергії на основі сучасних спостережуваних даних. Показано здатність квінтесенційних і фантомних скалярних полів практично з однаковою точністю узгоджуватися з ключовими спостережуваними даними. Обговорюються можливості розрізнення космологічних моделей з різними типами темної матерії та темної енергії за даними сучасних галактичних оглядів неба, вимірювань анізотропії реліктового випромінювання та томографії епохи Темних Віків в лінії нейтрального водню 21 см, зміщеною в декаметровий діапазон довжин хвиль.

## 1. ВЛАСТИВОСТІ ТА ПАРАМЕТРИ СКАЛЯРНО-ПОЛЬОВИХ МОДЕЛЕЙ ТЕМНОЇ ЕНЕРГІЇ

Припускається, що Всесвіт заповнений нерелятивістськими частинками (холодна темна матерія та баріони), релятивістськими (фотони, активні безмасові нейтрино) та динамічною темною енергією (ТЕ) у формі скалярного поля із заданим лагранжіаном, яке взаємодіє з іншими компонентами тільки гравітаційно (мінімально зв'язане). Властивості останньої компоненти задавались наступним чином: а) лагранжіан поля канонічний, $L_{de}=\pm X - U(\varphi)$ ($X$ – позитивно визначений кінетичний член, $U$ – потенціал самодії поля $\varphi$), із знаком "+" для квінтесенційного скалярного поля (КСП) і "–" для фантомного скалярного поля (ФСП), так що ефективна швидкість звуку $c_s^2{}_{(de)} \equiv \delta p_{de}/\delta\rho_{de}$ (швидкість поширення збурень скалярного поля) дорівнює швидкості світла $c$; б) рівняння стану $p_{de}=w_{de}c^2\rho_{de}$, де параметр рівняння стану $w_{de}$ є функцією часу; в) відношення швидкостей зміни тиску і густини енергії поля є сталою величиною, яку позначаємо $c_a^2$.

Рівняння динаміки розширення однорідного та ізотропного Всесвіту отримуються з рівнянь Айнштайна та рівнянь збереження енергії-імпульсу, а рівняння еволюції збурень метрики та густини усіх компонент – з рівнянь Айнштайна-Больцмана [1-6]. Всі викладки та розрахунки, приведені в [2-6], здійснювались в моделях з метрикою Фрідмана-Робертсона-Уокера (фонової) космологічної моделі з нульовою кривиною 3-простору $ds^2 = g_{ij}dx^i dx^j = a^2(\eta)(d\eta^2 - \delta_{\alpha\beta}dx^\alpha dx^\beta)$, де $\eta$ – конформний час, $a$ – масштабний множник, нормований на 1 в сучасну епоху. Далі ми покладатимемо $c=1$, так що час має розмірність довжини. В такому підході диференціальні закони збереження дають прості аналітичні залежності $w_{de}$ і $\rho_{de}$ від масштабного множника $a$,

$$w_{de} = \frac{(1+c_a^2)(1+w_0)}{1+w_0-(w_0-c_a^2)a^{3(1+c_a^2)}} - 1, \quad \rho_{de} = \rho_{de}^{(0)}\frac{(1+w_0)a^{-3(1+c_a^2)}+c_a^2-w_0}{1+c_a^2}, \quad (1)$$

а рівняння Айнштайна – аналітичні вирази для параметра Габбла і параметра сповільнення

$$H = H_0\sqrt{\Omega_r a^{-4}+\Omega_m a^{-3}+\Omega_{de}f(a)}, \quad q = \frac{1}{2}\frac{2\Omega_r a^{-4}+\Omega_m a^{-3}+(1+3w_{de})\Omega_{de}f(a)}{\Omega_r a^{-4}+\Omega_m a^{-3}+\Omega_{de}f(a)}, \quad (2)$$

де $f(a)=\rho_{de}/\rho_{de}^{(0)}$, $\rho_{de}^{(0)}$ - сучасне значення густини ТЕ, $w_0$ – параметр рівняння стану $w_{de}$ в сучасну епоху, $H_0$ – стала Габбла в сучасну епоху, $\Omega_r=\rho_r^{(0)}/\rho_{tot}^{(0)}$, $\Omega_m=\rho_m^{(0)}/\rho_{tot}^{(0)}$, $\Omega_{de}=\rho_{de}^{(0)}/\rho_{tot}^{(0)}$, $\Omega_{dm}=\rho_{dm}^{(0)}/\rho_{tot}^{(0)}$, $\Omega_b=\rho_b^{(0)}/\rho_{tot}^{(0)}$, $\Omega_m=\Omega_b+\Omega_{dm}$. У космологічній моделі з нульовою кривиною 3-простору $\Omega_{de}+\Omega_{dm}+\Omega_{b}+\Omega_{r}=1$, а $\rho_{tot}^{(0)}=3H_0^2/8\pi G$. Динаміка розширення Всесвіту визначається його складовими, зокрема, якщо $|(1+3w_{de})\Omega_{de}f(a)|>\Omega_m a^{-3}$ і $w_{de}<-1/3$, то він розширюється з прискоренням, що випливає із другого виразу у (2). Темп зміни швидкості розширення в минулому і майбутньому, час переходу від сповільненого розширення до прискореного, зрештою, майбутня доля і час існування Всесвіту із заданими $H_0$, $\Omega_m$ і кривиною 3-простору цілком визначаються значеннями трьох величин: $\Omega_{de}, w_0$ і $c_a^2$.

У випадку квінтесенційного поля ($-1<w_0<-1/3$, $-1< c_a^2<0$) з постійним параметром рівняння стану ($w_0=c_a^2$) Всесвіт у майбутньому розширюватиметься вічно за степеневим законом, $a\sim t^{2/3(1+w_0)}$. У випадку $w_0<c_a^2$ в майбутньому $w_{de}\to -1$ і Всесвіт розширюватиметься вічно за експоненціальним законом $a\sim exp(H_\infty t)$, де $H_\infty =[\Omega_{de}(c_a^2-w_0)/(1+c_a^2)]^{1/2}H_0$ - асимптотичне значення сталої Габбла. Цікавим є також протилежний випадок, коли $w_0>c_a^2>-1$. В такій моделі квінтесенційного скалярного поля прискорене тепер розширення Всесвіту в майбутньому зміниться сповільненим, Всесвіт досягне максимуму розширення, почне стискатися і закінчить своє існування сингулярністю Великий Тріск (Big Crunch). Детально властивості таких моделей проаналізовано в [2].

У випадку фантомного поля ($w_0$, $c_a^2<-1$) у майбутньому Всесвіт розширюватиметься надшвидко так, що досягне сингулярності Великий Розрив (Big Rip) за скінчений час

$$t_{BR} - t_0 \approx \frac{2}{3}\frac{1}{H_0}\frac{1}{|1+c_a^2|}\sqrt{\frac{1+c_a^2}{(1+w_0)\Omega_{de}}},$$

який визначається параметрами поля $\Omega_{de}$, $w_0$ і $c_a^2$. Використовуючи співвідношення (1) і (2), для моделі із заданими $\Omega_{de}$, $\Omega_{dm}$, $\Omega_b$, $w_0$, $c_a^2$ і $H_0$ можна розрахувати залежності "відстань за світністю – червоне зміщення" та "відстань за кутовими розмірами – червоне зміщення" для визначення величин цих параметрів оптимізаційними методами чи методами Монте Карло шляхом порівняння розрахованих відстаней для різних червоних зміщень із спостережуваними для "стандартних свічок" (наднові типу Ia, гама-барстери чи ін.) та "стандартних лінійок" (акустичні піки в спектрах потужності збурень температури реліктового випромінювання та неоднорідностей просторового розподілу галактик, рентгенівські скупчення галактик чи ін.).

Важливим є аналіз еволюції збурень густини ТЕ та їх впливу на формування великомасштабної структури Всесвіту. Він здійснений в рамках традиційної парадигми формування великомасштабної структури Всесвіту та її елементів: а) галактики та неоднорідності їх просторового розподілу сформувались шляхом гравітаційного росту початкових адіабатичних збурень метрики простору-часу, густини та швидкості речовини, згенерованих в дуже ранньому Всесвіті; б) спектр потужності початкових збурень густини релятивістської та нерелятивістської складових є степеневим: $P_i(k)=A_s k^{ns}$, де $A_s$ і $n_s$ – початкова амплітуда і спектральний індекс скалярної моди збурень густини ($k$ – хвильове число). Скалярне поле у Всесвіті із збуреною матеріальною складовою (фонове теплове електромагнітне випромінювання, темна матерія та баріонна складова) не може бути ідеально однорідним, воно є збуреним як внаслідок гравітаційного впливу збурень інших складових, так і власних збурень, що були згенеровані у ранньому Всесвіті і за амплітудою та спектром у загальному випадку можуть бути відмінними від збурень матеріальної складової. Система лінійних диференціальних рівнянь еволюції збурень квінтесенційного/фантомного скалярного поля в синхронній супутній до частинок темної матерії (ТМ) системі відліку приведена в [2-6]. В цих роботах вивчена еволюція збурень квінтесенційного і фантомного скалярних полів та їх вплив на еволюцію збурень спостережуваної матеріальної складової. Показано, що спостережувані дані про великомасштабну структуру Всесвіту в різних просторово-часових масштабах можна використовувати разом з даними про відстані до "стандартних свічок" та "стандартних лінійок" для визначення параметрів космологічних моделей і параметрів ТЕ та довірчих інтервалів їх значень. Крім того, показано, що в моделях з фантомною ТЕ лінійні збурення густини матерії на великих масштабах в майбутньому загасатимуть [5].

Ключовими спостережуваними даними для визначення космологічних параметрів крім модулів відстані до наднових типу Ia є анізотропія реліктового випромінювання, розрахунок якої за допомогою програми CAMB [19,20] потребує задання ще одного параметра – оптичної глибини за томсонівським розсіюванням $\tau_{dec}$. Таким чином, модель має 9 параметрів, $\Omega_{de}$, $\Omega_{dm}$, $\Omega_b$, $w_0$, $c_a^2$, $H_0$, $A_s$, $n_s$ і $\tau_{dec}$, значення яких повинні бути визначені шляхом узгодження теоретично розрахованих характеристик динаміки розширення та великомасштабної структури Всесвіту із спостережуваними даними. Такими на сьогодні є: спектри потужності флюктуацій температури та поляризації реліктового випромінювання, отримані в експерименті WMAP [7,8,9]; баріонні акустичні осциляції (ВАО) в спектрі потужності просторового розподілу галактик, отриманому в автоматичному огляді галактичного неба SDSS [10]; визначення сталої Габбла за допомогою космічного телескопа імені Габбла (HST) [11]; визначення вмісту баріонної компоненти за вмістом дейтерію в міжгалактичному середовищі на основі моделі космологічного нуклеосинтезу [12,13]; модулі відстаней до наднових типу Ia з компіляції SDSS (SN SDSS) [14], визначені за допомогою методів підгонки кривих блиску SALT2 [15] чи MLCS2k2 [16]. Оскільки обговорення цих методів не виявило явних переваг одного з них, ми визначали параметри для двох наборів даних: WMAP7 + HST + BBN + BAO + SN SDSS SALT2 (набір даних 1) та WMAP7 + HST + BBN + BAO + SN SDSS MLCS2k2 (набір даних 2). Для цього було використано пакет програм CosmoMC [17,18], який включає програму CAMB, модифікований нами для врахування особливостей наших моделей, відображених в (1)-(2) та рівняннях для еволюції збурень ТЕ [2,5]. Для знаходження найбільш оптимальних параметрів $w_0$ і $c_a^2$ було використано умови $-1<w_0<0$ і $-1<c_a^2<0$ для КСП та $-2<w_0<-1$ і $-2<c_a^2<-1$ для ФСП. Слід відзначити, що кількість ступенів свободи ($N_{DoF}$) для одного і того ж набору даних є однаковою для обох полів, що є важливим для їх порівняння з точки зору кращого узгодження теоретично розрахованих і спостережуваних величин.

Таблиця 1.

| Параметри | $q_1$ | $p_1$ | $q_2$ | $p_2$ |
|---|---|---|---|---|
| $\Omega_{de}$ | $0.730^{+0.022}_{-0.035}$ | $0.723^{+0.030}_{-0.027}$ | $0.702^{+0.029}_{-0.036}$ | $0.692^{+0.040}_{-0.026}$ |
| $w_0$ | $-0.996^{+0.089}_{-0.004}$ | $-1.043^{+0.043}_{-0.14}$ | $-0.83^{+0.087}_{-0.17}$ | $-1.002^{+0.002}_{-0.095}$ |
| $c_a^2$ | $-0.022^{+0.022}_{-0.978}$ | $-1.12^{+0.12}_{-0.46}$ | $-0.88^{+0.59}_{-0.12}$ | $-1.19^{+0.19}_{-0.37}$ |
| $10\omega_b$ | $0.226^{+0.011}_{-0.010}$ | $0.223^{+0.011}_{-0.009}$ | $0.226^{+0.012}_{-0.010}$ | $0.223^{+0.010}_{-0.009}$ |
| $\omega_{cdm}$ | $0.110^{+0.007}_{-0.009}$ | $0.115^{+0.008}_{-0.007}$ | $0.108^{+0.011}_{-0.008}$ | $0.119^{+0.007}_{-0.008}$ |
| $h$ | $0.702^{+0.019}_{-0.035}$ | $0.704^{+0.032}_{-0.021}$ | $0.663^{+0.031}_{-0.027}$ | $0.678^{+0.035}_{-0.017}$ |
| $n_s$ | $0.974^{+0.025}_{-0.027}$ | $0.965^{+0.024}_{-0.024}$ | $0.971^{+0.031}_{-0.022}$ | $0.965^{+0.020}_{-0.027}$ |
| $\log(10^{10} A_s)$ | $3.085^{+0.067}_{-0.071}$ | $3.089^{+0.069}_{-0.061}$ | $3.069^{+0.084}_{-0.055}$ | $3.113^{+0.049}_{-0.078}$ |
| $\tau_{rei}$ | $0.091^{+0.024}_{-0.025}$ | $0.085^{+0.025}_{-0.021}$ | $0.089^{+0.027}_{-0.023}$ | $0.086^{+0.021}_{-0.024}$ |
| $-\log L$ | 3865.01 | 3864.86 | 3857.21 | 3859.30 |

Результати визначень найбільш оптимальних значень параметрів та їх довірчих інтервалів для обох наборів даних та типів скалярних полів представлені в таблиці 1, де $q_1$ і $q_2$ - параметри космологічної моделі з КСП для 1-го і 2-го наборів даних відповідно, а $p_1$ і $p_2$ - параметри космологічної моделі з ФСП для тих самих наборів даних. Верхня і нижня цифри справа біля найбільш оптимальних значень космологічних параметрів відображають $2\sigma$ маргіналізовані межі кожного з них. В нижньому рядку наведені логарифми максимумів функції правдоподібності, взяті зі знаком мінус, що дає $\chi^2_{min}/2$. Бачимо, що для 1-го набору $\chi^2_{min}$ є дещо меншим для моделі з фантомним скалярним полем ($p_1$), ніж з квінтесенційним ($q_1$). Для 2-го набору даних навпаки: $\chi^2_{min}(q_2)<\chi^2_{min}(p_2)$.

Тобто, набір даних 1 віддає перевагу фантомному скалярному полю, а набір даних 2 – квінтесенційному. Слід відзначити, що різниця -$log(L_{max})$ для моделей з фантомним і квінтесенційним скалярними полями для обох наборів даних є статистично незначною і можна було б знехтувати цією різницею і стверджувати, що цими даними вони не розрізняються. Але є ще один аргумент: в роботі [1] ми показали, що для моделей з космологічною сталою ($w_0=c_a^2=-1$) величина -$log(L_{max})$ знаходиться між відповідними величинами для КСП і ФСП для обох наборів даних. Ця стійка тенденція вказує на те, що різниця не є випадковою, а найбільш оптимальні параметри дійсно відповідають максимумам функції правдоподібності. Розраховані модулі відстані для наднових типу Ia, параметри баріонних осциляцій, спектр збурень густини матерії та спектр флюктуацій температури реліктового випромінювання для всіх 4-х моделей практично однаково добре узгоджуються із відповідними спостережуваними даними. Різниці розрахованих величин усіх характеристик в космологічних моделях з параметрами **q**$_1$, **q**$_2$, **p**$_1$ і **p**$_2$ є значно меншими, ніж точності відповідних спостережуваних даних, що вказує на можливе існування модельних вироджень: для різних типів ТЕ (в нашому випадку – квінтесенційної і фантомної), які суттєво різняться як своїми параметрами, так і фізичними властивостями, можуть існувати такі набори найбільш оптимальних значень космологічних параметрів, що їх розрізнення на основі космологічних спостережуваних даних стає неможливим. Це створює підстави для сумнівів щодо принципової можливості встановлення природи ТЕ на основі даних спостережувальної космології. Очевидно, що це твердження потребує ще перевірки як на змодельованих синтетичних даних, так і на спостережуваних даних із суттєво більшим числом спостережуваних точок зі значно меншими систематичними похибками та статистичними невизначеностями. В роботі [4] ми показали, що подібне виродження є і в класі квінтесенційних моделей ТЕ, наприклад, із спадним і зростаючим параметром рівняння стану. Все це зобов'язує розвивати інші космологічні тести ТЕ, віднесені до різних червоних зміщень та масштабів.

## 2. ФОРМУВАННЯ КУТОВОЇ АНІЗОТРОПІЇ В ЛІНІЇ 21 СМ В ЕПОХУ ТЕМНИХ ВІКІВ

Для визначення раннього значення параметра рівняння стану ТЕ необхідно мати спостережувані дані з якомога ширшого діапазону червоних зміщень. Епоху між космологічною рекомбінацією та появою перших світних об'єктів називають епохою Темних Віків. Єдиним джерелом інформації про те, що відбувалося у Всесвіті в цей проміжок часу, є зміщена в декаметровий діапазон довжин хвиль лінія поглинання нейтрального водню 21 см [21]. Дані про кутовий спектр потужності цієї лінії можуть використовуватися як тест як для моделей динамічної ТЕ, так і для моделей розпадної теплої ТМ, типовим представником якої є стерильні нейтрино.

На основі проведених досліджень встановлено, що розпади стерильних нейтрино впливають на форму та амплітуду спектра потужності зміщеної лінії 21 см через фонову еволюцію концентрацій іонізованих фракцій, збурення концентрацій іонізованих фракцій, модифікований тиск баріонів, еволюцію збурень густини ТМ. Форма і амплітуда спектра є також чутливими до типу ТЕ, яка впливає на нього через динаміку космологічного фону та вплив на еволюцію збурень густини ТМ.

Ми розрахували кутові спектри потужності лінії 21 см на різних червоних зміщеннях (томографія за червоним зміщенням) для δ-подібної функції вікна (в діапазоні частот) в космологічних моделях з ТЕ у формі Λ-сталої і класичного скалярного поля з баротропним рівнянням стану з $c_a^2=0$ та різними типами ТМ (рис. 1). Результати показали, що рання динамічна ТЕ суттєво впливає на амплітуду спектру та дещо впливає на форму спектру, у той час як розпадна ТМ впливає на форму та амплітуду спектра потужності на високих сферичних гармоніках, і вплив тим більший, чим менший час життя стерильних нейтрино. Отримані відносні відхилення спектрів, розрахованих для моделей з динамічною ТЕ та розпадною теплою ТМ, від спектрів в ΛCDM моделі становлять від кількох до кількох десятків відсотків, що дозволяє припустити, що в майбутньому спектр потужності лінії 21 см в епоху Темних Віків може стати потужним тестом для моделей темних компонент Всесвіту, ТМ та ТЕ.

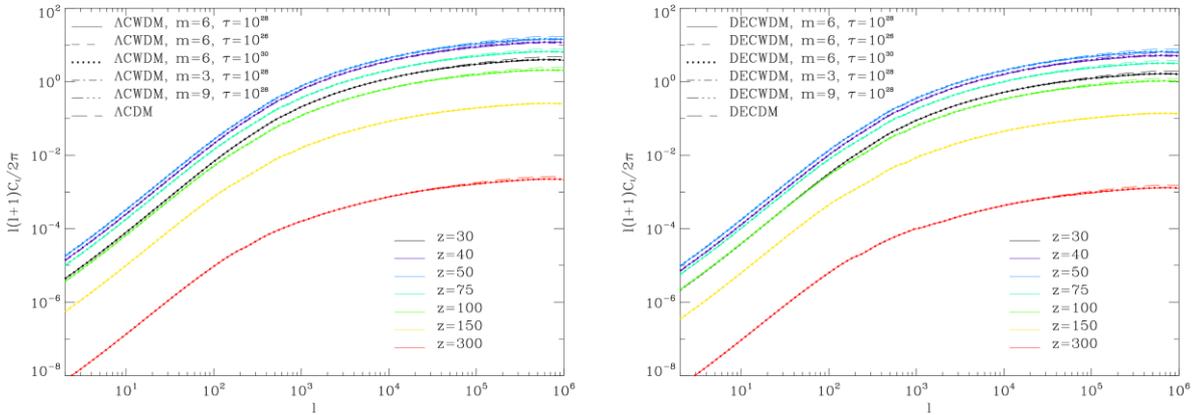

Рис. 1. Кутовий спектр потужності лінії 21 см на різних червоних зміщеннях (томографія за червоним зміщенням) для δ-подібної функції вікна в космологічних моделях з Λ-сталою (зліва) і ТЕ з баротропним рівнянням стану з $c_a^2=0$ (справа) та різними типами ТМ.

## 3. ФОРМУВАННЯ СТРУКТУР ІЗ СВІТНОЇ РЕЧОВИНИ ТА ТЕМНОЇ МАТЕРІЇ НА МАЛИХ МАСШТАБАХ

Ми розвинули тест для встановлення типу ТМ (холодна чи тепла), заснований на розрахунку параметра концентрації гало, функції мас гало та спектра потужності неоднорідностей просторового розподілу світної речовини та ТМ на малих масштабах. Його реалізація в програмах досліджень структури груп і скупчень галактик дасть можливість також надійніше встановити тип ТЕ.

Нами було досліджено процес формування гало ТМ у космологічних моделях Всесвіту із холодною та теплою ТМ [22]. Було розвинуто напів-аналітичний метод розрахунку середнього значення параметра концентрації гало в залежності від параметрів космологічної моделі, початкового спектра неоднорідностей, червоного зміщення та маси гало. Знайдено значення вільних параметрів наближення Чета-Тормена $β=0.7$ та $g=5.4$, при яких розрахунки напів-аналітичним методом узгоджуються із даними числових моделювань формування великомасштабної структури в моделях з теплою та холодною ТМ з роботи [23]. Аналіз розрахунків показує, що середнє значення параметра концентрації росте із часом, тому старіші гало мають більші його значення. Відповідно до ієрархічного сценарію формування великомасштабної структури в моделях із холодною ТМ спочатку формуються гало із меншими масами, що пояснює збільшення параметра концентрації при зменшенні маси гало. В моделях із теплою ТМ збурення малих масштабів є пригніченими внаслідок теплових швидкостей беззіткнювальних частинок, так що гало малих мас починають формуватись пізніше, коли ТМ охолодиться за рахунок розширення Всесвіту. Це призводить до того, що параметри концентрацій гало малих мас для моделей із теплою ТМ є меншими, ніж для моделей з холодною ТМ, і спадають із зменшенням маси.

Розвинутий метод дозволяє розрахувати для заданої космологічної моделі спектр потужності просторового розподілу галактик на малих масштабах [22]. Вирази для спектра отримані в припущенні, що гало великих мас формуються переважно в результаті бурхливої релаксації, внаслідок якої галактики-сателіти, що належать скупченню, здебільшого є розподілені так само, як і ТМ, що узгоджується зі спостережуваними даними [24]. Метод розрахунку спектра потужності просторового розподілу галактик можна застосовувати на масштабах $k<20$ hМпк$^{-1}$. Для отримання спектру потужності на менших масштабах $k>20$ hМпк$^{-1}$ "модель гало" слід удосконалити на основі досліджень формування груп галактик, для яких припущення про однаковий розподіл галактик і ТМ не є вірним.

## 4. ВИСНОВКИ

Квінтесенційне скалярне поле із спадним параметром рівняння стану еволюціонує від динамічного поля в ранню епоху із спадною густиною енергії до вакуумноподібного з $\rho_{de}$=const ($w_{de}$=-1) та вічним експоненціальним розширенням. Квінтесенційне скалярне поле із зростаючим параметром стану в майбутньому зумовить зупинку розширення, стискання та досягнення сингулярності Великий Тріск (Big Crunch) за скінчений час. Фантомне скалярне поле з баротропним рівнянням стану ($\rho_{de}$ зростає з часом) на початку мімікрує Λ-сталу, а в майбутньому зумовлює еволюцію Всесвіту до сингулярності Великий Розрив (Big Rip). Методом Марківських ланцюжків Монте Карло знайдено найбільш оптимальні значення параметрів космологічної моделі і параметрів скалярного поля для двох наборів спостережуваних даних. Встановлено існування найбільш оптимальних моделей як з квінтесенційним, так і фантомним скалярним полем із статистично незначущою різницею значень максимумів функції правдоподібності. Виявлена виродженість моделей потребує перевірки на основі змодельованих синтетичних даних з однієї сторони, з іншої, пошуків нових чутливих до ТЕ космологічних та астрофізичних тестів.

Досліджено вплив динамічної ТЕ та розпадів стерильних нейтрино як частинок теплої ТМ на кутовий спектр потужності зміщеної в декаметровий діапазон довжин хвиль лінії поглинання нейтрального водню 21 см, що сформувалася в епоху Темних Віків. Показано, що в майбутньому дані щодо анізотропії лінії 21 см можуть служити тестом для моделей розпадної ТМ та ТЕ зі змінним параметром рівняння стану.

Досліджено процес формування гало ТМ у космологічних моделях Всесвіту із холодною та теплою ТМ. Запропоновано напів-аналітичний метод розрахунку параметра концентрації гало, функції мас та спектра потужності неоднорідностей просторового розподілу галактик на малих просторових масштабах аж до $k>20$ hМпк$^{-1}$. Отримання таких даних в програмах досліджень структури груп і скупчень галактик дасть можливість надійніше встановити тип темної матерії та темної енергії.

## СПИСОК ЛІТЕРАТУРИ

# Scalar-field models of dark energy: properties and observational constraints

B.S. Novosyadlyj[1], O.N. Sergijenko[1], Yu.A. Kulinich[1], V.O. Pelykh[2]

[1]*Astronomical Observatory of Ivan Franko National University of Lviv,
Kyryla i Methodia str., 8, Lviv, 79005, Ukraine*
[2]*Ya. S. Pidstryhach Institute for Applied Problems of Mechanics and Mathematics,
Naukova str., 3-b, Lviv, 79060, Ukraine*

The properties of quintessential and phantom scalar fields with barotropic equation of state, their effect on dynamics of expansion and formation of large scale structure of the Universe are studied. The best-fit parameters of cosmological and dark energy models are estimated on the basis of current observational data. The ability of quintessential and phantom scalar fields to fit the key observational data with nearly the same accuracy is demonstrated. The possibilities to distinguish between cosmological models with different types of dark matter and dark energy are discussed for the data of current galaxy sky surveys, the measurements of CMB anisotropies and the redshift tomography of Dark Ages in the line 21 cm of neutral Hydrogen (in Ukrainian).

# Дослідження нейтрино і пошуки темної матерії

Р.С. Бойко, Ф.А. Даневич, В.В. Кобичев, Б.М. Кропив'янський, В.М. Мокіна,
А.С. Ніколайко, Д.В. Пода, Р.Б. Подвіянюк, О.Г. Поліщук, В.І. Третяк, Д.М. Черняк

*Інститут ядерних досліджень НАН України, МСП 03680 Київ, Україна*

Властивості нейтрино відіграють ключову роль у фізиці елементарних частинок і космології. Пошуки взаємодій темної матерії з баріонною матерією є одним з шляхів з'ясувати її природу. Представлені результати досліджень нейтрино за допомогою детектора Borexino, експериментів з пошуку подвійного бета-розпаду атомних ядер, розробки детекторів подвійного бета-розпаду і темної матерії, пошуку аксіонів в рамках програми фундаментальних досліджень НАН України «Астрофізичні і космологічні проблеми прихованої маси і темної енергії (Космомікрофізика-2)».

## 1. ВСТУП

Спостереження осциляцій нейтрино (явища, яке не можна пояснити в рамках стандартної моделі частинок), вимагають подальших досліджень властивостей нейтрино для розробки більш досконалої теорії [1]. Про необхідність розширення стандартної моделі свідчать також спостереження темної матерії і темної енергії [2]. Слабковзаємодіючі масивні частинки (зокрема, запропоновані в рамках суперсиметричних моделей), аксіони та стерильні нейтрино розглядаються як найбільш вірогідні кандидати на роль темної матерії. Реєстрація взаємодії таких частинок у лабораторних експериментах була б важливим кроком у розумінні природи темної матерії. Тут представлено результати досліджень нейтрино та пошуків темної матерії, отримані в Інституті ядерних досліджень НАН України в рамках програми фундаментальних досліджень НАН України «Астрофізичні і космологічні проблеми прихованої маси і темної енергії (Космомікрофізика-2)» у 2010 – 2012 роках.

## 2. ДОСЛІДЖЕННЯ ВЛАСТИВОСТЕЙ НЕЙТРИНО В ЕКСПЕРИМЕНТІ BOREXINO

### 2.1. Вимірювання нейтрино від розпадів $^7$Be в Сонці

Низькофоновий сцинтиляційний детектор Borexino на основі рідкого сцинтилятору у підземній лабораторії Гран-Сассо був збудований для вимірювань потоку моноенергетичних нейтрино з енергією 862 кеВ, що випромінюються в Сонці ядрами $^7$Be в реакціях *pp*-циклу [3]. Результатом експерименту є реєстрація подій пружного розсіяння цих нейтрино на електронах у детекторі з інтенсивністю:

$$46.0 \pm 1.5 \text{(стат.)} ^{+1.5}_{-1.6} \text{(сист.) відліків/(доба} \times 100 \text{ тон)},$$

що відповідає потоку „берилієвих" нейтрино $(3.10 \pm 0.15) \times 10^9$ см$^{-2}$ с$^{-1}$ і, у припущенні осциляцій нейтрино, дає ймовірність виживання електронного нейтрино з енергією 862 кеВ $0.51 \pm 0.07$. Гіпотеза про відсутність осциляцій відкинута на рівні $5.0\sigma$.

Глобальний аналіз потоків сонячних нейтрино дозволив визначити величину потоку *pp*-нейтрино:

$$\Phi_{pp} = 6.06 ^{+0.02}_{-0.06} \times 10^{10} \text{ см}^{-2} \text{ с}^{-1} \text{ (з довірчою ймовірністю 95\%)},$$

та нейтрино від CNO-циклу:

$$\Phi_{CNO} < 1.3 \times 10^9 \text{ см}^{-2} \text{ с}^{-1} \text{ (з довірчою ймовірністю 95\%)}.$$

Із цих величин можна отримати значення відношення виміряних потоків до передбачених стандартною моделлю Сонця (з високою металічністю): $f_{pp} = 1.013 ^{+0.003}_{-0.010}$ та $f_{CNO} < 2.5$ з довірчою ймовірністю 95%. Обмеження на $f_\text{CNO}$ відповідає внеску нейтрино CNO-циклу у світимість Сонця < 1.7 %. Точність визначення $f_\text{pp}$ та величина обмеження на $f_\text{CNO}$ у 2 рази перевищують попередні дані. Ці вимірювання дозволили значно підвищити точність, з якою експериментально перевірено модель нейтринних осциляцій Міхєєва – Смирнова – Вольфенштейна з великим кутом змішування при низьких енергіях нейтрино [4].

### 2.2. Дослідження потоків антинейтрино від Сонця та інших джерел

Детектор Borexino був застосований для пошуку антинейтрино невизначеного походження, зокрема від Сонця. Антинейтрино реєструються завдяки реакції $\tilde{\nu}_e + p \rightarrow n + e^+$. У припущенні про неспотворений спектр нейтрино від $^8$B, встановлено обмеження на потік антинейтрино від Сонця: 760 см$^{-2}$ с$^{-1}$. Це дозволило встановити верхню межу на вірогідність перетворення сонячних нейтрино в антинейтрино $1.3 \times 10^{-4}$ (усі результати наведено з довірчою ймовірністю 90%) для нейтрино з енергією більше 1.8 МеВ. Також встановлено обмеження на ймовірність перетворення сонячних нейтрино в антинейтрино: $p_{\nu \rightarrow \tilde{\nu}} < 1.7 \times 10^{-4}$ для $E_\nu > 7.3$ МеВ та $p_{\nu \rightarrow \tilde{\nu}} < 0.35$ для $E_\nu = 862$ кеВ. Кращі диференціальні обмеження на потоки антинейтрино від невизначених джерел були отримані для інтервалу енергій 1.8 – 17.8 МеВ [5].

### 2.3. Нейтрино з глибин Землі (гео-нейтрино)

Вимірювання потоків антинейтрино від β-розпадів радіоактивних ізотопів у Землі (гео-нейтрино) є унікальною можливістю досліджувати склад і розподіл радіонуклідів в надрах нашої планети. Вказівку на реєстрацію гео-нейтрино із статистичною достовірністю ~2.5σ було отримано в експерименті KamLAND [6]. В експерименті Borexino гео-нейтрино вперше спостережено з достовірністю більш ніж 3σ [7]. Установка Borexino чутлива до антинейтрино завдяки реакції оберненого β-розпаду на протоні $\tilde{\nu}_e + p \rightarrow n + e^+$ з порогом 1.806 МеВ. Позитрони, що випромінюються в реакції, анігілюють з емісією двох γ-квантів з енергією 511 кеВ, у той час як вільний нейтрон захоплюється протоном за середній час близько 256 мкс з випромінюванням γ-кванту з енергією 2.22 МеВ. Такі події, враховуючи умови відбору за часом та енергією, реєструються з ефективністю 85%. Усього знайдено 21 подію від антинейтрино в даних з експозицією 252.6 тон×рік. Відомими джерелами антинейтрино, окрім ядерних розпадів всередині Землі, є ядерні реактори. Очікуваний сигнал від реакторних антинейтрино (з урахуванням нейтринних осциляцій) оцінено на рівні $10.7^{+4.3}_{-3.4}$ подій. З урахуванням фону від реакторних нейтрино та усіх критеріїв відбору, кількість зареєстрованих гео-нейтрино становить $N_\text{geo} = 9.9^{+4.1}_{-3.4}(^{+14.6}_{-8.2})$ з довірчою ймовірністю 68.3% (99.73%). Швидкість лічби гео-нейтрино складає $3.9^{+1.6}_{-1.3}(^{+5.8}_{-3.2})$ подій/(100 тон×рік). Спостережений спектр антинейтрино вище енергії 2.6 МеВ відповідає очікуваному від європейських ядерних реакторів (середня відстань до детектора складає близько 1000 км) і виключає гіпотезу про відсутність осциляцій нейтрино на рівні довірчої ймовірності 99.60%. Гіпотеза про існування всередині Землі активного реактора [8] потужністю більше 3 терават виключена з довірчою ймовірністю 95%.

### 2.4. Вимірювання швидкості мюонних нейтрино на пучку ЦЕРН – Гран-Сассо за допомогою детектора Borexino

У вересні 2011 частина колаборації OPERA повідомила про спостереження нейтрино, що проходять відстань 730 км від ЦЕРНа до Гран-Сассо на ~60 нс швидше, ніж пройшло б таку відстань світло у вакуумі [9]. Детектор Borexino також реєструє нейтрино з пучка CNGS за допомогою вторинних мюонів, які виникають при взаємодії нейтрино з ядрами у шарі скельної породи товщиною кілька десятків метрів перед детектором. Мюони, що проходять через шар води, дають черенковський спалах, у той час як у буферному об'ємі та в об'ємі детектора, наповненому рідким сцинтилятором, виникає сцинтиляційний сигнал. Це дозволяє відновити напрямок руху та час прильоту частинки [10]. Показано, що затримка мюонного нейтрино з середньою енергією 17 ГеВ на його шляху до детектора по відношенню до світла у вакуумі становить

$$\delta t = 0.8 \pm 0.7(\text{стат.}) \pm 2.9(\text{сист.}) \text{ нс},$$

що є сумісним з нулем. Встановлено також обмеження на відносну різницю між швидкістю нейтрино та світла:

$$|v - c|/c \leq 2.1 \times 10^{-6} \text{ (з довірчою ймовірністю 90\%)}.$$

Таким чином, не було знайдено порушення релятивістської кінематики для нейтрино [11].

### 2.5. Відсутність асиметрії "день-ніч" потоку нейтрино від розпадів $^7$Be в Сонці

Переріз взаємодії нейтрино на електроні залежить від аромату нейтрино. Це дає можливість перевірити ефект Міхєєва – Смірнова – Вольфенштейна, пов'язаний з когерентною регенерацією електронних нейтрино завдяки механізму підсилення осциляцій при проходженні нейтрино через речовину. В експерименті Borexino ефект міг би проявитися у різниці швидкості лічби нейтрино вдень та вночі. Пошуки асиметрії у швидкості лічби нейтрино від розпадів $^7$Be в Сонці були здійснені за допомогою аналізу даних вимірювань вдень (загалом 360.25 діб) і вночі (380.63 діб). Виміряний параметр асиметрії (його визначено як $A_{dn}=2(R_N–R_D)/(R_N+R_D)$, де $R_N$ і $R_D$ – швидкості лічби нейтрино вночі і вдень, відповідно) дорівнює $A_{dn} = 0.001 \pm 0.012$ (стат.) $\pm 0.007$ (сист.). Цей результат узгоджується із передбаченнями теорії Міхєєва – Смірнова – Вольфенштейна з великим кутом змішування між станами нейтрино у матриці Понтекорво – Макі – Накагави – Сакати. У той же час, цей результат відкидає осциляції з малими кутами змішування зі статистичною достовірністю 8.5 $\sigma$ [12].

### 3. ДОСЛІДЖЕННЯ ВЛАСТИВОСТЕЙ НЕЙТРИНО І СЛАБКОЇ ВЗАЄМОДІЇ У ПРОЦЕСАХ ПОДВІЙНОГО БЕТА-РОЗПАДУ АТОМНИХ ЯДЕР

Пошуки безнейтринного подвійного бета-розпаду (0ν2β) виглядають єдиним реалістичним шляхом визначення маси і схеми масових станів нейтрино, встановлення природи нейтрино (частинка Дірака або Майорани), перевірки закону збереження лептонного заряду і ряду ефектів за межами стандартної моделі [13-18]. У той час як двонейтринна мода (2ν) 2β-розпаду зареєстрована у кількох ядрах, 0ν2β-розпад не спостережений. У найбільш чутливих експериментах лише обмежено періоди напіврозпаду, звідки слідують обмеження на масу нейтрино на рівні 0.3 – 3 еВ. Разом з тим, H.V. Klapdor-Kleingrothaus і кілька його співавторів стверджують про спостереження 0ν2β-розпаду $^{76}$Ge [19, 20]. Незважаючи на скептичне ставлення до цього повідомлення наукової спільноти (див. [21], де є також посилки на інші роботи), лише експерименти з більшою чутливістю можуть його спростувати або підтвердити. У випадку підтвердження результату в експерименті GERDA [22], важливо буде спостерігати ефект у кількох інших ядрах на рівні чутливості до періоду напіврозпаду $T_{1/2} \sim 10^{25}$ років. Якщо результат [19, 20] не підтвердиться, необхідно підвищити чутливість до маси нейтрино до рівня, що відповідає інвертованій схемі мас нейтрино $\langle m_\nu \rangle \approx 0.02 – 0.05$ еВ. Дослідження подвійного електронного захоплення (2ε), електронного захоплення з випромінюванням позитрону (εβ$^+$) та подвійного позитронного розпаду (2β$^+$) важливі для розробки теорії та оцінки внеску гіпотетичних правих струмів у слабкій взаємодії [23]. Крім того, у деяких ядер можливе резонансне підсилення 0ν2ε-розпаду у випадку, коли перехід відбувається на збуджений рівень з енергією, близькою до енергії переходу мінус дві енергії зв'язку електронів в атомній оболонці дочірнього елементу [24-28].

### 3.1. Дослідження за допомогою напівпровідникових гамма-спектрометрів

У деяких каналах подвійного β-розпаду можуть випромінюватися γ-кванти, що дозволяє застосовувати γ-спектрометри для дослідження таких процесів. Перевагою таких детекторів є висока енергетична роздільна здатність (близько 2 кеВ на енергії 1.3 МеВ).

Експеримент з метою точного вимірювання періоду напіврозпаду ядра $^{100}$Mo на перший збуджений $0^+$ рівень дочірнього ядра був здійснений за допомогою 1.2 кг оксиду молібдену, збагаченого ізотопом $^{100}$Mo до 99.5%, встановленого у наднизькофоновий напівпровідниковий детектор з чотирма кристалами надчистого германію (об'ємом близько 225 см$^3$ кожен), розміщеними в одному кріостаті [29]. Це дозволило реєструвати γ-кванти у збігах у часі для підтвердження того, що зареєстровано саме каскад двох γ-квантів, як це і очікується для шуканого процесу. З аналізу сумарного спектру, накопиченого впродовж 18120 годин, отримано значення періоду напіврозпаду відносно 2ν2β-розпаду ядра $^{100}$Mo на збуджений рівень $0^+_1$ ядра $^{100}$Ru з енергією 1131 кеВ:

$$T_{1/2} = 6.9^{+1.0}_{-0.8} \text{(стат.)} \pm 0.7 \text{(сист.)} \times 10^{20} \text{ років.}$$

Це значення узгоджується з результатами попередніх експериментів, але має найменші похибки, що важливо для розвитку теорії 2β-розпаду.

За допомогою наднизькофонових напівпровідникових γ-детекторів із надчистого германію були також здійснені пошуки 2β-розпаду ізотопів диспрозію ($^{156,158}$Dy [30]), осмію ($^{184,192}$Os [31]) та платини ($^{190,198}$Pt [32]). Слід підкреслити особливий інтерес до ядер $^{156,158}$Dy, $^{184}$Os, $^{190}$Pt, у яких можливе резонансне подвійне електронне захоплення. Варто відмітити, що в ході експерименту з пошуку подвійного бета-розпаду платини вперше був спостережений α-розпад ізотопу $^{190}$Pt на перший збуджений рівень дочірнього ядра [33].

### 3.2. Сцинтиляційні експерименти

Важливою перевагою сцинтиляційних детекторів для пошуку 2β-розпаду є можливість використання сцинтиляторів, до складу яких входять елементи з 2β-активними ізотопами. Таким чином ефективність реєстрації процесів 2β-розпаду може бути наближена до 100%. Пошук 2β-розпаду в ізотопах цинку ($^{64,70}$Zn) і вольфраму ($^{180,186}$W) був здійснений за допомогою сцинтиляційних кристалів вольфрамату цинку (ZnWO$_4$) [34], які відрізняються високим рівнем радіочистоти [35]. Пошук процесів подвійного бета-розпаду в ізотопах церію ($^{136,138,142}$Ce) був здійснений із сцинтилятором хлориду церію [36], дослідження $^{84}$Sr були проведені з кристалом йодиду стронцію SrI$_2$(Eu) [37]. Чутливість сцинтиляційних експериментів можна суттєво збільшити завдяки використанню сцинтиляторів із збагачених ізотопів [38]. Було розроблено радіоактивно чисті сцинтилятори вольфрамату кадмію з високими сцинтиляційними характеристиками із збагачених ізотопів кадмію 106 [39, 40] та 116 [41]. Експерименти з пошуку подвійного бета-розпаду ядер $^{106}$Cd і $^{116}$Cd тривають [42, 43]. Чутливість експерименту з пошуку 2β-розпаду $^{106}$Cd з кристалом $^{106}$CdWO$_4$ у низькофоновій установці з чотирма HPGe детекторами об'ємом 225 см$^3$ кожен, що функціонують як у режимі збігів, так і антизбігів з основним детектором, близька до теоретичних передбачень для двонейтринних процесів ($T_{1/2} \sim 10^{20} - 10^{21}$ років). Чутливість експерименту з ядром $^{116}$Cd очікується на рівні $T_{1/2} \sim 10^{24}$ років, що відповідає масі нейтрино $\langle m_\nu \rangle \approx 0.4 - 1.4$ еВ. Єдиним експериментом, здатним вимірювати характеристики випромінюваних у 2β-розпаді електронів (енергію і треки), буде SuperNEMO. Для цього проекту розроблено калориметр [44] та метод вимірювання радіоактивної забрудненості фольги на рівні чутливості кілька мкБк/кг [45].

### 3.3. Низькотемпературні сцинтиляційні болометри

Чутливість експериментів з пошуку 0ν2β-розпаду залежить від енергетичної роздільної здатності детектора (див. наприклад, [46]). Більше того, фон, обумовлений двонейтринною модою розпаду, неможливо уникнути ніяким чином, крім використання детекторів з високою енергетичною роздільною здатністю $R$. Ідентифікація ефекту 0ν2β-розпаду також вимагає якомога вищої $R$ [47]. Експерименти, спрямовані на дослідження схеми масових станів нейтрино, повинні мати якомога вищу $R$, але не гірше кількох %. Низькотемпературні сцинтиляційні болометри мають $R$ близько 0.2% [48]. Одним з найбільш перспективних для пошуку 0ν2β-розпаду є ізотоп $^{100}$Mo, завдяки великій енергії переходу 3034.4 кеВ, високій ізотопічній поширеності 9.7%, сприятливим теоретичним розрахункам [17]. Було показано можливість застосування кристалів молібдату кальцію [49] та цинку [50-53] як низькотемпературних сцинтиляційних болометрів для пошуку 0ν2β-розпаду $^{100}$Mo. Вперше проаналізовано фон від випадкових збігів подій 2ν2β-розпаду у 0ν2β-експериментах у випадку низької часової роздільної здатності детектора та запропоновано шляхи зниження такого фону [54]. Ведуться дослідження оптимальної форми та способу обробки поверхні сцинтиляторів для збільшення світловиходу у кріогенних сцинтиляційних болометрах [55].

### 4. ПОШУКИ АКСІОНІВ

Аксіони – гіпотетичні псевдоскалярні нейтральні частинки, які виникають при спонтанному порушенні глобальної симетрії, запропонованої в 1977 році Печеі і Квін для вирішення проблеми збереження CP-симетрії у квантовій хромодинаміці [56, 57]. Аксіони розглядаються як одні з найбільш ймовірних кандидатів на роль темної матерії [58].

#### 4.1. Пошук аксіонів від розпадів $^7$Li в Сонці за допомогою напівпровідникових γ-спектрометрів

Сонце може бути інтенсивним джерелом аксіонів. Зокрема, аксіони можуть випромінюватися у M1-переходах із збуджених рівнів ядер. Таким є γ-перехід у ядрі $^7$Li з енергією 477.6 кеВ. Ядра $^7$Li у збудженому стані 477.6 кеВ виникають у $pp$-циклі в реакціях $^3$He + $\alpha \to$ $^7$Be + γ, $^7$Be + $e^- \to$ $^7$Li* + ν$_e$ з ймовірністю 10.5%. Аксіони, що виникають у процесі зняття збудження ядер $^7$Li в Сонці, можуть резонансно поглинатись ядрами $^7$Li на Землі. Частинки, які випромінюються у процесі розрядки (γ- та рентгенівські кванти, конверсійні електрони), можуть бути зареєстровані детектором, що має в своєму складі ядра $^7$Li, або розміщеним близько до мішені з ядрами $^7$Li. При цьому в енергетичному спектрі детектора буде спостерігатись характерний пік з енергією 477.6 кеВ. Експеримент з пошуку сонячних аксіонів був здійснений за допомогою наднизькофонових HP Ge γ-спектрометрів зі зразками фториду літію в якості мішені. Детектори були розміщені у Національній підземній лабораторії Гран-Сассо. Аналіз даних, отриманих за час вимірювання 4044 годин з кристалом фториду літію масою 553 г, дозволив встановити нове обмеження на масу аксіону < 8.6 кеВ (90% C.L.) [59, 60].

#### 4.2. Пошук сонячних аксіонів за допомогою детектора BOREXINO

Пошук аксіонів з енергією 5.5 МеВ, що можуть утворюватися в протон-протонному циклі на Сонці в реакції $p + $D$ \to $ $^3$He + $A$, був здійснений за допомогою детектора Borexino [61]. Були розглянуті комптонівська конверсія аксіонів у фотони $A + e \to e + γ$; аксіо-електричний ефект на електронах атомних оболонок $A + e + Z \to e + Z$; двофотонний розпад аксіону $A \to 2γ$; зворотний ефект Примакова в полі ядра $A + Z \to γ + Z$. Для всіх цих реакцій в енергетичному спектрі детектора очікується пік з енергією 5.5 МеВ. Ніяких статистично значимих ознак взаємодії аксіонів з речовиною детектора не знайдено. Для аксіонів масою менше 1 МеВ були встановлені нові модельно незалежні верхні обмеження на постійні зв'язку аксіонів з електронами, фотонами та нуклонами:

$$|g_{Ae} \times g_{3AN}| \leq 5.5 \times 10^{-13}, |g_{A\gamma} \times g_{3AN}| \leq 4.6 \times 10^{-11} \text{ ГеВ}^{-1}.$$

Якщо припустити, що постійна зв'язку аксіону с нуклоном $g_{3AN}$ залежить від маси аксіону за моделлю KSVZ [62, 63], з отриманих даних випливають верхні обмеження на зв'язок аксіону з електроном і з фотоном як функція маси аксіону:

$$|g_{Ae} \times m_A| \leq 2.0 \times 10^{-5} \text{ еВ}, |g_{A\gamma} \times m_A| \leq 1.7 \times 10^{-12}.$$

Отримані нові обмеження виключають велику частину можливого простору параметрів ($g_{Ae}, m_A$) та ($g_{A\gamma}, m_A$) і на 2-4 порядки сильніші за отримані в попередніх лабораторних експериментах з використанням ядерних реакторів і прискорювачів.

## 5. РОЗРОБКА ДЕТЕКТОРІВ ДЛЯ ДОСЛІДЖЕННЯ СЛАБКОВЗАЄМОДІЮЧИХ МАСИВНИХ ЧАСТИНОК ТЕМНОЇ МАТЕРІЇ

### 5.1. Розробка сцинтиляційних кристалів для пошуку темної матерії

Пошуки темної матерії ведуться за допомогою детекторів, що реєструють сигнали різного походження: іонізаційні, теплові, сцинтиляційні, або ж одночасно сигнали двох різних типів. Одними з найбільш перспективних детекторів для пошуку взаємодій, які мають призводити до виникнення ядер віддачі, є кріогенні сцинтиляційні болометри. У детекторах цього типу ядра віддачі відрізняються від фонових електронів, гамма та рентгенівських квантів завдяки значно меншому світловому виходу при реєстрації іонів у порівнянні з гамма-квантами та електронами та завдяки іншій формі сцинтиляційних сигналів. Важливо, як це планується у проекті EURECA [64], щоб у складі детектора були різні елементи. Тому актуальною задачею є розробка сцинтиляційних кристалів, до складу яких входили б елементи з різним атомним номером, з якомога нижчим рівнем радіоактивної забрудненості і високим світловим виходом при низьких (близько 10 – 20 мК) температурах. Досліджені сцинтиляційні властивості (у тому числі при низьких температурах) та рівень радіоактивної чистоти перспективних матеріалів для пошуків темної матерії – кристалів вольфраматів кальцію [65], цинку [35] та свинцю [66], а також молібдату свинцю [67] та йодиду літію [67].

### 5.2. Відгук сцинтиляційних детекторів до ядер віддачі

Важливою характеристикою сцинтиляційних детекторів темної матерії, що націлені на реєстрацію ядер віддачі, є відгук до іонів. Як правило, сцинтиляційні сигнали від іонів менші за сигнали від гамма-квантів такої ж енергії. Виміряти відгук сцинтиляційного детектора до іонів малої енергії (кілька кеВ – десятки кеВ) доволі складно. Тому було запропоновано метод оцінки відгуку детектора до ядер віддачі, використовуючи дані про відгук цього ж детектора до, наприклад, до α-частинок від розпаду дочірніх урану і торію, які завжди присутні у сцинтиляторі як домішки [68]. Метод був перевірений у вимірюваннях з кристалом вольфрамату кадмію з іонами малих енергій [69].

### 5.3. Пошук добових модуляцій потоку темної матерії за допомогою анізотропних сцинтиляторів ZnWO$_4$ (проект ADAMO)

Ознаками сигналів від темної матерії є енергетичний спектр ядер віддачі, річні та добові варіації інтенсивності сигналів, залежність сигналу від маси ядер. Очікуваний енергетичний спектр ядер віддачі близький до експоненти і тому його надзвичайно важко відрізнити від фону, який також має монотонний характер. Деякі з діючих детекторів темної матерії здатні зареєструвати річні варіації (зокрема, такі варіації спостерігаються в експерименті DAMA/LIBRA [70]). Поряд з тим, немає експерименту, здатного зареєструвати добові варіації потоку темної матерії. У роботі [71] вперше запропоновано детектор на основі анізотропних сцинтиляторів вольфрамату цинку, чутливий до добових модуляцій потоку темної матерії. Важливо відмітити, що ідею застосувати сцинтилятори $ZnWO_4$ для пошуку добових модуляцій потоку темної матерії було вперше запропоновано у 2005 році, на основі досліджень, виконаних у Солотвинській підземній лабораторії [72].

## ВИСНОВКИ

В експерименті Borexino вперше виміряно у реальному часі потік низькоенергетичних нейтрино від Сонця (від розпадів $^7Be$). Встановлено нові обмеження на потік антинейтрино низьких енергій (від Сонця та інших можливих джерел), на різницю потоків нейтрино "день-ніч", спостережено гео-нейтрино, показано, що швидкість нейтрино з великою точністю співпадає зі швидкістю світла у вакуумі.

З високою точністю досліджено 2ν2β-розпад $^{100}Mo$ на збуджений рівень $^{100}Ru$ та процеси подвійного бета-розпаду ядер $^{64, 70}Zn$, $^{84}Sr$, $^{106,116}Cd$, $^{136, 138, 142}Ce$, $^{156, 158}Dy$, $^{180, 186}W$, $^{184,192}Os$, $^{190, 198}Pt$ на новому рівні чутливості (дослідження $^{156, 158}Dy$, $^{184,192}Os$ і $^{190, 198}Pt$ виконано вперше). Розроблено унікальні низькофонові сцинтиляційні детектори вольфрамату кадмію зі збагачених ізотопів кадмію 106 та 116 і розпочато експериментальні дослідження 2β-розпаду ядер $^{106}Cd$ та $^{116}Cd$. Вперше досліджено кристали $CaMoO_4$ і $ZnMoO_4$ як сцинтиляційні болометри для пошуку 0ν2β-розпаду $^{100}Mo$ і обґрунтовано експерименти з детекторами $CaMoO_4$ та $ZnMoO_4$ з чутливістю до 0ν2β-розпаду $^{100}Mo$ на рівні $T_{1/2} \sim 10^{26}$–$10^{27}$ років. Вперше проаналізовано фон від випадкових збігів подій 2ν2β-розпаду у низькотемпературних експериментах з пошуку 0ν2β-розпаду. Розроблено калориметр та метод вимірювання радіоактивної забрудненості фольги $^{82}Se$ для пошуку 0ν2β-розпаду в експерименті SuperNEMO.

Отримано нове лабораторне обмеження (< 8.6 кеВ) на масу аксіонів від розпадів збудженого $^7Li$ в *pp*-циклі на Сонці; встановлено нове обмеження на потік аксіонів, які можуть виникати в Сонці в реакціях взаємодії ядер дейтерію з ядрами гелію.

Досліджено сцинтиляційні властивості та рівень радіоактивної чистоти сцинтиляторів $PbWO_4$, $PbMoO_4$, $ZnWO_4$, $CaWO_4$ та $SrI_2(Eu)$, які є перспективними для розробки детекторів темної матерії. Запропоновано напівемпіричний підхід до оцінок світловиходу в сцинтиляційних детекторах від важких іонів, який перевірено в експерименті зі сцинтилятором $CdWO_4$ на прискорювачі іонів. Запропоновано детектор на основі анізотропних сцинтиляторів $ZnWO_4$, чутливий до добових модуляцій взаємодій зі слабковзаємодіючими масивними частинками темної матерії, який є єдиним реалістичним проектом пошуку добових модуляцій потоку темної матерії.

## СПИСОК ВИКОРИСТАНИХ ДЖЕРЕЛ

72. *F.A. Danevich et al.*, ZnWO$_4$ crystals as detectors for 2β decay and dark matter experiments. Nucl. Instr. Meth. A 544 (2005) 553.

# Study of neutrino properties and search for dark matter

R.S. Boiko, D.M. Chernyak, F.A. Danevich, V.V. Kobychev, B.N. Kropivyansky,
V.M. Mokina, A.S. Nikolaiko, D.V. Poda, R.B. Podviyanuk, O.G. Polischuk, V.I. Tretyak

*Institute for Nuclear Research, MSP 03680 Kyiv, Ukraine*

The properties of neutrino play a key role in developing extensions of the standard model of elementary particles and cosmology. The search for dark matter interactions with baryon matter is considered as a way to determine its nature. The neutrino investigations with the Borexino detector, experiments to search for double β decay and for axions, development of double beta decay and dark matter detectors are presented. The researches were performed in the frame of the "Kosmomikrofizyka-2" Project (Astroparticle Physics) of the National Academy of Sciences of Ukraine (in Ukrainian).

# Астрофізика космічних променів: космологічний аспект

## Б.І. Гнатик

*Астрономічна обсерваторія, Київський національний університет імені Тараса Шевченка, 04053 Київ, Україна*

Механізми прискорення та джерела космічних променів надвисоких енергій (КПНВЕ) залишаються досі нез'ясованими. Галактичний компонент, ймовірно, прискорюється в залишках Надневих зір. Позагалактичний компонент спостережуваного потоку космічних променів несе цінну інформацію про фізичні умови як в областях прискорення (ймовірно, це активні ядра галактик (АЯГ)), так і в міжгалактичному середовищі. Отримані останнім часом дані спостережень детекторів HiRes, Auger, Telescope Array щодо енергетичного спектру та хімічного вмісту КПНВЕ підтверджують теоретично передбачене обрізання енергетичного спектру на енергіях понад $10^{19.7}$ еВ (ГЗК-ефект), однак хімічний вміст на найвищих енергіях залишається невизначеним. В роботі приводяться результати теоретичного моделювання спостережуваного спектру КПНВЕ та показується, що протонна модель пояснює всі особливості енергетичного спектру космічних променів в широкому інтервалі енергій $10^{17}$ - $10^{20}$ еВ, включно з областю переходу від галактичного до позагалактичного компонентів в загальному потоці космічних променів. Аналізуються нові результати, отримані у вивченні залишків Надневих зір. Обговорюються перспективи використання даних планованих космічних місій Ломоносов (експеримент ТУС) та ГАММА-400 для встановлення природи КПНВЕ.

## 1. ВСТУП

В 2012 році минуло сто років з часу відкриття космічних променів, однак досі головні питання – де і як вони прискорюються – залишаються відкритими [1,2]. Зараз дослідження космічних променів стали одними з найактуальніших в астрофізиці високих енергій, з ними пов'язані такі актуальні проблеми як ранні етапи еволюції Всесвіту, фізичні процеси в околі релятивістських об'єктів, природа та астрофізичні прояви темної матерії тощо, їх прояви вивчаються широкою мережею детекторів різної природи: це космічні місії рентгенівської та гама-астрономії XMM Newton, Chandra, INTEGRAL, FERMI, наземні черенковські телескопи ТеВного діапазону H.E.S.S., VERITAS, MAGIC, детектори космічних променів AGASA, HIRES, AUGER, Telescope Array, нейтрино IceCube та багато інших. Зараз експериментальні дані в галузі астрофізики високих енергій можна отримувати і в Україні - на фізичному факультеті Київського національного університету імені Тараса Шевченка в 2006 р. відкрито Віртуальну рентгенівську та гама-обсерваторію (проект ВІРГО) - науковий комп'ютерний центр доступу до даних космічних місій в рентгенівському та гама-діапазонах [3].

В нашій роботі приведено огляд досліджень джерел та механізмів прискорення космічних променів, виконаних в Астрономічній обсерваторії Київського національного університету імені Тараса Шевченка в 2007-2012 роках та описані перспективи подальших досліджень, зокрема, в рамках участі в космічних проектах Ломоносов та ГАММА-400.

## 2. КОСМІЧНІ ПРОМЕНІ НАДВИСОКИХ ЕНЕРГІЙ

Аналіз поширення космічних променів в магнітному полі Галактики показує, що космічні промені - протони чи ядра із зарядовим числом Z та з енергіями понад $10^{18}Z$ еВ слабо відхиляються галактичним магнітним полем, тому із ізотропії спостережуваного потоку на цих енергіях слідує, що джерела КПНВЕ є позагалактичними.

Експериментальні дані щодо енергетичного спектру КПНВЕ, отримані різними детекторами – AGASA, HiRes, Yakutsk, Auger, TA – узгоджуються в загальних рисах, але суттєво відрізняються за величиною потоків [2], [4]. В циклі робіт, виконаних в групі проф. В.С. Березінського, зокрема, в нашій роботі [4] запропонована інтерпретація спостережуваного енергетичного спектру КПНВЕ в рамках протонної моделі, де основним компонентом в потоці КПНВЕ є протони (доля гелію не повинна перевищувати 10 відсотків). В цій моделі перехід від галактичного (головно, ядра заліза) до позагалактичного компонента відбувається в області другого коліна ($10^{17.5}$ еВ), а спостережуване суттєве пониження потоку на енергіях понад $10^{18}$ еВ пояснюється пороговою реакцією народження електрон-позитронних пар при розсіянні високоенергетичних протонів на фотонах мікрохвильового (реліктового) випромінювання. Важливим підтвердженням реалістичності протонної моделі є співпадіння всіх спостережуваних різними детекторами спектрів, якщо їх прокалібрувати на теоретично розрахований профіль в області народження пар.

Енергетичні оцінки свідчать, що найбільш ймовірними джерелами КПНВЕ є активні ядра галактик та космологічні гама-спалахи [1-2]. У такому разі можна очікувати, що в загальному потоці КПНВЕ буде помітним вклад найближчих АЯГ, зокрема, Cen A. В наших роботах [5-7] проведено чисельне моделювання поширення космічних променів у галактичному та міжгалактичному магнітних полях та показано, що ряд подій, зафіксованих детектором Auger з напрямків, близьких до Cen A, справді можуть прискорюватись в цьому АЯГ.

АЯГ – кандидати на джерела КПНВЕ – повинні володіти достатньо потужними релятивістськими струменями, де відбуватиметься прискорення [2], [4]. В наших роботах [8], [9] проведені дослідження фізичних умов в релятивістських струменях блазарів з одиночною та подвійною чорною дірою.

КПНВЕ можуть також прискорюватись в околі космічних струн – топологічних дефектів, які можуть виникати під час фазових переходів в ранньому Всесвіті [10]. Рух космічних струн та їх петель в міжгалактичному просторі може супроводжуватись рядом астрофізичних проявів, зокрема, як показано в наших роботах, генерацією характерних синхронних флуктуацій блиску лінзованих зображень квазарів [11], генерацією нетеплового синхротронного та оберненого комптонівського випромінювання від надпровідних космічних струн та їх петель [10], [12-13].

## 3. ДЖЕРЕЛА ГАЛАКТИЧНИХ КОСМІЧНИХ ПРОМЕНІВ

Космічні промені з енергіями менше $10^{18}$ еВ прискорюються в нашій Галактиці. Основним кандидатом на їх джерела, що задовольняє вимоги на максимальні енергії та загальний спостережуваний потік, є залишки Наднових зір [1]. Нетеплове випромінювання залишків, що генерується релятивістськими частинками (головно, електронами), інтенсивно вивчається в усіх спектральних діапазонах: від радіо- до ТеВного гама-діапазону. Зокрема в ТеВному випромінюванні повинен проявитись протонно-ядерний компонент прискорених в залишку космічних променів.

Еволюція залишків Наднових зір в міжзоряному середовищі (МЗС) описується в рамках магнітогідродинамічного опису динаміки плазми при локальному енерговиділенні в багатокомпонентному (розріджений міжзоряний газ та конденсації-хмарки) середовищі з можливим великомасштабним градієнтом густини та магнітними полями. Як правило, дані спостережень залишків в різних спектральних діапазонах моделюються в рамках відомих аналітичних розв'язків Сєдова задачі про ізотропний точковий вибух в однорідному МЗС. Однак отримані останніми роками результати спостережень залишків в рентгенівському та гама-діапазонах потребують для свого пояснення більш складних моделей із врахуванням неоднорідностей МЗС в області вибуху Наднової, впливу магнітних полів і космічних променів на динаміку та свічення залишків тощо. В наших роботах розроблені наближені аналітичні методи дослідження неавтомодельних задач про точковий вибух, які застосовуються для моделювання недостатньо вивчених досі аспектів еволюції залишків Наднових. Зокрема, нами досліджено особливості еволюції залишків при переході від адіабатичної до радіаційної стадії [14-16]. Показано, що цей період має характерні особливості, що дозволяють виділити його в окрему – перехідну – стадію, яка в спостережувальному плані проявляється підвищеною яскравістю в ТеВному діапазоні.

При спалаху Наднової в двокомпонентному (газ та хмарки) МЗС із домінуванням хмарок в загальному балансі маси еволюція залишку на адіабатичній стадії також суттєво відрізняється від сєдовської. В наших роботах [19-20] показано, що залишок Наднової Вітрила (Vela) належить до класу таких залишків та розроблена гідродинамічна модель залишку Вітрила та його оточення - комплексу взаємодіючих об'єктів: залишок Вітрила, зоряна вітрова бульбашка навколо зорі $\gamma^2$ Вітрил та туманність Гама.

Останніми роками відкрито новий клас Наднових зір – Гіпернові зорі, енергія вибуху яких в десятки разів перевищує енергію вибуху звичайних Наднових, та які породжують космологічні гама-спалахи. Тоді як релятивістські струмені в гама-спалахах є кандидатами на прискорення КПНВЕ, помірно релятивістська сферична ударна хвиля, що супроводжує початкові стадії розльоту оболонки Гіпернової, здатна прискорювати космічні промені до енергій порядку $10^{18}$-$10^{19}$ еВ. В наших роботах [17-18] показано, що спалахи Гіпернових в нашій Галактиці можуть дати помітний вклад в загальний потік космічних променів на цих енергіях, при цьому, внаслідок збагачення вітрів зір – попередників Гіпернових важкими елементами (зорі Вольфа-Райє – попередники Гіпернових – володіють потужними вітрами з домінуванням вуглецю та кисню), середнє масове число частинок – космічних променів в цьому діапазоні зростатиме.

## 4. ПЕРСПЕКТИВИ ДОСЛІДЖЕННЯ КОСМІЧНИХ ПРОМЕНІВ МАЙБУТНІМИ КОСМІЧНИМИ МІСІЯМИ

Значний прогрес у дослідженні космічних променів за останні десятиліття в значній мірі зумовлений використанням космічних місій як для прямої реєстрації космічних променів, так і для детектування породженого ними електромагнітного випромінювання. Так, тільки недавно космічний експеримент PAMELA виявив суттєве зростання позитронної фракції (відношення долі позитронів до сумарного електрон-позитронного потоку) на енергіях понад 10 ГеВ [21], що зумовило ряд робіт з інтерпретацією цих спостережень проявом розпаду частинок темної матерії. Недавні дані орбітального гама-телескопа Fermi щодо фонового гама-випромінювання суттєво обмежують характеристики спектру КПНВЕ та породжених ними космогенних нейтрино [22].

В 2012 році наукові команди Київського національного університету імені Тараса Шевченка стали офіційними учасниками двох міжнародних космічних експериментів.

Перший з них – експеримент ТУС (Трековая Установка) – орбітальний телескоп для детектування флуоресцентного свічення атмосфери, викликаного широкими атмосферними зливами при входженні КПНВЕ в атмосферу Землі [23]. Детектор буде встановлений на борту космічного апарата Ломоносов, підготовленого Московським державним університетом ім. М.В. Ломоносова (Росія) і запланованим до запуску на 2013 рік. Порогова чутливість детектора – $10^{19.7}$ еВ, ефективна площа – порядку сучасного наземного детектора Auger. Детектор ТУС – одночасно підготовча стадія більш ефективного детектора КЛПВЭ (Космические Лучи Предельно Высоких Энергий), який планується встановити на російському модулі Міжнародної космічної станції (МКС). Детектор КЛПВЭ буде найбільшим за ефективною площею детектором КПНВЕ в світі (до запуску в експлуатацію міжнародного проекту JEM EUSO на японському модулі МКС). Науковці КНУ братимуть участь в виконанні різних завдань проекту, включно з обробкою та інтерпретацією даних спостережень.

Другий проект – міжнародна космічна обсерваторія ГАММА-400 для високоточних вимірювань потоків гама-випромінювання, електронів та позитронів в діапазоні енергій 0.1-3000 ГеВ. Його особливістю буде здатність проводити одночасні вимірювання потоків гама-квантів, електронів та позитронів, що може мати вирішальне значення для детектування можливих процесів анігіляції чи розпаду темної матерії. Головною організацією, відповідальною за створення комплексу наукової апаратури ГАММА-400 та проведення наукового експерименту, є Федеральна державна бюджетна установа науки Фізичний інститут ім. П.Н. Лебедєва Російської академії наук (ФІАН, Москва). Науковці КНУ відповідальні в проекті за отримання теоретичних (прогнозних) оцінок очікуваних сигналів в космічній обсерваторії ГАММА-400 від потенційних джерел, участь у виборі об'єктів-мішеней для первинних спостережень, а також за створення і аналіз бази даних одночасних спостережень ГАММА-400-об'єктів в рентгенівському та гама-діапазонах, отриманих іншими космічними місіями.

## 5. ВИСНОВКИ

Недивлячись на 100-літню історію, дослідження космічних променів залишаються одними з найактуальніших в сучасній астрофізиці. З ними пов'язані можливості вивчати структуру матерії та особливості фундаментальних взаємодій при енергіях, що значно перевищують можливості земної експериментальної бази. Є всі підстави вважати, що космічні промені тісно пов'язані із властивостями темної матерії, зокрема, можуть бути продуктами розпаду останньої. Тому подальше вивчення космічних променів, вже на основі нових можливостей XXI століття, пов'язаних із широким використанням космічних місій, є багатообіцяючим не тільки для астрофізики, але й для фізики фундаментальних взаємодій.

## ЛІТЕРАТУРА

# Cosmic ray astrophysics: cosmological aspect

**B.I. Hnatyk**

*Astronomical Observatory, Taras Shevchenko National University of Kyiv, 04053 Kyiv, Ukraine*

Acceleration mechanisms and sources of ultra high energy cosmic rays (UHECRs) are still unclear. Galactic component is, probably, accelerated in Supernova remnants. Extragalactic component of the observed flux of cosmic rays gives valuable information about the physical conditions in places of acceleration (it's probably active galactic nuclei) and in the intergalactic medium. Recent observational data from detectors HiRes, Auger, Telescope Array on the energy spectrum and chemical composition of UHECRs confirm the theoretically predicted cut-off of energy spectrum at energies above $10^{19.7}$ eV (GZK effect), but the chemical composition at the highest energies remains uncertain. This paper presents the results of theoretical modeling of the observed UHECR spectrum and shows that proton model explains all the features of the energy spectrum of cosmic rays in a wide energy range $10^{17}$-$10^{20}$ eV, including the region of transition from galactic to extragalactic component in the total flux of cosmic rays. New results of Supernova remnant investigation are presented. The prospects of using planned space missions Lomonosov (experiment TUS) and GAMMA-400 to establish the nature of UHECRs are discussed (in Ukrainian).

# Исследования космических лучей, свойств частиц и ядер, верхних слоев атмосферы в спутниковых экспериментах

**В.В. Кобычев, Ф.А. Даневич, В.И. Третяк**

*Институт ядерных исследований НАН Украины, Киев, Украина*

Предложен ряд экспериментов с целью исследования космических лучей, свойств частиц и ядер, поиска гипотетических частиц и эффектов, исследования верхних слоев атмосферы, которые могут быть осуществлены с помощью искусственных спутников Земли (на низкоорбитальном научном мини-спутнике, с использованием космической радиоинтерферометрии) в рамках Украинской программы космических исследований.

## 1. ВВЕДЕНИЕ

Измерения потоков, энергии, состава, направления космических лучей позволяют исследовать не только космические объекты, которыми они генерируются, но и свойства частиц. Информация, получаемая от космических источников электромагнитных волн, космических лучей и нейтрино, может быть использована для исследования свойств элементарных частиц и ядер, проверки фундаментальных физических теорий. Мы приводим предварительные оценки экспериментов, которые могли бы быть проведены в рамках Украинской программы космических исследований: как с использованием наблюдений на уже существующих инструментах (проект «Радиоастрон»), так и с разработкой специально сконструированного низкоорбитального спутника. Кроме того, с помощью такого спутника можно исследовать свойства ядер (поиски вариаций скорости радиоактивных распадов) и атмосферу Земли, в частности, измерения радиоактивности верхних слоев атмосферы и потоков частиц от высотных молний.

## 2. ИССЛЕДОВАНИЯ КОСМИЧЕСКИХ ЛУЧЕЙ

### 2.1. Временная структура гамма-всплесков

Гамма-всплески (Gamma-ray bursts) являются одними из наиболее загадочных явлений в астрофизике, поскольку все еще нет приемлемой модели, которая бы объясняла столь огромное выделение энергии. Считается, что длинные всплески (более 2-х секунд - часы) возникают в результате взрывов массивных сверхновых типа Ib и Ic (hyper-nova). Еще менее исследована и понятна природа коротких γ-всплесков, которые продолжаются около секунды [1]. Несмотря на успешные результаты ряда уже завершенных космических экспериментов (BATSE [2], BeppoSAX [3], HETE II [4]), а также функционирующих в настоящее время (в частности, Swift [5] и Gamma-ray Burst Monitor на спутнике Fermi [6]), есть потребность в исследованиях временной структуры таких всплесков и послесвечения. Исследования временной структуры фронта всплесков могло бы дать важную информацию об их природе, в частности, размерах и пространственной структуре объектов, которые их излучают [1].

В качестве детектора для регистрации коротких γ-всплесков предлагается сцинтилляционный детектор с пластмассовыми сцинтилляторами площадью около 1 м$^2$ в виде части полусферы с радиусом около 0.4 м, просматриваемый фотоэлектронным умножителем (ФЭУ) с полусферическим фотокатодом диаметром около 20 см. Полусфера состоит из отдельных элементов гексагональной формы толщиной около 2 – 3 см. Для отделения γ-квантов от заряженных частиц, пластмассовый сцинтиллятор покрыт тонким слоем сцинтиллятора (несколько десятых мм) с более длинной вспышкой (предварительно, вольфрамат кадмия или цинка со временем высвечивания 14 мкс и 25 мкс, соответственно). Детектор будет иметь временное разрешение не хуже 10 нс, система регистрации будет записывать время начала сигнала и его форму в течение от 10 с до нескольких минут. Послесвечение после γ-всплеска может регистрироваться с помощью сцинтилляционного γ-детектора, описанного в подразделе 2.2. Энергетическое разрешение и чувствительность детектора к направлению γ-всплесков не имеют большого значения, поскольку ожидается, что прибор будет работать в составе сети экспериментов Gamma-ray Burst Coordinates Network, измеряющих направление на источник всплеска.

### 2.2. Химический состав космических лучей и поиск сверхтяжелых элементов

В эксперименте PAMELA зарегистрирован избыток позитронов с энергиями 10-90 ГэВ, что может свидетельствовать об аннигиляции частиц тёмной материи [7]. Вместе с тем, детектор не зарегистрировал сопоставимого по величине потока антипротонов, которые также должны излучаться в аннигиляции тёмной материи. Были зарегистрированы «атмосферные» антипротоны в энергетическом интервале 60 – 750 МэВ с отношением к потоку протонов $10^{-4} – 10^{-6}$ [8], возникшие после распада родившихся в верхней атмосфере антинейтронов, а также антипротоны галактических космических лучей с энергиями до 180 ГэВ с отношением к потоку протонов $10^{-4} – 10^{-5}$ [9], спектр которых совместим с гипотезой об их происхождении от вторичных реакций на ядрах межзвёздного газа. Поэтому необходимы дополнительные исследования потоков античастиц с энергиями $10^{1} – 10^{3}$ ГэВ (шкала масс нейтралино – гипотетических слабо взаимодействующих массивных частиц тёмной материи).

Детектор для исследования потоков антипротонов (анти-ядер) может быть построен на основе сцинтилляционных кристаллов вольфрамата цинка (или кадмия), просматриваемых ФЭУ. Детектор будет размещен рядом с вышеописанным детектором γ-всплесков, что позволит с помощью последнего регистрировать γ-кванты, которые будут излучаться в случае аннигиляции антипротонов (анти-гелия и других анти-ядер) в кристалле вольфрамата. Именно наличие γ-излучения будет признаком регистрации античастицы. Детекторы на основе вольфраматов имеют высокую способность различать частицы по форме сцинтилляционных сигналов [10, 11], что позволит исследовать химический состав лучей и отличать события с излучением γ-квантов. Кроме того, прибор сможет вести поиск сверхтяжелых элементов, т.е. предположительно долгоживущих ядер с числом нейтронов N ~ 170 и атомным номером Z ~ 106 – 120, в космических лучах.

### 3. СПУТНИКОВЫЕ ИССЛЕДОВАНИЯ ВАРИАЦИЙ КОНСТАНТ РАДИОАКТИВНОГО РАСПАДА

В ряде работ утверждается [12], что у некоторых изотопов наблюдаются слабые, но статистически значимые временные вариации констант радиоактивного распада, причём эти вариации имеют период, равный одному году, и коррелируют с расстоянием от Земли до Солнца. В этих работах проявляющийся эффект связывается в основном с вариациями потока солнечных электронных нейтрино, который, если пренебречь вакуумными осцилляциями, обратно пропорционален квадрату расстояния до Солнца. Однако зависимость от потока нейтрино представляется сомнительной; в частности, мощный поток антинейтрино вблизи ядерных реакторов не оказывает на скорость распада ощутимого влияния [33]. Предполагая, что наблюдающиеся вариации констант распада не являются экспериментальным артефактом и что они действительно имеют описанную периодичность и фазу, можно отметить, что с тем же периодом и фазой изменяется не только расстояние до Солнца, но и скорость Земли относительно Солнца и относительно Галактики (амплитуда около 29.76 км/с по направлению и около 0.50 км/с по абсолютной величине – 30.27 км/с в перигелии и 29.27 км/с в афелии). На поиск годичных вариаций нацелены эксперименты по прямому детектированию частиц тёмной материи. Кроме орбитального движения вместе с Землёй вокруг Солнца, детектор совершает суточное движение с амплитудой скорости в несколько сотен метров (в зависимости от его широты); суточные вариации также предложены как возможная экспериментальная сигнатура для подобных экспериментов. Таким образом, не только годичные вариации расстояния от Земли до Солнца, но и годичные вариации скорости движения детектора в галактоцентрической и гелиоцентрической системах отсчёта также могут рассматриваться как причина наблюдающихся вариаций скорости распада, пока не прояснена физическая связь последних с орбитальным положением Земли.

Проверку зависимости констант распада от скорости движения (относительно Солнца и Галактики) можно выполнить, разместив радиоактивный источник и детектор на спутнике Земли. Выбрав такое наклонение орбиты спутника, чтобы плоскость орбиты совпадала с галактическим экватором, можно достигнуть периодического изменения скорости детектора в плоскости Галактики с амплитудой до 7,8 км/с и периодом от 1,5 часа. Прибор для измерения вариаций распада может включать в себя

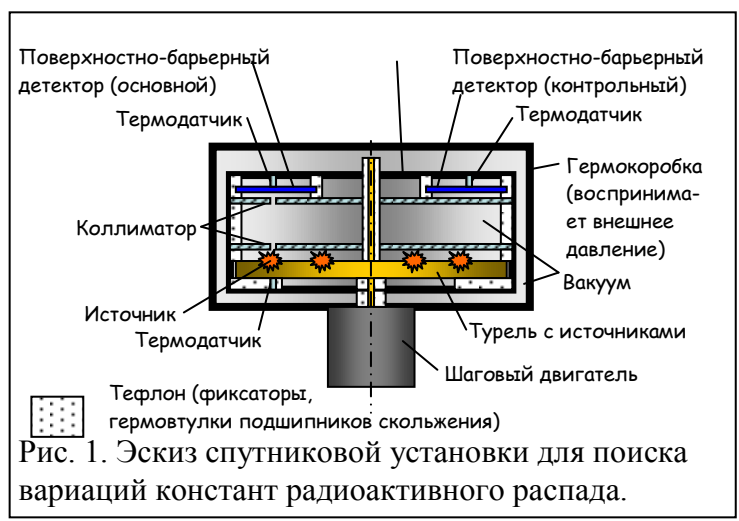

Рис. 1. Эскиз спутниковой установки для поиска вариаций констант радиоактивного распада.

несколько радиоактивных источников с соответствующими детекторами (Рис. 1). Для подавления фона и снижения массы установки можно в качестве детекторов взять поверхностно-барьерные кремниевые детекторы α-частиц, нечувствительные к β- и γ-излучению; в этом случае β-активный изотоп, активность которого измеряется, должен распадаться в дочерний нуклид, испытывающий быстрый α-распад.

### 4. ИЗМЕРЕНИЯ РАДИОАКТИВНОСТИ ВЕРХНИХ СЛОЕВ АТМОСФЕРЫ И ИЗЛУЧЕНИЯ ОТ ВЫСОТНЫХ МОЛНИЙ

#### 4.1. Радиоактивность верхних слоев атмосферы

Радиоактивность верхних слоев атмосферы определяется главным образом процессами взаимодействия космических лучей с ядрами атомов газов, приводящих к образованию трития, $^7$Be, $^{14}$C и $^{37}$Ar, в то время как $^{222}$Rn является продуктом распада урана [13]. Исследование радиоактивности верхних слоев атмосферы даст информацию о процессах взаимодействия космических лучей, наличии тяжелых элементов в верхней атмосфере, процессах обмена веществ в верхних слоях атмосферы и др.

Для измерений радиоактивности верхних слоев атмосферы может быть применен тот же сцинтилляционный детектор с кристаллом вольфрамата кадмия или цинка, который планируется использовать для исследования космических лучей (подраздел 2.2). Энергетическое разрешение такого детектора может составлять от 7% – 8% (вольфрамат кадмия) до 9% – 11% (вольфрамат цинка) для гамма-квантов $^{137}$Cs с энергией 662 кэВ. Детектор сможет измерять гамма-спектр в диапазоне 50 кэВ – 10 МэВ, функционируя в режиме антисовпадений с детектором гамма-всплесков.

## 4.2. Гамма, рентгеновское и нейтронное излучение от высотных молний

Ливни вторичных космических частиц, движущихся снизу вверх через области с повышенной грозовой активностью, могут вызывать высотные атмосферные разряды направленные вверх (Upper-atmospheric lightning или Transient luminous event) [14, 15]. В свою очередь, в таких разрядах могут возникать условия для появления потоков гамма-квантов и других ядерных излучений, направленных вверх. Измерения гамма-квантов могут вестись с помощью гамма-детектора на основе сцинтиллятора вольфрамата кадмия или цинка (подраздел 2.2), а регистрация нейтронов может быть осуществлена с помощью сцинтилляционных детекторов (например, иодида лития [16]) и (или) газовых (наполненных гелием-3) пропорциональных счетчиков. Кроме того, на спутнике должен быть оптический прибор, способный регистрировать высотные атмосферные разряды. Прибор должен посылать сигнал триггера на систему регистрации детекторов с целью передачи данных, которые совпали во времени с наблюдением разрядов. На актуальность спутниковых исследований потоков частиц от высотных атмосферных разрядов указал ведущий сотрудник Института космических исследований РАН И.Г. Митрофанов [17].

## 5. ИССЛЕДОВАНИЯ С ПОМОЩЬЮ КОСМИЧЕСКОЙ РАДИОИНТЕРФЕРОМЕТРИИ И ГАММА-ДЕТЕКТОРОВ

Ряд свойств фотонов и других частиц может быть изучен с помощью наблюдений распространения электромагнитных волн на внегалактических расстояниях. Такие наблюдения позволяют получить ограничения на параметры и свойства частиц, предсказываемые теориями, выходящими за пределы Стандартной модели.

Если распространение света в вакууме отличается от стандартного (предсказываемого Стандартной моделью + ОТО), может наблюдаться отклонение волнового фронта от плоского, частотная дисперсия и расплывание точечных источников при их наблюдении с высоким угловым разрешением, в частности, методом радиоинтерферометрии со сверхдлинной базой (РСДБ). Большинство нижеизложенных эффектов можно рассматривать как распространение света сквозь случайно-неоднородную среду с хаотически меняющимся в пространстве и/или во времени показателем преломления (фазовый экран). В результате изначально плоский волновой фронт становится искажённым, что приводит к видимому расплыванию точечного источника. Вакуумная дисперсия (ВД) приводит к смещению направления на источник внутри полосы частот интерферометра. Из минимального наблюдаемого углового диаметра источника, превышающего инструментальное угловое разрешение, можно ограничить параметры эффекта, приводящего к расплыванию.

ВД приводит также к расплыванию коротких фронтов электромагнитных импульсов. Этот эффект может стать значимым на другом конце спектра электромагнитных колебаний, исследуемом методами гамма-астрономии. Ограничения на ВД в этой области энергий может дать тонкая временная структура гамма-всплесков. Для её исследования предлагается создание спутникового быстрого детектора гамма-квантов с большой апертурой, собирающего несколько миллионов гамма-квантов в секунду от короткого гамма-всплеска. Требования к угловому и энергетическому разрешению такого детектора минимальны. Детектор мог бы представлять открытую полусферическую или цилиндрическую конструкцию из пластикового сцинтиллятора с характерными размерами около метра, обеспечивающую сбор сцинтилляционного света на один или несколько ФЭУ (в случае сегментированного детектора возможно также исследование поляризации гамма-всплесков). Масса такого детектора с толщиной сцинтиллятора, позволяющей поглотить большинство гамма-квантов с энергией до 200 кэВ, оценивается в 100-150 кг. Электронный тракт детектора должен иметь большой динамический диапазон и обеспечивать оцифровку формы сцинтилляционного импульса с высоким временным разрешением.

### 5.1. Электрический заряд и нестандартные электромагнитные моменты фотона

В Стандартной модели электрический заряд фотона и его дипольный момент равны нулю. Впервые ограничения на заряд фотона методом РСДБ были получены в 1992 году [18]. В работе [19] эти верхние пределы были улучшены до $3\times10^{-33}\,e$ с помощью анализа РСДБ наблюдений источника сквозь известное магнитное поле предлежащего скопления галактик. В более поздних работах [20, 21], близких по методологии к нашей, ещё более высокие ограничения на заряд были получены путём учёта эффекта Аронова-Бома при распространении фотона в межгалактических магнитных полях, был также ограничен магнитный дипольный момент фотона. Все эти результаты (см. их список в Review of Particle Physics [22]) были получены для наблюдений с угловым разрешением в несколько миллисекунд дуги. Значительное увеличение базы интерферометра в проекте «Радиоастрон» [23] позволит выполнить анализ наблюдений с угловым разрешением до 7 мкс дуги и уточнить полученные ограничения.

Изучение временной структуры гамма-всплесков при наблюдении сквозь магнитное поле позволяет ограничить эффективный заряд фотона (и другие нестандартные свойства) в области высоких энергий. В работе [19] были получены ограничения на эффективный заряд гамма-кванта порядка $10^{-21}\,e$ в области энергий 100 кэВ для минимальной продолжительности временного фронта короткого гамма-всплеска порядка 250 мс, однако исследование временной структуры гамма-всплесков с высоким разрешением (до ~0,1 мкс) и выявление микросекундных транзиентов позволило бы улучшить эти ограничения и, кроме того, исследовать мелкомасштабную пространственную структуру источников.

### 5.2. Параметры смешивания фотона с аксионом и другими гипотетическими частицами

Аксионом называется гипотетическая массивная псевдоскалярная частица, введённая для решения так называемой «сильной CP-проблемы» в КХД. Аксион рассматривается как один из лучших кандидатов на роль частицы тёмной материи, в связи с чем поиск аксионов выполняется многими группами экспериментаторов, как на ускорителях, так и неускорительными методами. Важным для нас свойством аксионов (и ряда других гипотетических лёгких частиц, объединяемых общим названием ALPs — axion-like particles) является наличие диаграмм Фейнмана с вершинами, связывающими аксион и два фотона. Такие диаграммы описывают как распад аксиона на два фотона, так и конверсию между аксионом и фотоном в статическом магнитном или электрическом поле (эффект Примакова). Магнитное поле смешивает фотон с аксионом, в связи с чем возникает ВД света в магнитном поле. Как указано выше, такая дисперсия может приводить к наблюдаемому уширению точечных источников, рассматриваемых сквозь внегалактическое магнитное поле в радиотелескоп с входной полосой частот конечной ширины. Все замечания относительно наблюдений временной структуры гамма-всплесков, высказанные в предыдущем подпункте, действуют и в этом случае. Следует отметить, что многие заключения, сделанные для аксионов и ALPs, справедливы также для всей группы гипотетических лёгких частиц, слабо взаимодействующих с веществом (фотоны скрытого сектора, хамелеоны, минизаряженные частицы), которые часто объединяются под названием WISPs – Weakly Interacting Sub-eV Particles (см. недавние обзоры [24, 25]); однако ниже за недостатком места мы рассмотрим только аксионы.

Параметры смешивания аксиона с фотоном в магнитном поле зависят от взаимной ориентации поля, волнового вектора и вектора поляризации. Это приводит к специфическим эффектам: вакуумному дихроизму и вакуумному двулучепреломлению. В первом случае мнимая часть вакуумного коэффициента преломления $n_{vac}$ зависит от поляризации света; часть

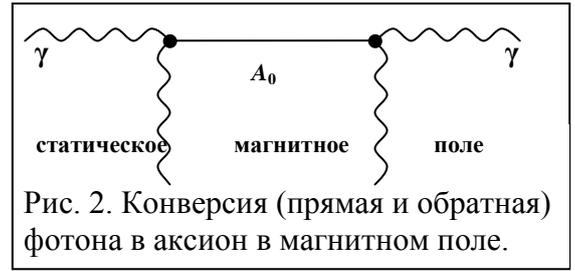

Рис. 2. Конверсия (прямая и обратная) фотона в аксион в магнитном поле.

поперечно (по отношению к полю) поляризованного света конвертируется (Рис. 2) в реальные аксионы и становится ненаблюдаемой, тогда как продольно (вдоль поля) поляризованный свет с полем не взаимодействует. В результате световая волна с круговой поляризацией, пройдя сквозь область магнитного поля, становится эллиптически поляризованной, а плоскость поляризации линейно поляризованного света поворачивается, если её начальный наклон к направлению поля отличен от 0 и 90°.

Во втором случае поле оказывает влияние на действительную часть $n_{vac}$: фазовые скорости распространения света, поляризованного поперечно и продольно по отношению к полю, различны из-за примешивания виртуальных аксионов к поперечной волне. В результате относительная фаза этих волн изменяется, линейно поляризованный свет превращается в эллиптически поляризованный, а цуги волн приходят к наблюдателю с различной задержкой. (Эти эффекты используются для лабораторных поисков аксионов – например, эксперимент PVLAS). В случае наблюдения внегалактических источников (гамма-всплесков, квазиточечных радиоисточников) при распространении электромагнитных волн сквозь области с магнитным полем эти эффекты приведут к пространственному (в картинной плоскости) и временному расплыванию, подобно эффекту, рассмотренному в предыдущем подразделе. Таким образом, наблюдение тонкой структуры источника в пространственной (угловой) и временной областях позволит установить ограничения на константу связи фотона с аксионом, а также на массу аксиона.

### 5.3. Квантовая гравитация, нарушения Лоренц- и CPT-инвариантности

В ряде работ (см., например, [26]) высказаны предположения, что квантовая гравитация приводит к возникновению ВД (а также к малым нарушениям Лоренц- и/или CPT-инвариантности). В одном из вариантов предложенных дисперсионных соотношений $E^2 = p^2 \pm 2\xi p^3/M_{Pl}$, знак перед поправочным членом зависит от поляризации фотонов; параметр $\xi$ был ограничен в работе [26], где изучалась поляризация гамма-всплесков. В радиодиапазоне такие исследования пока не выполнялись. Вызванная указанными эффектами ВД радиоволн должна вести к расплыванию изображения точечного источника, наблюдаемого сквозь гравитационную линзу. Исследования тонкой временной структуры гамма-всплесков также позволят ограничить зависимость скоростей гамма-квантов от их энергии и поляризации.

### 5.4. Спектр масс частиц тёмной материи

Пространство-время с малым отклонением метрики от плоской может рассматриваться как среда с показателем преломления $n_{vac} = 1 + \varepsilon(r, t)$, где $\varepsilon$ — малая положительная величина; $\varepsilon(r, t) = -2\Phi/c^2$, $\Phi$ — ньютоновский гравитационный потенциал. Последний связан уравнением Пуассона с плотностью вещества в данной точке. Статистические флуктуации плотности движущихся с вириальными скоростями частиц тёмной материи вызывают флуктуации коэффициента преломления и создают случайно-неоднородный фазовый экран между удалённым источником и радиоинтерферометром. Расплывание точечного источника позволило бы определить статистические параметры неоднородностей и получить отсюда информацию о массе частиц тёмной материи (или о спектре масс мелкомасштабных флуктуаций её плотности). В качестве экрана удобно использовать протяжённые области пространства с доказанной высокой плотностью тёмной материи и малой плотностью барионного вещества (гало галактик и их скоплений). Задача о статистических флуктуациях наклона волнового фронта при распространении радиоволны сквозь газ гравитирующих частиц была рассмотрена (для звёзд в качестве частиц) в работе [28], где было показано, что средний квадрат угла наклона $\Delta\theta^2$ за время $T$ пропорционален $T^2 r_g n_0 \log n_0$ (здесь $r_g = 2GM/c^2$ — гравитационный радиус частицы, $n_0$ — концентрация частиц), причём за время порядка года $\Delta\theta$ достигает микросекунд дуги, что близко к угловому разрешению эксперимента «Радиоастрон». При увеличении объёмной концентрации частиц и уменьшении их массы (с сохранением общей массовой плотности $n_0 M = \mathrm{const}$) угол $\Delta\theta^2$ растёт как логарифм концентрации. Поэтому фундаментальное ограничение на угловое разрешение РСДБ, установленное в [28], оказывается более существенным, если скрытая масса образована меньшими частицами – звёзды оказывают меньшее влияние, чем вимпы, при той же средней плотности массы.

Газ вимпов образует статистические микрофлуктуации плотности и, соответственно, флуктуации гравитационного потенциала и эффективного коэффициента преломления, которые, в свою очередь, создают случайно-неоднородный фазовый экран для света. Микролинзирование на некомпактных небарионных объектах и мелкомасштабных флуктуациях плотности тёмной материи рассмотрено в обзоре [29].

### 5.5. Гравитационные волны

Как уже было отмечено выше, $(n_\mathrm{vac} - 1)$ в слабом гравитационном поле пропорционально ньютоновскому потенциалу (т.е. элементу $g_{00}$ метрического тензора). Таким образом, пространство, сквозь которое распространяется гравитационная волна (ГВ), может рассматриваться в качестве фазового экрана для света. Например, можно проанализировать фронт электромагнитной волны, движущийся мимо галактики с активным ядром или другими предположительными источниками ГВ. Как регулярные, так и стохастические искажения, создаваемые фоновыми ГВ на световом фронте, приведут к видимому расплыванию точечных радиоисточников. Ожидается, что такой метод будет чувствителен к длинам ГВ порядка базы радиоинтерферометра.

Однако в ряде работ [30, 31] показано, что распространение света в поле стохастической гравитационной волны не приводит к накоплению фазового сдвига, пропорционального корню квадратному из длины пробега, как это было бы для случайно-неоднородного фазового экрана. Тем не менее, в [30] даются довольно оптимистические оценки для возможности детектирования как реликтовых ГВ, так и вспышек ГВ, связанных с гамма-всплесками, сверхновыми и т.д. Так, фазовый сдвиг, получаемый электромагнитной волной с длиной $\lambda_e$, проходящей сквозь цуг низкочастотных ГВ с амплитудой $h$ и длиной $\lambda_g$, оценивается как $\sim (h/2)\cdot(\lambda_g/\lambda_e)$. Для радиоинтерферометров сантиметрового диапазона, гравитационная волна с амплитудой $10^{-18}$ и периодом 1 сутки создаст фазовый сдвиг порядка $10^{-3}$. В [31] предложено использовать указанное почти полное исчезновение отклонения света в поле стохастической гравитационной волны, вызванное общерелятивистскими симметриями, для тестирования ОТО: ограничения на угловой диаметр точечных радиоисточников можно свести к ограничениям на отклонения от уравнений ОТО для ГВ.

### 5.6. Электромагнитные свойства частиц тёмной материи

Наличие у частиц тёмной материи незначительных электромагнитных моментов различных порядков не исключено наблюдениями. Анализ распространения электромагнитной волны сквозь газ из массивных нейтральных частиц, обладающих малыми дипольными магнитными моментами (или другими электромагнитными моментами) и сопоставление этого анализа с уже упоминавшимися наблюдательными угловыми и временными параметрами сигнала (в радио- или гамма-диапазоне) может позволить установить ограничения на эти электромагнитные моменты.

## ЗАКЛЮЧЕНИЕ

Мы предлагаем исследования и эксперименты для изучения верхней атмосферы, космических лучей, проверки ряда гипотетических явлений, относящихся к физике частиц и ядра, которые могут быть выполнены в рамках Украинской программы космических исследований. Часть из них сводится к анализу РСДБ наблюдений в проекте «Радиоастрон». Детектор гамма-всплесков с высоким временным разрешением, прибор для поиска вариаций радиоактивного распада, детектор для измерений состава античастиц (анти-ядер) в космических лучах (который может быть также задействован для изучения радиоактивности верхних слоев атмосферы и регистрации гамма и нейтронного излучения от высотных атмосферных разрядов, направленных вверх) должны быть размещены на борту малого научного спутника на низкой околоземной орбите.

## СПИСОК ЛИТЕРАТУРЫ

# Satellite experiments to study cosmic rays, properties of particles and nuclei, to search for effects beyond the Standard Model of particles, and to investigate upper atmosphere

**V.V. Kobychev, F.A.Danevich, V.I.Tretyak**

*Institute of Nuclear Research, MSP 03680 Kyiv, Ukraine*

We propose experiments to study cosmic rays, radioactivity of upper atmosphere and irradiation from transient luminous events, to investigate properties of particles and variation of nuclear decay, to search for effects beyond the Standard Model of particles, which can be performed with satellites (a low-orbit scientific mini-satellite, space radio interferometry) in the frame of the Ukrainian Space Research Program (in Russian).

# Inhomogeneity growth in the one-dimensional pressureless matter: direct integration in the Fourier space

**V.M. Sliusar [1], V.I. Zhdanov [2]**

*National Taras Shevchenko University of Kyiv, Astronomical Observatory*
[1]*vitaliy.slyusar@gmail.com,* [2]*ValeryZhdanov@gmail.com*

We propose a method of investigation of the power spectrum of cosmological perturbations by means of a direct numerical integration of hydrodynamic equations in the Fourier space for a random ensemble of initial conditions with subsequent averaging procedure. This method can be an alternative to the cosmological N-body simulations. We test realizability of this method in case of one-dimensional motion of gravitating pressureless shells. In order to test the numerical simulations, we found an analytical solution which describes one-dimensional collapse of the plane shells. The results are used to study a nonlinear interaction of different Fourier modes.

## 1. INTRODUCTION

Most developed computational techniques to study the cosmological structure formation involve N-body simulations combined with the smoothed particle hydrodynamics [1-3]. Currently performed simulations involve up to $10^9$ particles. These methods should be tested in independent simulations. On the other hand, analytical and semi-analytical schemes dealing with the Fourier-transformed hydrodynamical variables were worked out [4-9]; they typically involve approximations and/or some *apriori* assumptions after transition to the Fourier space. A comparison of different approaches can be found in [9].

In this report, we propose a method that uses a direct integration of hydrodynamical equations in the Fourier space. In order to estimate workability of the method, as a first step, we consider a one-dimensional problem of hydrodynamical evolution for a non-relativistic pressureless gravitating matter. The integration is performed for each realization from a random ensemble of initial data with subsequent averaging procedure.

## 2. NUMERICAL SIMULATIONS

For simplicity, we deal with a symmetric (with respect to the point $x=0$) one-dimensional motion of pressureless fluid. In case of homogeneous initial conditions for the density $\rho(x,0)=\rho_{h0}(x)$ and velocity $V(x,0)\equiv V_{h0}(x)=H_0 x$ ($H_0$ stands for the one-dimensional "Hubble constant"), we get:

$$\rho_h(x,t)=\frac{\rho_{h0}}{R(t)}, \quad V_h(x,t)=x\frac{\dot R(t)}{R(t)}, \quad R(t)=1+H_0 t - 2\pi G\rho_{h0} t^2. \tag{1}$$

We denote $\mathcal{H}(\tau)\equiv dR/dt$, $\tau$ is the "conformal time": $dt=R(\tau)d\tau$, and $y=x/R(\tau)$ is the comoving spatial coordinate. Then

$$\tau=\frac{1}{2H_0}ln\left[\frac{1+3H_0 t}{3(2-H_0 t)}\right], \quad \mathcal{H}(\tau)=\frac{dR}{dt}=\frac{1}{R}\frac{dR}{d\tau}=-2H_0\tanh(H_0\tau).$$

Below $\delta$ stands for the density contrast, $\theta=\partial v/\partial y$, $v$ is the peculiar velocity.

Further we work with the Fourier coefficients in case of a spatially-periodic motion with period $L$: $\delta(x,\tau) = \sum_{n=-\infty}^{\infty} b_n(\tau)\exp(ik_n x)$, where $k_n = 2\pi n/L$, $\theta(x,\tau) = \sum_{n=-\infty}^{\infty} a_n(\tau)\exp(ik_n x)$. We assume $a_0 = 0$, $b_0 = 0$ at $t = 0$; one can show that these equalities are fulfilled for all $t > 0$. The equations for the Fourier coefficients take on the form (analogous to that of papers [4-8]):

$$\frac{da_n}{d\tau} + \mathcal{H}(\tau)a_n + \alpha R(\tau)b_n + n\sum_{\substack{p=-\infty \\ p\neq 0}}^{\infty} \frac{a_p a_{n-p}}{p} = 0, \quad n = \pm 1, \pm 2,... \quad (2)$$

$$\frac{db_n}{d\tau} + a_n + n\sum_{\substack{p=-\infty \\ p\neq 0}}^{\infty} \frac{a_p b_{n-p}}{p} = 0, \quad n = \pm 1, \pm 2,..., \quad (3)$$

where we have chosen a special case $\alpha = 3H_0^2/4$. The numerical solution of the equations (2), (3) was performed using the 4-th order Runge-Kutta method after a truncation of the infinite chain of coefficients $a_n, b_n$. Calculations were carried out by the specially written GPGPU code using OpenCL SDK by AMD. The time, required to calculate $a_n$ and $b_n$ on a personal computer, in case of 256 values of $n$ (points over $k$) for single realization of initial conditions, is about 10 seconds. To test the results we found a semi-analytical solution of the hydrodynamic problem in terms of the Lagrange variables.

In order to look how the perturbation propagates from small wavenumbers to larger ones, we considered the following initial conditions: $b_n(0) = 0$ where integer $n$ varies $-128$ to $128$ except $n = \pm 1$; $b_{\pm 1}(0) = b_{\pm 1}^0/2 = 0.1$; all $a_n(0) = 0$; we put $H_0 = 1$ that defines the units of the time. The results (for semi-analytical and numerical methods) are presented in Fig. 1 for two values of $t$. For larger $t$ we observe an infinite growth (for finite time) that corresponds to collapse of plane gravitating shells. The error of calculation increases near the points of the collapse; it is larger for larger wavenumbers.

Then we calculated the power spectrum obtained by averaging of the solutions for the ensemble of initial data with uniform distribution of $b_{\pm 1}(0)$, $<b_{\pm 1}^2(0)> = 0.5$. We observe the significant growth of the variance, which is explained as follows: as $t$ grows, some of realizations of the ensemble of solutions (with larger $|b_{\pm 1}(0)|$) for given $t = 0.8$ enter the region yielding the singularity. This example is shown in Fig. 2.

We expect that our approach will be especially effective in the weakly nonlinear regime. The trial runs of our method allow us to think that it could be really used for the power spectrum calculations in the 3-D case with realistic requirements to the computer time.

## ACKNOWLEDGEMENTS

This work has been supported in part by Swiss National Science Foundation (SCOPES grant 128040).

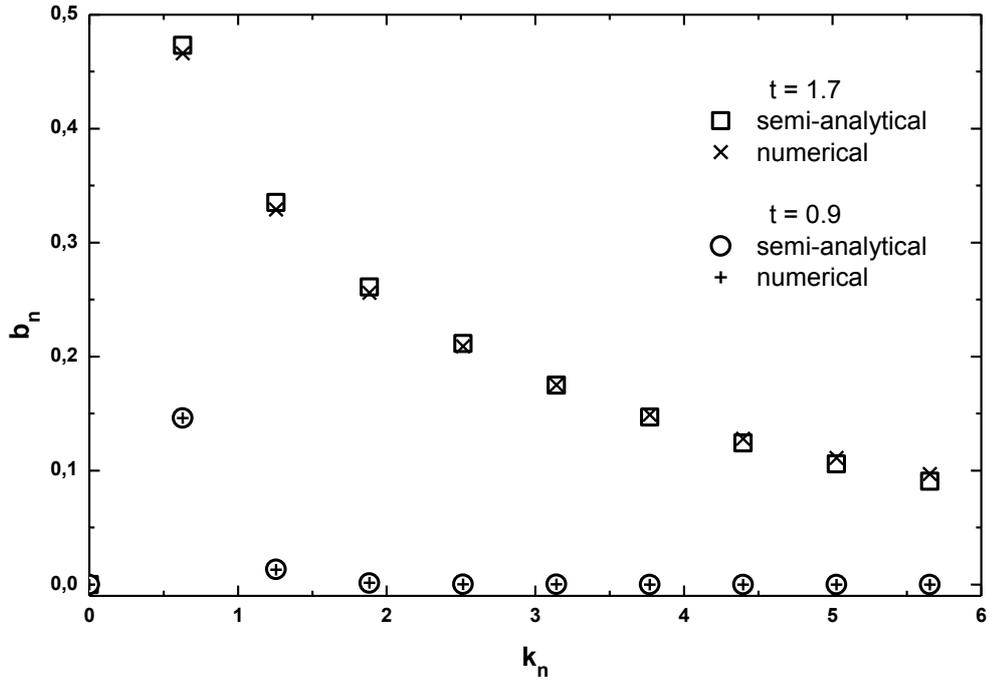

Fig. 1. Coefficients $b_n(t)$ for $t = 0.9$ (bottom) and $t = 1.7$ (top) determined by semi-analytical and numerical methods with the only nonzero initial values $b_{\pm 1}(0) = 0.1$.

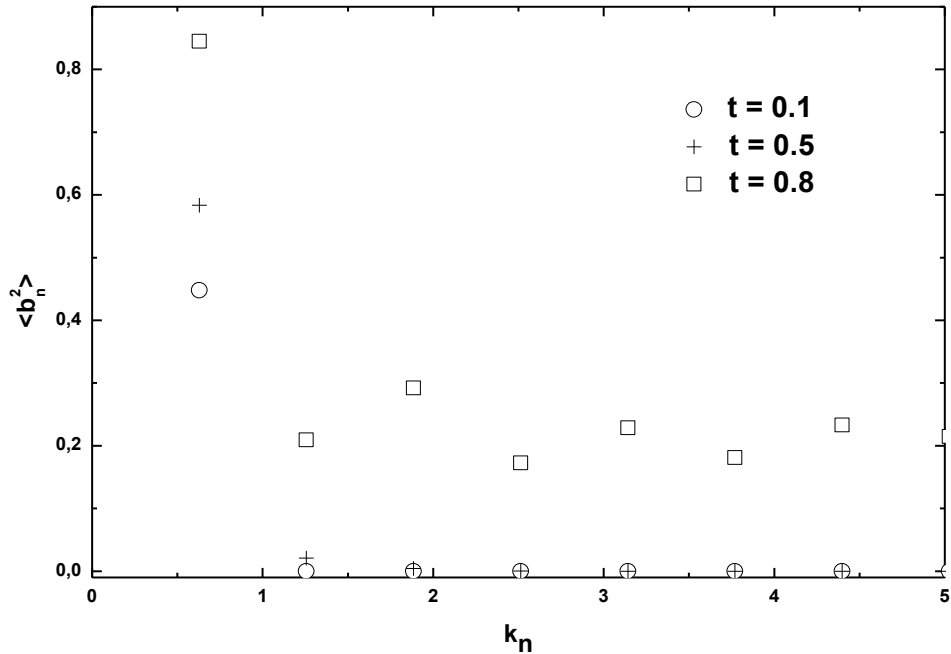

Fig. 2. Power spectrum $<b_n^2>$ calculated for 100 realizations of randomly generated initial conditions: uniform distribution of $b_{\pm 1}(0)$, $<b_{\pm 1}^2(0)> = 0.5$; the other initial coefficients equal to zero. We see that for $t = 0.8$ the perturbation becomes nonzero for larger $k_n$ due to the non-linear interaction.

# Luminosity dependence of the quasar clustering from SDSS NBCKDE catalogue

O. Vasylenko[1], G. Ivashchenko[2]

[1]*Faculty of Physics, Taras Shevchenko National University of Kyiv, 03127, Kyiv, Ukraine*
[2]*Astronomical Observatory, Taras Shevchenko National University of Kyiv, 04058, Kyiv, Ukraine*

We study the clustering of quasars from the SDSS NBCKDE catalogue of photometrically selected quasar candidates (SDSS DR6). Dividing our sample with $0.8 < z_{phot} < 2.2$ onto three luminosity bins we have found no evidence for luminosity dependence of the quasar clustering. It is consistent with the models of the quasar formation, in which bright and faint quasars are assumed to be similar sources, hosted by dark matter halos of similar masses, but observed at different stages of their evolution.

## 1. INTRODUCTION

Formation of quasars, and active galactic nuclei (AGN) in general, is still an open question, which is connected to their clustering. Several models of quasar formation (e. g. [3, 6, 12]) were proposed, and based on studying the luminosity dependence of the quasar clustering, some of models were completely ruled out, while some others are partially supported by observations. The luminosity dependence of the quasar clustering was previously studied mainly for the 2QZ and the SDSS surveys, the large size of which allows to divide a sample onto redshift and luminosity bins simultaneously. Some authors (e. g. [1, 2, 13]) report no evidence for luminosity dependence, while some others (e. g. [10]) found marginal evidence for it.

## 2. COMPILATION OF SAMPLES

Our sample is taken from the SDSS NBCKDE Catalogue of Photometrically Classified Quasar Candidates [11] that contains 1 015 082 quasar candidates selected from the photometric imaging data of the SDSS using a non-parametric Bayesian classification kernel density estimator (NBC-KDE). We selected only the objects with photometric redshifts within the range $0.8 \leq z_{phot} \leq 2.2$ and photometric redshift range probability $z_{photprob} > 0.5$. To check the luminosity dependence of the quasar clustering we selected three samples, which we called *bright*, *medium* and *faint* samples. They contain 73 307, 77 891 and 73 262 objects with absolute magnitudes in *i*-band within the ranges $-27.0 \leq M_i \leq -25.3$, $-25.3 \leq M_i \leq -24.5$, and $-24.5 \leq M_i \leq -22.5$. Corresponding mean values of $M_i$ are -25.89, -24.90 and -23.33.

## 3. PARAMETERS OF ANGULAR AND SPATIAL CLUSTERING

To analyse the quasar clustering we used the angular and spatial 2-point correlation functions (2pCF). The former was calculated using the Landy-Szalay estimator [7]:

$$\omega_{LS}(\theta) = \frac{DD(\theta)}{RR(\theta)} - 2\frac{DR(\theta)}{RR(\theta)} + 1, \qquad (1)$$

where $DD(\theta)$ and $RR(\theta)$ are the numbers of pairs with separations $\theta$ in initial (data-data) and random (random-random) samples, respectively, and $DR(\theta)$ is a number of cross-pairs between data and random samples (data-random). In each case 20 random catalogues were generated with own technique, the idea of which was proposed in [4]. Detailed description of this technique will be presented elsewhere. The values of *RR* and *DR* were calculated as the arithmetic means of 20 corresponding values from random catalogue. The angular and spatial 2pCF of quasars within some limited scales are fitted well with the power law:

$$\omega(\theta) = \left(\frac{\theta_0}{\theta}\right)^\alpha, \quad \xi(r) = \left(\frac{r}{r_0}\right)^\gamma. \tag{2}$$

where $\theta_0$ and $r_0$ are the angular and spatial correlation lengths, $\alpha$ and $\gamma$ are corresponding slopes, and $\gamma = 1 + \alpha$. Assuming $\xi(r)$ to be described by the power law, and following Myers [9], we used the Limber's equation [8] to reconstruct $r_0$ from obtained $\theta_0$, $\alpha$ and corresponding spectroscopic redshift distribution for each sample. Adopting the extrapolation of the dark matter 2pCF to high redshift according to the linear growth model, and assuming the correlation length of the dark matter to be $r_{0;m}(z = 0) = 5\ h^{-1}$ Mpc [5], we also calculated the quasar linear bias, which relates the 2pCFs of the dark matter, $\xi_m(r)$, and that of the luminous matter, e. g. quasars, $\xi(r)$, as $\xi(r) = b^2 \xi_m(r)$.

## 4. RESULTS AND DISCUSSION

In Fig. 1 the obtained angular 2pCF for three samples with different luminosity along with the best fits within 1' – 40' range are shown. The corresponding parameters of the fit along with the spatial 2pCF parameters and $b$ are presented in Table 1. As one can see, the values of $r_0$, $\gamma$ and $b$ for *faint* sample differ from those for *bright* and *medium* samples, but nevertheless, the values for all three samples with different luminosity are consistent within 1σ. The discrepancy can be attributed to the difference in redshift distribution for these samples. Hence, we cannot argue for any luminosity dependence of the quasar clustering. This result is consistent with previous studies of spectroscopically confirmed quasars (e. g. [1, 2, 10, 13]), as far as with study of similar samples of photometrically classified quasars by [9].

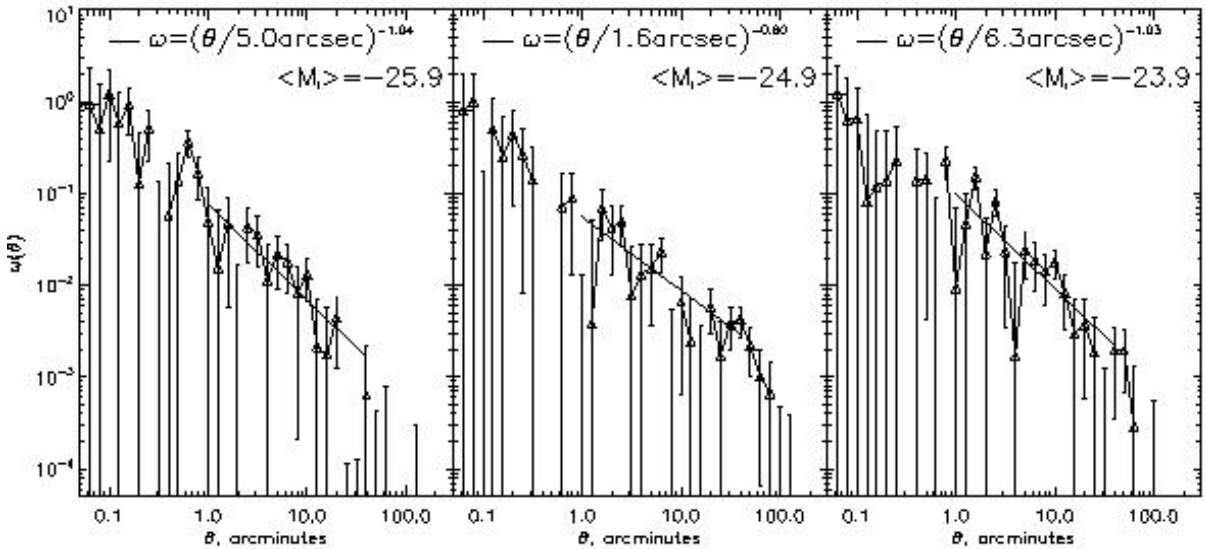

Figure 1: The angular 2pCF for (left to right) *bright*, *medium* and *faint* quasars. The solid line is the best fit within 1' – 40' angular range.

Table 1: Parameters of the angular and spatial 2pCF, and the bias parameter for the samples of quasars with different luminosity within $1' \leq \theta \leq 40'$ angular range.

| sample | $\alpha$ | $\theta_0$, arcsec | $\chi^2$/d.o.f. | $\gamma$ | $r_0$, $h^{-1}$ Mpc | $b$ |
|---|---|---|---|---|---|---|
| *bright* | $1.04^{+0.16}_{-0.09}$ | $5.04^{+2.41}_{-2.45}$ | 1.8 | $2.04^{+0.16}_{-0.09}$ | $11.26^{+2.48}_{-3.24}$ | $3.27^{+0.74}_{-0.95}$ |
| *medium* | $0.80^{+0.08}_{-0.05}$ | $1.62^{+0.82}_{-0.74}$ | 2.8 | $1.80^{+0.08}_{-0.05}$ | $9.08^{+1.82}_{-2.15}$ | $2.35^{+0.41}_{-0.51}$ |
| *faint* | $1.03^{+0.10}_{-0.07}$ | $6.32^{+2.12}_{-2.15}$ | 5.0 | $2.03^{+0.10}_{-0.07}$ | $7.85^{+1.24}_{-1.49}$ | $2.26^{+0.36}_{-0.44}$ |

It is believed that the main mechanism, that drives the formation of supermassive black holes and triggers quasar activity in ultraviolet and X-ray range, is galaxy mergers. Some of the first models of quasar formation (e. g. [6]) assume that quasars radiate at fixed luminosity for some characteristic lifetime, thus there should be a tight relation between the instantaneous quasar luminosity and the mass of the central black hole, which in its turn depends on the dark matter halo mass. The next generation of models (e. g. [3]) assumes that bright and faint quasars are similar sources, observed at different stages of their evolution, thus a broad range of the quasar luminosity corresponds to a narrow range in the masses of quasar host halos, and quasar clustering should depend only weakly on luminosity. Predictions of the latter models are consistent, at least qualitatively, with the results on optical luminosity dependence of the quasar clustering, obtained by different authors, and in the present work.

## ACKNOWLEDGEMENTS

The authors are thankful to the Sloan Digital Sky Survey team. Funding for the SDSS has been provided by the Alfred P. Sloan Foundation, the Participating Institutions, the National Aeronautics and Space Administration, the National Science Foundation, the US Department of Energy, the Japanese Monbukagakusho, and the Max Planck Society. This work has been supported by Swiss National Science Foundation (SCOPES grant No 128040).

# Studying the X-ray properties of Seyfert 1.9 galaxy NGC 1194 with XMM-Newton and INTEGRAL observational data

A. Vasylenko, E. Fedorova, V.I. Zhdanov

*Astronomical Observatory of Kiev National Taras Schevchenko University, 04053 Kyiv, Ukraine*

We analyze the X-ray properties of Compton thick Seyfert 1.9 Radio quiet AGN in NGC 1194 from both the INTEGRAL/ISGRI and XMM-Newton observational data, paying the special attention to the set of iron emission lines between 6 and 7.5 keV. No statistically significant flux variability was found. Three emission lines can be resolved in this range, near 6.37, 6.5 and 7.0 keV, and two of them follow the relativistic profile of Fe Kα line typical for the accretion disk surrounding the Schwarzshild black hole. This suggests an idea about the double black hole system; corresponding parameters of accretion disks are presented.

## 1. INTRODUCTION

NGC 1194 is the S0-spiral galaxy in Cetus constellation, with a redshift of z=0.013596. Alternative names of this object are UGC 2514, MCG 0-8-78, ZWG 389.68, IRAS03012-0117, PGC 11537. The controversial radio-quiet active nucleus is nested here. The AGN type here was being a subject of discussions during more than decade: initially it was identified as a LINER [1], then as S1 type AGN basing on infrared observations [2, 3], despite the suspicions that it's a HBLR S2 [4]. Optical spectroscopic results asserted it as S1.5. The final decision was made only recently, after the investigations of the megamaser emission of this object, that NGC 1194 contain the Seyfert 1.9 type active nucleus. Then the new method of type determination using the X-ray and IR data both, developed by Severgnini et al. [5], finally confirmed that this object is Compton-thick. It was also included in the Palermo Swift-BAT hard X-ray catalogue [6] and Swift BAT Survey of AGNs [7].

Another interesting property of NGC 1194 is its water megamaser emission from the subnuclear area. Following the results obtained by Kuo et al. [8] from the VLBI observations of several megamaser sources, the inclination of the maser disk is ≈85°, the inner and the outer radii are 0.54 and 1.33 pc, respectively, and the measured BH mass is $(6.5 \pm 0.3) \times 10^7 M$.

Here we present the results of our treatment of the INTEGRAL/IBIS data on NGC 1194 (these data are published for the first time) together with the results of XMM-Newton observations. In Section 2 we describe the data reduction and the spectral analysis. In the Section 3 we investigate the iron emission lines in details.

## 2. DATA REDUCTION AND ANALYSIS

The data analyzed here include all publicly available INTEGRAL data on the source as of July 2012, i.e. 208 science windows (SCW) for the IBIS/ISGRI instrument. The total exposure time of the data set used for the IBIS/ISGRI spectrum is 1,6 Ms, including all observations when NGC 1194 was within $10^o$ off-axis angle. ISGRI data analysis was performed using standard procedures of OSA 9.0 (Offline Standard Analysis Software), the task *ibis_science_analysis*. The energy range where the resulting spectrum can be detected is 15-300 keV.

NGC 1194 was observed with XMM-Newton 19.02.2006 (ID 0307000701) during the exposure time 16 ks. During this time it was in the field-of-view of all the three EPIC cameras which were operating in the full-frame mode. EPIC data were processed with XMM-Newton Standard Analysis System (SAS, public release version 11.0, see http://xmm.vilspa.esa.es/sas/). After filtering of bad-flagged data, the effective exposure times were: 12.5 ks for PN-camera, and 15.6 ks for both the MOS-cameras. The energy range of EPIC cameras is 0.2-12 keV.

To obtain the spectra and the cleaned light curves, we extracted full count rates from circular areas with different radii around the source. To estimate the background effects we also used areas of different sizes near the source on the same CCD plates. Flux variations from different background areas agree with the corresponding estimated Poisson variance. These variations have been taken into account in the final error estimates of the resulting net signal. The background counts were subtracted (with corresponding factors taking into account different background areas) from the total counts from the source regions to obtain the cleaned light curves, which were added together in one synthetic curve. No variability was found in this background-subtracted light curve. EPIC spectra were obtained through standard SAS procedures: *evselect, arfgen*, and *rmfgen*. We had not extracted soft proton flares with *tabgtigen* due to the absence of significant flares during the time of the observation. After all, the three EPIC spectra from MOS1, MOS2 and PN cameras were merged together using the HEASOFT 12.6 standard procedure *addspec*. This background subtracted total EPIC spectrum together with the INTEGRAL/IBIS spectrum are shown in Fig.1.

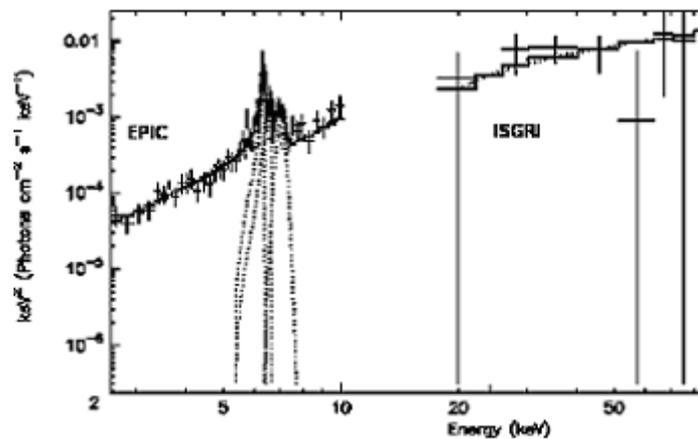

Fig. 1. Broadband merged of XMM EPIC & INTEGRAL ISGRI unfolded spectrum of NGC 1194.

For the spectral analysis we have used the specialized program XSPEC, version 12.6, which is a part of the HEASOFT 12.6 software package. To obtain the parameters of the continuum we had excluded from the consideration the energy range 6-7 keV and fitted the spectrum with a Compton-reflected power-law model (*pexrav* [9]), absorbed both in the AGN torus (*zphabs*) and in the Galaxy (*phabs*), using the frozen value of the hydrogen column density in the Galaxy of $6.03 \times 10^{20}$ cm$^{-2}$, determined by Kalberla et al. [10].

The best-fit parameters, obtained for the continuum within the range 0.1-5.5 keV and 7.5-200 keV, are: photon index $\Gamma=1.09\pm0.03$, reflection parameter $R=1.2\pm0.9$, hydrogen column density $N_H=(1.08\pm0.05)\times10^{24}$ cm$^{-2}$, no high-energy cut-off found.

### 3. IRON EMISSION LINES

Between 6.0 keV and 7.5 keV three emission lines in the EPIC spectrum can be clearly seen. We had modeled them with several models (i.e. unrelativistic Gaussian profile *zgauss*, relativistic profile *diskline* emitted by the matter accreting onto the Schwarzschild black hole, *relline* profile for Kerr black hole and *laor* profile for hyper-spinning black hole). The best-fit model for the first two lines (at 6.37 and 6.51 keV) is *diskline*, with the null-hypothesis probability of 99% vs. Gaussian model and 90% vs. *laor* one. The Kerr black hole model gave the results marginally compatible with zero value of a black hole spin. The line near 7.0 keV appears to be a Gaussian. The line parameters within the best-fit model are:

- Fe K$_{\alpha1}$ line, energy $6.38\pm0.04$ keV, equivalent width $560^{+250}_{-90}$ eV, disk parameters: inner radius $(12\pm4)R_{Sh}$, inclination angle $6\pm3$;

- Fe $K_{\alpha 2}$ line, energy 6.51±0.05 keV, equivalent width $460^{+90}_{-70}$ eV, disk parameters: inner radius $(16\pm 6)R_{sh}$, inclination angle 20±4;
- Gaussian line at the energy of 7.01±0.05 keV, equivalent width $310^{+200}_{-120}$ eV, line width 0.16±0.05 keV.

The latter line can be interpreted rather as the recombination line of hydrogen-like $Fe^{24}$ (6.97 keV rest frame), than Fe-$K_\beta$, due to its unrelativistic profile. This line cannot be formed in the same media as previous two lines (that is why it hardly can be Fe-$K_\beta$); it can be emitted rather in the BLR area or even in the highly-ionized, spherically falling in medium of the hot corona surrounding the central black hole. To explain the other two line shapes and other parameters, we can propose three possible versions:

- The double black hole system; following Yu & Lu [13], in this case we can see the two lines with energies slightly higher and slightly lower than initial one as we see here. We consider this interpretation of doubled Fe $K_\alpha$ lines as the most probable.
- The first (6.37 keV) line is produced in the process of the Compton-scattering of the second one (6.51 keV) into the line-of-the-sight from some surface (for instance, the torus wall), as it was described by Wang et al. [11].
- The lines (or rather one complex-shape two-horned line) are produced in the warped Bardeen-Petterson disk [12].

These two latter interpretations seem to be less probable. In case of the reflection or scattering, the form of the secondary line will be distorted due to velocities of the reflecting material. To obtain the exact form and position of the line needs the fine tuning in the velocity distribution of the reflecting matter. Analogous precaution refers to the case of the warped disk.

Unfortunately, existing data are not enough to make a final choice between these models. In perspective, long-term observations of this object in X-rays can give us a possibility to test these models by analyzing the line shapes as well as the continuum reflection. Thus, we can conclude that NGC 1194 appears to be rather interesting object with ambiguous and complicate properties. There is at least one known object, NGC 4151, which is in some sense similar to it. The iron emission lines in NGC 4151 follow the similar pattern, moreover, it is also a double black hole candidate with orbital period of the black holes of ~ 16 years, following Bon E. et.al. [14]. So we can finally stress out that to understand the nature of the spectral peculiarities in NGC 1194, we need to have longer series of observations in X-rays, as well as in optical and radio wavebands.

# The mean transmission of the Lyα-forest at 2<z<4 from a sample of 33 high-resolution quasar spectra

O. Torbaniuk[1], G. Ivashchenko[2]

[1]*Faculty of Physics, Taras Shevchenko National University of Kyiv, 03127, Kyiv, Ukraine*
[2]*Astronomical Observatory, Taras Shevchenko National University of Kyiv, 04058, Kyiv, Ukraine*

We present our measurements of the mean transmission $\bar{F}(z)$ of the neutral intergalactic medium in the Lyα line. For this purpose we used the sample of 33 high-resolution (R=5300-45000) spectra obtained from the public available archive of the European Southern Observatory (ESO) with the help of the Virtual Observatory tools. The obtained values of $\bar{F}(z)$ are consistent with those of previous studies of high- and low-resolution quasar spectra.

## 1. INTRODUCTION

The Lyα-forest is a set of Lyα absorption lines in quasar spectra originated from the neutral hydrogen, HI, the main component of intergalactic medium (IGM) in filaments. Intensity (*f*) and observed wavelength ($\lambda_{obs}$) of these lines are determined by the density of HI in a given absorbing region and the distance to this region, respectively. Hence, the opacity of IGM in the Lyα line, $F(\lambda_{obs}) = f(\lambda_{obs}) C(\lambda_{rest}, z)$ (*C* is the intrinsic quasar spectrum), which is called transmission, traces the thermal evolution of the Universe and the grows of structures in it. It is related to the optical depth $\tau_{eff}$ via $\tau_{eff} = -\ln(F)$. A number of studies have been carried out for determination of the redshift dependence of the mean transmission F(z), using different techniques and different samples of high- and medium-resolution spectra (e. g. [1,3-5]). We tried to summarize the current state of these studies in Fig. 1.

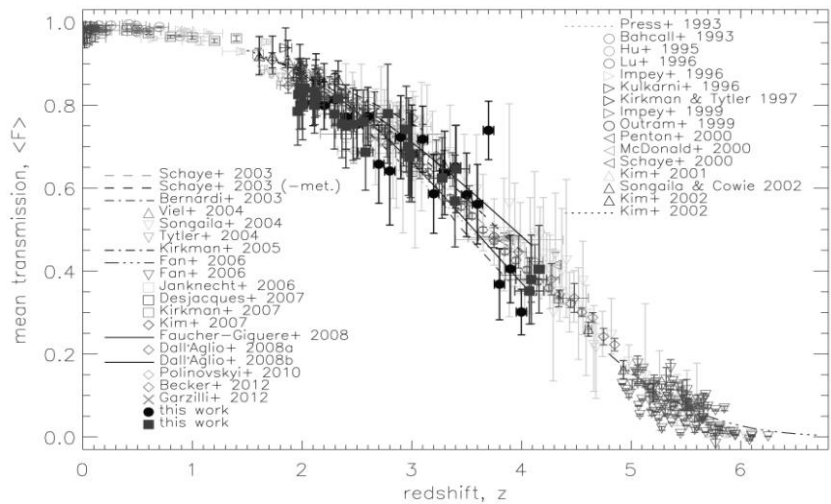

Figure 1: Redshift dependence of the mean transmission: results of previous studies.

## 2. THE DATA

To compile our sample, we searched in the literature for quasars with *z*>2, the spectra of which were obtained with high-resolution spectrographs of ESO and which were previously used for studying the Lyα forest (e. g. [2,4,5]). All those spectra available in the ESO public database were obtained using the Virtual Observatories spectroscopic tool *VOSpec*.

Our sample contains 33 spectra obtained with ESI/Keck (4 spectra, R~5300), MIKE/Magellan (2 spectra, R~28000), HIRES/Keck (8 spectra, R~36000), UVES/VLT (17 spectra, R~45000), and FORS2/VLT (2 spectra, R~10000) spectrographs.

## 3. THE METHOD

Firstly, the spectra were (i) averaged over several exposures and re-binned with 0.1 Å step, and (ii) cleaned from telluric lines and bad pixels with the *DECH* software. The noise set for each new pixel is the rms of all pixels in the given bin. To fit continuum in the Lyα-forest region (we consider the wavelength range of 1050<$\lambda_{rest}$<1180 Å) we interpolated the most absorption-free regions with the *IDL spline* subroutine. Two spectra with fitted continuum are presented for example in Fig. 2.

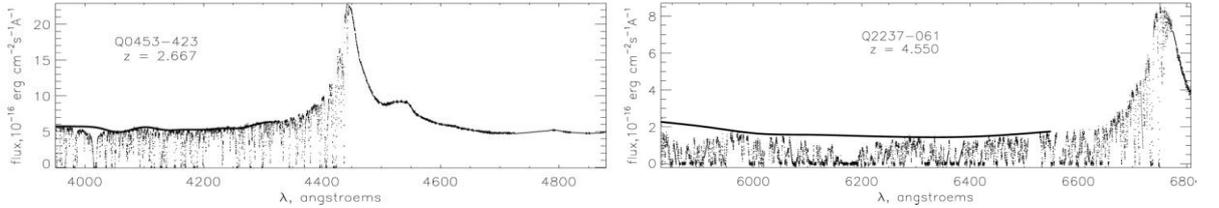

Figure 2: Spectra of quasars Q0453-423 and Q2237-061 with fitted continuum.

If the data (spectrum) are given in the form of pixels with (observed) wavelength labeled $\lambda_i$, the flux density value $f_i$ and the noise $n_i$, we can present the observed flux density $f_i^j$ in *i*-th pixel within the Lyα-forest region of *j*-th quasar as

$$f_i^j = C(\lambda_{rest})\bar{F}(z)(1 + \delta_{F,i}^i) + n_i^j \quad (1)$$

where the wavelength $\lambda_i$ of the absorption Lyα feature produced by the `cloud' of intergalactic HI is related to its redshift $z_i$ as $\lambda_i = 1215.67(1 + z_i)$, $C(\lambda_{rest})$ is the `continuum' level (i.e. the intrinsic quasar spectrum), *n* is the noise, $\bar{F}$ is the mean transmission of IGM in the Lyα line for a given *z*, and $\delta F$ is the variance of *F*. To obtain $\bar{F}(z)$ we divided (1) onto fitted *C* and averaged it over a number of pixels (of one or more spectra) within a given z bin. In this case the values $\langle \delta_{F,i}^j \rangle$ and $\langle n_i^j \rangle$ are equal to zero.

## 4. RESULTS AND CONCLUSIONS

The mean transmission $\bar{F}(z)$ of IGM in Lyα line was calculated averaging over (i) the redshift bins Δz=0.1, and (ii) over each quasar. The obtained redshift dependence of $\bar{F}$ for both cases is presented in upper panel of Fig. 3. We fitted z-dependence of the optical depth $\tau$ with $\lg \tau = a + b \lg((1 + z)/4)$, and obtained the following values of parameters: $a = -0.383^{+0.045}_{-0.082}$, $b = 3.27^{+0.69}_{-0.99}$ for case (i), $a = -0.359^{+0.045}_{-0.078}$, $b = 2.69^{+0.51}_{-0.83}$ for case (ii). The best fits are shown in the lower panel of Fig. 3. These results are in a good agreement with previous results by other authors. This sample and obtained results will be used in further studies for testing the own technique of the continuum fitting in medium-resolution spectra, which is based on utilization of composites stacked from spectra with similar spectral index.

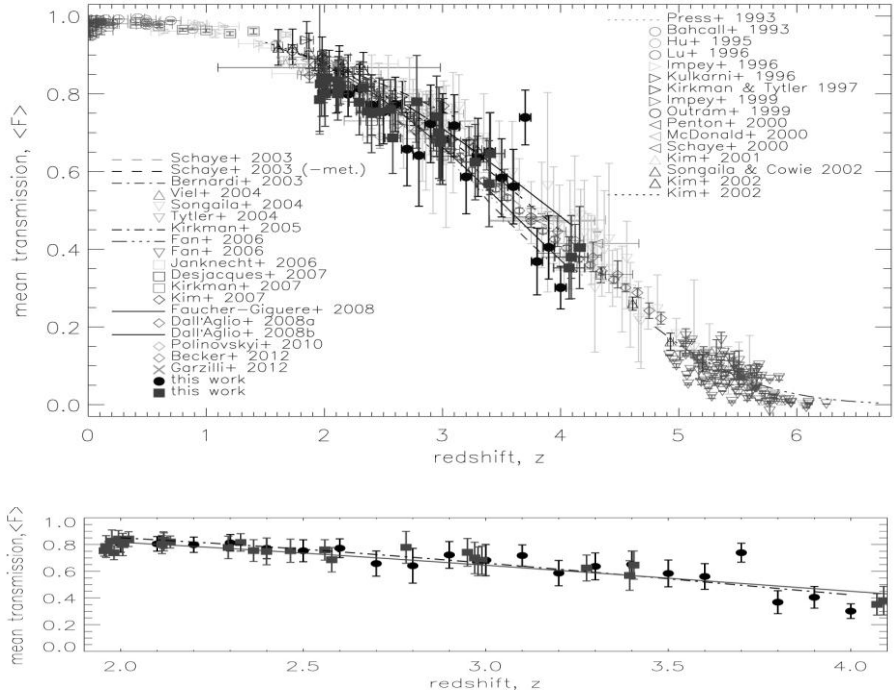

Figure 3: The redshift dependence of $\bar{F}$ averaged over redshift bins (circles) and over each quasar (squares), compared to previous results (upper panel), and with the best fits (lower panel). Error-bars in $\bar{F}$ and $z$ are jack-knife errors and redshift ranges over which $\bar{F}$ was averaged, respectively.

## ACKNOWLEDGEMENTS

The authors are thankful to the ESAVO Team. This work has been supported by Swiss National Science Foundation (SCOPES grant No 128040).

# Влияние положения микролинзы-субзвезды на блеск удаленного источника

## Л.А. Бердина

*Радиоастрономический институт НАН Украины, 61002 Харьков, Украина*

Проведен анализ фокусировки излучения удаленного источника, распространяющегося в гравитационных полях с различными пространственными масштабами. Для изучения гравитационной фокусировки предложен обобщенный метод фазового экрана (ОМФЭ). В рамках ОМФЭ, на примере микролинз-субзвезд, погруженных в гравитационное поле галактики, исследована степень влияния микролинзы на фокусировку излучения. Получена зависимость коэффициента усиления от удаления микролинзы от центра галактики и сделаны оценки влияния микролинзы на блеск удаленного источника.

## 1. ВВЕДЕНИЕ

Явление гравитационного линзирования является одним из прямых средств, дающих информацию о строении самых отдаленных объектов и распределении материи во Вселенной. Отдельный интерес представляют явления гравитационного линзирования компактными телами с субзвездными массами – эффект микролинзирования (ЭМЛ). Из анализа событий ЭМЛ можно получить информацию о массах микролинз и их скорости, о структуре источников излучения, оценить вклад темной материи в общую массу галактик. ЭМЛ проявляется в изменении блеска изображения источника в зависимости от массы микролинзы и ее положения относительно луча зрения. При этом нельзя исключать влияние на наблюдаемое усиление макроизображения положения микролинзы вдоль траектории распространения луча. Создание специального алгоритма для анализа гравитационной фокусировки в полях субзвездных объектов, произвольным образом расположенных внутри галактики, может дать результаты, которые не удается отследить в приближении других методов.

## 2. ОБОБЩЕННЫЙ МЕТОД ФАЗОВОГО ЭКРАНА

ОМФЭ получен путем усовершенствования известной в математической физике формулы Соболева [1]. Основной алгоритм получения формулы Соболева является аналогом выведения формулы Грина. За исключением одного существенного отличия: в формуле Грина рассматривается распространение волны в вакууме, тогда как метод Соболева решает задачу о распространении волны в среде с заданным показателем преломления. Строится решение однородного волнового уравнения, записанного для постоянной во времени и плавно изменяющейся в пространстве среды. В рассмотрение вводятся две вспомогательных функции, которые находятся из уравнения эйконала $\varphi$ и уравнения переноса для амплитуды $\sigma$ [2]. Дальнейшее решение задачи сводится к нахождению эйконала и амплитуды в линейном приближении метода геометрической оптики. С использованием стандартной процедуры построения решения с помощью принципа Гюйгенса-Френеля, после некоторых упрощений формула Соболева преобразуется к виду:

$$U(\vec{R}) \approx \frac{k\tilde{\sigma}}{2\pi i}\int_\Sigma \frac{d\vec{R}'}{|\vec{R}-\vec{R}'|} U(\vec{R}')\exp\{ik\,\varphi(\vec{R},\vec{R}')\}. \quad (1)$$

Основным отличием ОМФЭ (1) от классического является то, что он позволяет построить искомое распределение поля или функции взаимной когерентности (ФВК) в случае, когда пространство заполнено средой, в то время как МФЭ рассматривает распространение излучения в вакууме.

На примере сложной гравитационно линзовой системы, которая состоит из источника излучения и микролинзы, погруженной в галактику, рассмотрен алгоритм решения задачи в приближении ОМФЭ. Анализ задачи производится в два этапа: 1 – строится решение для микролинзы, находящейся между источником излучения и центром масс галактики, 2 – между центром масс галактики и наблюдателем. На рис. 1 схематически представлено распространение излучения через гравитационное поле галактики и микролинзы. Начало системы координат $O$ помещается в центр масс макролинзы-галактики. Ось $oz$ соединяет точки $O$ и наблюдения $P$. В плоскости $z=-Z_s$ расположен протяженный источник излучения. Микролинза находится в плоскости $z=Z_m$ на некотором удалении $\vec{P}_m$ от оси линзы $oz$.

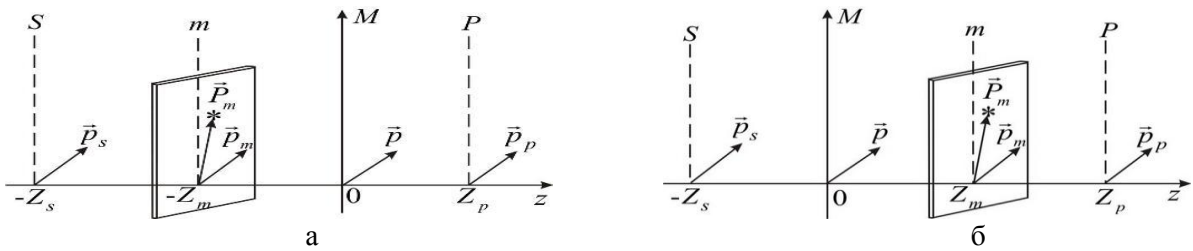

Рис. 1. Взаимное расположение микролинзы $m$ между а) - источником излучения $S$, и центром макролинзы $M$, б) - центром макролинзы $M$ и наблюдателем $P$.

Вся трасса распространения излучения разбивается на три участка, заключенные между плоскостью источника и микролинзы - 1-й участок; микролинзы и галактики - 2-й, галактики и наблюдателя - 3-й. Аналогичная процедура проводится и для второго этапа. При данном разбиении, в пределах каждого участка трассы эффект многолучевости отсутствует, что позволяет строить искомое решение задачи, последовательно перенося значение поля $U(z,\vec{p})$ или ФВК с плоскости на плоскость [3]. Таким образом, путем последовательного переноса решения с плоскости на плоскость можно получить ФВК в точке наблюдения. По найденным представлениям ФВК определяется распределение лучевой интенсивности и коэффициента усиления блеска изображения [3, 4].

## 3. ЧИСЛЕННЫЙ АНАЛИЗ ЭФФЕКТА ФОКУСИРОВКИ ИЗЛУЧЕНИЯ

При получении численных расчетов необходимо конкретизировать параметры моделей. Для изучения ЭМЛ наиболее интересным объектом является гравитационно-линзовая система (ГЛС) Q2237+0305 [5], прежде всего из-за близости линзирующей галактики. Источник-квазар проецируется на центральную часть спиральной галактики, которая состоит из достаточно компактного сферического ядра и протяженного диска [5]. Оба эти фактора в совокупности обеспечивают высокую частоту и сравнительно небольшие характерные времена событий микролинзирования, что позволило накопить достаточно представительные выборки кривых блеска микролинзирования. В дальнейшем использовалась модель ГЛС с основными параметрами, аналогичными ГЛС Q2237+0305, однако для упрощения расчетов рассматривалась лишь сферическая составляющей массы, связанной с ядром галактики. Для массы микролинзы была выбрана величина 0,01 массы Солнца [6].

На рис. 2 показано поведение коэффициента усиления в зависимости от удаления микролинзы от центра галактики. Коэффициент усиления системы нормирован на коэффициент усиления, определяемый только галактикой. Как видно из графика, влияние микролинзы на суммарное усиление возрастает при смещении микролинзы в сторону наблюдателя. Микролинза проецировалась на центр видимого макроизображения. Поведение кривой коэффициента усиления микролинзой также показано рис. 2 - кривая (2). Эту величину необходимо рассматривать как максимально допустимое значение коэффициента усиления излучения микролинзой.

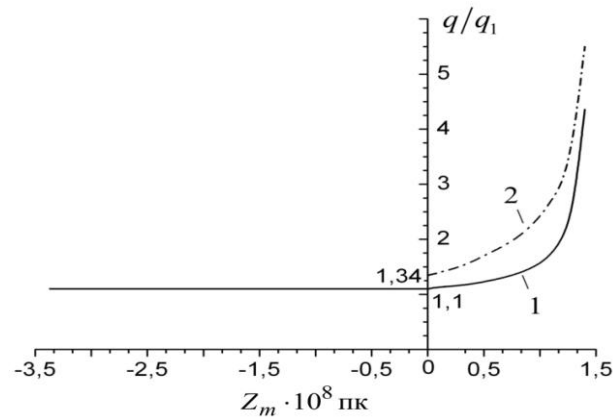

Рис. 2. Коэффициент усиления сложной линзы (кривая 1) и коэффициент усиления микролинзы (кривая 2), нормированные на коэффициент усиления макроизображения.

Заметные увеличения коэффициента усиления для системы с геометрией, подобной геометрии Креста Эйнштейна, начинаются на расстоянии примерно десяток мегапарсек, характерная толщина галактики в этой системе порядка десятков килопарсек. Т.е. для системы с наиболее близкой к наблюдателю галактикой влияние микролинзы становится ощутимым уже за пределами галактики. Для других ГЛС, в которых галактики удалены от наблюдателя на большие расстояния, рассмотренный эффект будет еще меньшим. В то же время при обработке результатов событий ближнего микролинзирования этот эффект должен приниматься во внимание.

**ВЫВОДЫ**

Обобщенный метод фазового экрана позволяет провести аналитическое исследование рефракции лучей на крупномасштабных неоднородностях (гравитационное поле галактики) и рассеяния лучей на мелкомасштабных неоднородностях (поле тяготения микролинзы). В ходе анализа установлено, что величина коэффициента усиления зависит от положения микролинзы вдоль трассы распространения излучения. С приближением микролинзы к плоскости наблюдателя коэффициент усиления возрастает. Наибольшее влияние микролинза оказывает в том случае, когда она расположена в промежутке между центром масс галактики и наблюдателем. Изменение коэффициента усиления при перемещении микролинзы в пределах толщины галактики невелики, однако, могут повлиять на оценку характеристик гравитационно-линзовой системы, таких как масса микролинзы, масса галактики, угловой размер квазара, скорость.

# Effect of the microlens-substar location on brightness of a remote source

## L.A. Berdina

*Institute of Radio Astronomy of NAS of Ukraine, 61022 Kharkiv, Ukraine*

Analysis of focusing of a remote source radiation propagating in the gravitational fields with different spatial scales was performed. To study gravitational focusing in such media, a generalized method of phase screen (GMPS) was proposed. With a microlens substar inside a galaxy as an example, the influence of the microlens location on radiation focusing was investigated. Dependence of amplification factor on the microlens distance from the galaxy plane was obtained and influence of a microlens location on remote source radiation was estimated (in Russian).

# Пошук сонячних аксіонів від $^7$Li

**П. Беллі[1], Р. Бернабей[1,2], Ф.А. Даневич[3], А. д'Анжело[4,5], , А. Інчікітті[4,5], Ф. Капелла[4,5], В.В. Кобичев[3], М. Лаубенштейн[6], О.Г. Поліщук[3,4], В.І. Третяк[3], Р. Черуллі[6]**

[1] *НІЯФ, відділення "Тор Верґата", Рим, Італія*
[2] *Фізичний факультет Римського університету "Тор Верґата", Рим, Італія*
[3] *Інститут ядерних досліджень, Київ, Україна*
[4] *НІЯФ, відділення "Ла Сапієнца", Рим, Італія*
[5] *Фізичний факультет Римського університету "Ла Сапієнца", Рим, Італія*
[6] *НІЯФ, Національна лабораторія Гран Сассо, Асерджі (Акв), Італія*

В Національній підземній лабораторії Гран Сассо (Італія) був проведений експеримент з пошуку сонячних аксіонів за допомогою наднизькофонових HP Ge γ-спектрометрів зі зразками фториду літію. Аналіз даних, отриманих за час вимірювання 4044 год. з кристалом LiF масою 553 г, дозволив встановити обмеження на масу аксіону < 8.6 кеВ (90% C.L.). Це значення є найкращим обмеженням на масу аксіонів, які можуть випромінюватися у розпадах $^7$Li на Сонці.

## 1. ВСТУП

Сонячні аксіони – псевдоскалярні нейтральні гіпотетичні частинки, які є наслідком запропонованого в 1977 році Печеі і Квін рішення проблеми відсутності порушення CP симетрії у квантовій хромодинаміці [1, 2]. Крім цього, аксіони розглядаються як можливі кандидати на роль частинок темної матерії [3].

Сонце може бути інтенсивним джерелом аксіонів, що можуть утворюватися в результаті конверсії γ-квантів у сонячних магнітних полях, а також у результаті розрядки збуджених ядерних рівнів (аксіони можуть з'являтись замість γ-квантів з певною ймовірністю, яка пов'язана з їх масою). Енергетичний спектр сонячних аксіонів повинен складатись з неперервної частини відповідно до ефекту Примакова, а також монохроматичних ліній, що є результатом розрядки внаслідок магнітних переходів збуджених рівнів ядер. Ці рівні можуть бути збуджені через високу температуру у сонячному ядрі або можуть заселятись в ядерних реакціях. Переходи типу М1 з першого збудженого рівня $^7$Li ($E_{зб.}$ = 477.6 кеВ) у сонячному ядрі можуть бути джерелом аксіонів [4]. В *pp*-ланцюжку на Сонці утворюється ядро $^7$Be (в результаті реакції $^3$He + α → $^7$Be + γ), яке потім розпадається у $^7$Li; при цьому рівень 477.6 кеВ заселяється з ймовірністю 10.5% (Рис. 1). Аксіон, що виникає в процесі зняття збудження ядра $^7$Li на Сонці, може бути резонансно поглинений ядром $^7$Li на Землі. Частинки, які випромінюються в наступному процесі розрядки (γ- та рентгенівські кванти, конверсійні електрони), можуть бути зареєстровані детектором, якщо він розміщений близько до мішені з літієм (або має літій у своєму складі); при цьому в енергетичному спектрі детектора буде спостерігатись характерний пік з енергією 477.6 кеВ.

## 2. ПОПЕРЕДНІ ДОСЛІДЖЕННЯ РЕЗОНАНСНОГО ПОГЛИНАННЯ СОНЯЧНИХ АКСІОНІВ

В першому експерименті, присвяченому пошуку сонячних аксіонів від $^{57}$Fe [5], було встановлено лише обмеження на масу сонячного аксіону: $m_a$ < 745 еВ з 95% C.L. Нещодавно було отримано нове обмеження 159 еВ [6]. Аксіони, що можливо випромінюються при розрядці термічно збуджених ядер $^{83}$Kr у сонячному ядрі ($E_{зб.}$ = 9.4 кеВ), досліджувались в роботі [7]; верхнє обмеження на масу адронного аксіону склало 5.5 кеВ.

В першій роботі по пошуку сонячних аксіонів від $^7$Li було встановлено лише обмеження на їх масу: $m_a < 32$ кеВ [8]. Пізніше було отримано обмеження 16 кеВ [9]. Хоча значення $m_a$ для $^7$Li вище, ніж для $^{83}$Kr та $^{57}$Fe, ці дані важливі, оскільки, якщо маса адронного аксіону вище, ніж енергія збудження ядер $^{83}$Kr ($E_{зб.} = 9.4$ кеВ) та $^{57}$Fe ($E_{зб.} = 14.4$ кеВ), аксіони просто не можуть утворюватись в переходах $^{83}$Kr та $^{57}$Fe. Також слід зазначити, що обмеження для $^7$Li пов'язані з $pp$-ланцюжком, що є головним джерелом сонячної енергії, в той час як обмеження для $^{57}$Fe та $^{83}$Kr мають деяку невизначеність, пов'язану з розповсюдженістю заліза та криптону в сонячному ядрі. Обмеження на масу сонячних аксіонів у графічному вигляді представлені на Рис. 1 [10].

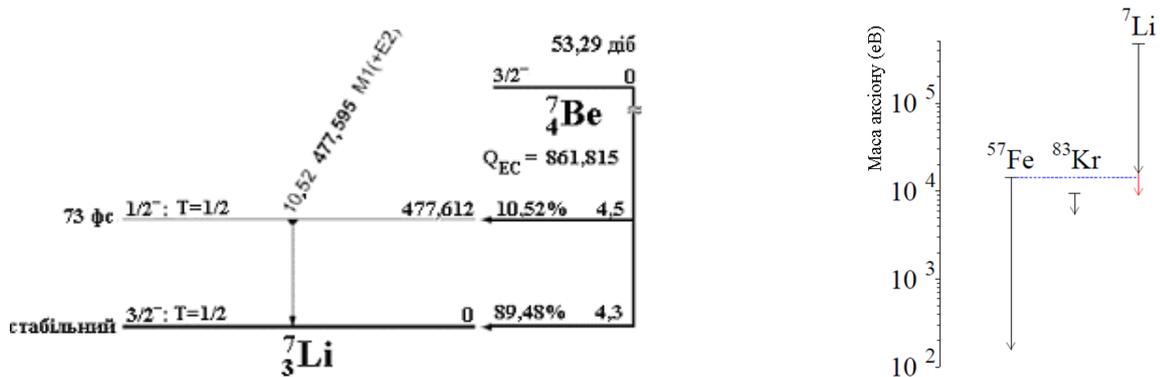

Рис. 1. Схема розпаду $^7$Be у відповідності до [11]. Праворуч: Обмеження на масу сонячних аксіонів, отримані у дослідженнях резонансного збудження ядер.

### 3. ВИМІРЮВАННЯ ЗРАЗКІВ LiF В ЛАБОРАТОРІЇ ГРАН-САССО

Експеримент з пошуку сонячних аксіонів був проведений у підземній лабораторії Гран-Сассо (Італія) за допомогою наднизькофонових HP Ge γ-спектрометрів зі зразками фториду літію (LiF). Оскільки порошки LiF, які використовувались на першому етапі експерименту, виявились забрудненими радіоактивними елементами уран-торієвих рядів, було вирішено для подальших досліджень використовувати кристали LiF, які мають значно нижчий рівень радіоактивної забрудненості. Було розроблено і виготовлено кристал масою 552.6 г. Цей зразок вимірювався за допомогою HP Ge детектора GeBer (224 см$^3$) протягом 4044 годин. На Рис. 2 показано енергетичний спектр, набраний з цим кристалом в діапазоні енергій до 3000 кеВ (ліворуч) та у діапазоні енергій, де ми маємо спостерігати очікуваний пік 477.6 кеВ (праворуч). За результатами вимірювань встановлені лише обмеження на забрудненість елементами уран-торієвих рядів на рівні $< \sim 0.01$ Бк/кг.

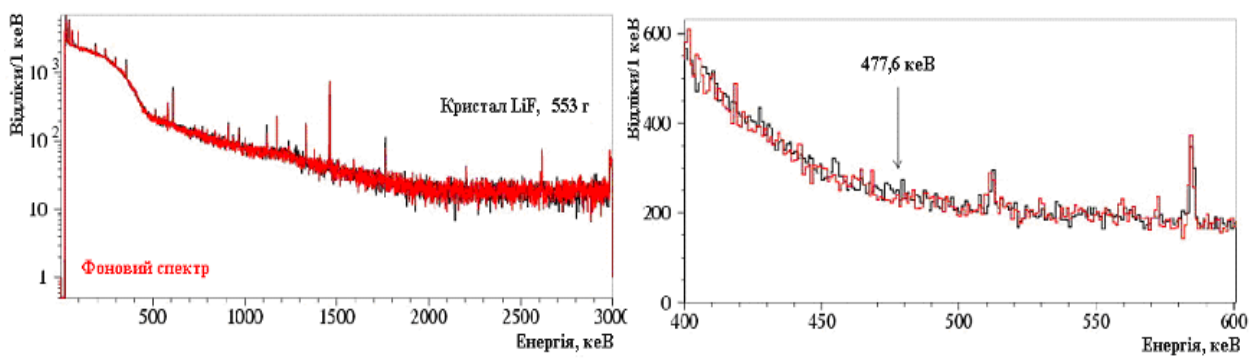

Рис. 2. Енергетичний спектр з кристалом LiF набраний за 4044 годин у діапазоні енергій 0–3000 кеВ (зліва) та діапазоні енергій, де очікується пік 477.6 кеВ (праворуч) у порівнянні з фоновим спектром, нормованим на час вимірювань із зразком.

## 4. ОБМЕЖЕННЯ НА МАСУ АКСІОНУ

Маса сонячних аксіонів від $^7$Li визначається наступним рівнянням [5]:

$$m_a = 1.55 \times 10^{11} \times \left(\frac{S}{\varepsilon N_7 t}\right)^{1/4} \text{ еВ}$$

де $S$ − площа піку 477.6 кеВ (або її обмеження), $\varepsilon$ − ефективність реєстрації для даного піку, $N_7$ − кількість ядер $^7$Li, $t$ − час вимірювання у секундах.

Отримано значення 8.6 кеВ, що на даний момент є найкращим обмеженням на масу сонячних аксіонів від $^7$Li. Це обмеження закриває вікно існуючих можливих мас аксіонів між попереднім обмеженням 16.0 кеВ [9] та енергією 14.4 кеВ, що відповідає іншому потенційному джерелу квазімонохроматичних сонячних аксіонів − $^{57}$Fe (Рис. 1).

## СПИСОК ЛІТЕРАТУРИ

# Search for $^7$Li solar axions

P. Belli[1], R. Bernabei[1,2], F. Cappella[3,4], R. Cerulli[5], F.A. Danevich[6], A. d'Angelo[3,4], A. Incicchitti[3,4], V.V. Kobychev[6], M. Laubenstein[5], O.G. Polischuk[3,6], V.I. Tretyak[6]

[1]*INFN, Sezione di Roma "Tor Vergata", Rome, Italy*
[2]*Dipartimento di Fisica, Universita di Roma "Tor Vergata", Rome, Italy*
[3]*INFN, Sezione di Roma "La Sapienza", Rome, Italy*
[4]*Dipartimento di Fisica, Universita di Roma "La Sapienza", Rome, Italy*
[5]*INFN, Laboratori Nazionali del Gran Sasso, Assergi (AQ), Italy*
[6]*Institute for Nuclear Research, Kyiv, Ukraine*

The resonance excitation of the $^7$Li nuclei in a LiF crystal with mass of 553 g by hypothetical axions emitted in the deexcitation of the $^7$Li nuclei in the Sun was searched for deep underground at the Gran Sasso National Laboratories (LNGS) of INFN (3600 m w.e.). The data collected with a low background HP Ge detector 244 cm$^3$ during 4044 h have allowed us to set the limit on the axion mass: $m_a < 8.6$ keV (90% C.L.) which is the best one for the $^7$Li solar axions to-date (in Ukrainian).

# Optimization of light collection from crystal scintillators to search for dark matter and 2β decay

V.M. Mokina [1], F.A. Danevich [1], V.V. Kobychev [1], H. Kraus [2], V.B. Mikhailik [2], L.L. Nagornaya [3]

[1] *Institute for Nuclear Research, MSP 03680, Kyiv, Ukraine*
[2] *Department of Physics, University of Oxford, Keble Road, Oxford OX1 3RH, UK*
[3] *Institute for Scintillation Materials, MSP 61001, Kharkiv, Ukraine*

Cryogenic scintillation bolometers are promising instruments to search for dark matter and neutrinoless double β decay. High light collection and excellent energy resolution are important requirements in such experiments. Energy resolutions and relative pulse amplitudes of scintillation detectors using $ZnWO_4$ scintillation crystals of different shapes (cylinder and hexagonal prism), reflector materials and shapes, optical contact and surface properties (polished and diffused) were measured. For experimental conditions typical for operating cryogenic scintillating bolometer the best energy resolution and relative pulse amplitude was obtained for a hexagonal shape scintillator with diffuse side and polished face surfaces, surrounded by a reflector with a gap between the scintillator and the reflector.

## 1. INTRODUCTION

Light collection plays an important role in dark matter experiments where a low energy threshold is essential, while in 2β experiments the light collection is crucial to achieve effective pulse-shape discrimination. This is particularly important in case of random coincidence events that was recently recognized as one of the problematic sources of background in bolometric detectors [1]. Zinc tungstate ($ZnWO_4$) crystals have optical properties rather similar to that of $CaWO_4$ [2, 3], $CaMoO_4$ [4], $CdWO_4$ [5] and $ZnMoO_4$ [6], which are promising targets for dark matter and/or 2β decay experiments. Therefore it seems to be natural to use one of these materials with the aim of finding optimal light collection conditions for a wide range of scintillators in cryogenic scintillating bolometers. The purpose of this work was to study the dependence of energy resolution (ER) and relative pulse amplitude (RPA) of $ZnWO_4$ scintillation detectors on crystal shape (hexagonal and cylindrical), design and material of the reflector, optical contact (OC), and optical condition of the crystal scintillator surface (polished and diffuse).

## 2. MEASUREMENTS AND RESULTS

Two $ZnWO_4$ crystal scintillators were produced from one $ZnWO_4$ crystal ingot grown by the Institute of Scintillation Materials (Kharkiv, Ukraine). One crystal was of cylindrical shape with dimensions ⌀ 20×20 mm (C), another was in the form of a hexagonal prism with diagonal 20 mm and height 20 mm (H). The similarity of the two crystals' optical properties was verified by measurements of transmission using a spectrophotometer (Shimadzu, UV-3600). Results of the measurements are presented in Fig. 1. Measurements of ER and RPA with γ-sources were carried out for the following conditions for a $ZnWO_4$ crystal (see Fig. 2): A) wrapped in 3 layers of PTFE tape and optically coupled to the PMT; B) surrounded by cylindrical reflector (3M) ⌀26×25 mm and optically coupled to the PMT; C) surrounded by cylindrical 3M reflector ⌀26×25 mm and placed on small Plexiglas supports (3 cubes with dimensions 2×2×2 mm), without OC between the crystals and the PMT. Optical contact in the geometries "A" and "B" was provided using Dow Corning Q2-3067 optical gel.

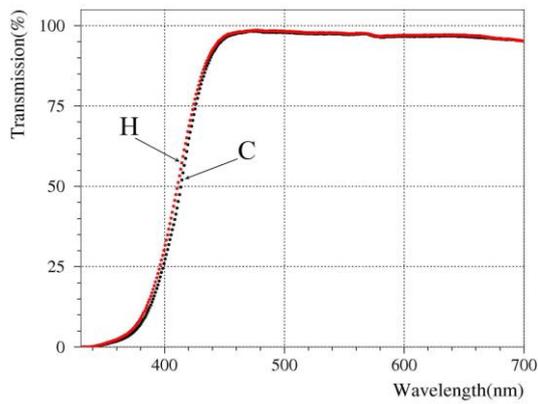 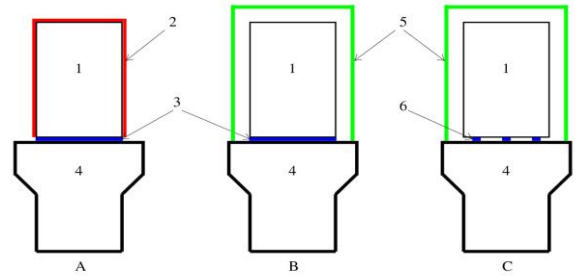

Fig. 1. The optical transmission spectra of cylindrical (C) and hexagonal (H) ZnWO$_4$ crystals. The spectra were measured with a 2 mm sample in the reference beam to account for reflections.

Fig. 2. Conditions of measurements with ZnWO$_4$ crystal scintillators (1 – crystal, 2 – 3 layers of PTFE tape, 3 – optical contact, 4 – PMT, 5 – 3M reflector, 6 – Plexiglas support).

The measurements were carried out for four conditions of the crystals' surfaces: 1) all surfaces of the ZnWO$_4$ crystal scintillator were polished; 2) the side surfaces of the crystals were diffuse, the face surfaces polished; 3) the side surfaces and top face of the crystals were diffuse, the face viewed by the PMT was polished; 4) all surfaces of the crystals were diffuse. Grinding the surfaces was done using the sanding paper with the grain size 18 ± 1 microns. The average roughness of the diffuse surface was estimated as ≈ 10 microns by using an optical microscope. The roughness of the polished surfaces is ≈ 0.2 micron.

To measure the scintillation properties, crystals were viewed by a 3" photomultiplier (PMT) Philips XP2412. The energy spectra of $^{137}$Cs and $^{207}$Bi γ quanta, exhibiting the best ERs are shown in Fig. 3. These were accumulated with the hexagonal ZnWO$_4$ scintillator in OC with the PMT, surrounded by the 3M reflector, with all surfaces diffuse. To check the stability of the spectrometer, a ZnWO$_4$ crystal scintillator 10×10×5 mm$^3$ was periodically installed on the PMT and irradiated by γ quanta from a $^{137}$Cs source. The measurements for surface configurations 1 – 3 were repeated three times and that for configuration 4 was carried out once. Average data on the ER and RPA measured with 662 keV γ quanta of $^{137}$Cs are presented in Fig. 4.

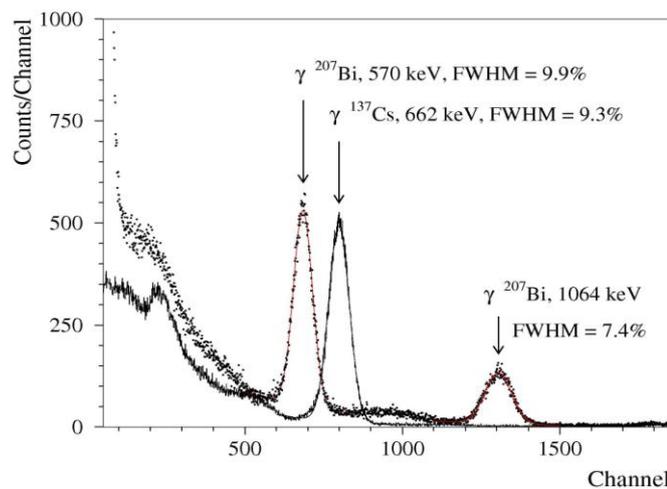

Fig. 3. Energy spectra of $^{137}$Cs and $^{207}$Bi γ rays measured for the ZnWO$_4$ crystal of hexagonal shape with all surface diffuse, in geometry "B" (see text and Fig. 2).

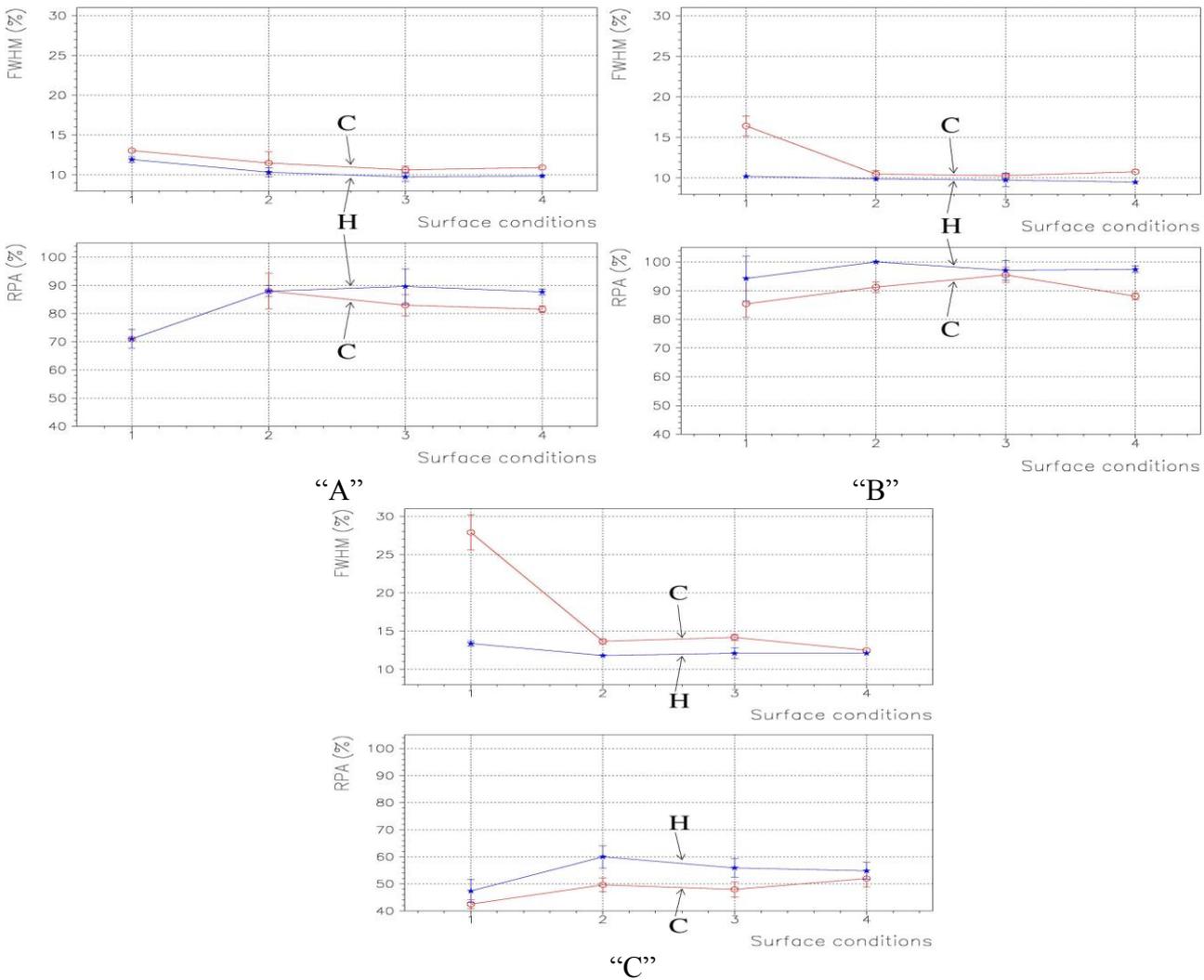

Fig. 4. ER and RPA for hexagonal (H) and cylindrical (C) ZnWO$_4$ scintillator crystals for geometries "A", "B", "C" (see Fig. 2) measured with a $^{137}$Cs source for different crystal surface conditions (1 − polished, 2 − diffuse side, 3 − diffuse side and top face, 4 − all surfaces diffuse).

## 3. CONCLUSIONS

Effects of crystal scintillator shape, reflector material and shape, OC and surface treatment were tested with cylindrical and hexagonal ZnWO$_4$ crystal scintillators. The hexagonal crystal shows better ER and RPA in all measurement conditions. The best ER (FWHM = 9.3% for 662 keV γ quanta of $^{137}$Cs) was obtained with the hexagonal scintillator with all surfaces diffuse, in OC with the PMT and surrounded by a cylindrical reflector (3M). In conditions applicable for light collection in a cryogenic scintillating bolometer the best RPA and ER (FWHM = 11.8 % for 662 keV γ quanta of $^{137}$Cs) were obtained for the hexagonal scintillator with diffuse side surfaces and polished end faces, surrounded by a cylindrical reflector (3M).

# Cryogenic zinc molybdate scintillating bolometers to search for neutrinoless double beta decay of $^{100}$Mo

D.M. Chernyak [1,2], F.A. Danevich [1], E.N. Galashov [3], A. Giuliani [2], V.V. Kobychev [1],
S. Marnieros [2], C. Nones [4], E. Olivieri [2], V.N. Shlegel [3], M. Tenconi [2],
V.I. Tretyak [1], Ya.V. Vasiliev [3]

[1] *Institute for Nuclear Research, MSP 03680 Kyiv, Ukraine*
[2] *Centre de Spectrometrie Nucleaire et de Spectrometrie de Masse, 91405 Orsay, France*
[3] *Nikolaev Institute of Inorganic Chemistry, 630090 Novosibirsk, Russia*
[4] *Service de Physique des Particules, CEA-Saclay, F-91191 Gif sur Yvette, France*

Scintillating bolometer with a 23.8 g ZnMoO$_4$ crystal was tested at temperature 18 mK. Intrinsic energy resolution of the heat channel was ≈ 1.3 keV, α/β rejection factor was achieved better than 99.9% in the energy region of interest (≈ 3 MeV). Pulse shape discrimination technique was developed to suppress background from the random coincidence of 2ν2β decay events in ZnMoO$_4$ cryogenic scintillating bolometer. Large ZnMoO$_4$ crystal scintillator (313 g of mass) was produced and tested as scintillating bolometer. A next generation experiment applying enriched Zn$^{100}$MoO$_4$ crystal with unprecedented sensitivity to neutrinoless 2β decay of $^{100}$Mo is discussed.

## 1. INTRODUCTION

Neutrinoless double beta (0ν2β) decay, if observed, will establish the Majorana nature of the neutrino and violation of the lepton number and, in case of transition mediated by light neutrino exchange, will determine the absolute scale of neutrino masses [1].

$^{100}$Mo is one of the most promising 2β isotopes because of its large transition energy $Q_{2\beta}$ = 3034.40(17) keV and a considerable natural isotopic abundance δ = 9.67(20)%. ZnMoO$_4$ crystals were developed recently [2]. An important advantage of ZnMoO$_4$ in comparison to other inorganic crystals which have Mo in their composition is the absence of heavy and radioactive elements, and high concentration of molybdenum (43% in mass).

Cryogenic bolometers can provide excellent energy resolution (a few keV) and high detection efficiency (80–90%). Furthermore, simultaneous detection of phonon and scintillation signals in cryogenic scintillating bolometers allows efficient particle discrimination, important to reject background caused by radioactive contamination of crystal scintillators.

## 2. A ZnMoO$_4$ SCINTILLATING BOLOMETER PROTOTYPE

High quality ZnMoO$_4$ crystals up to 25 mm in diameter and 60 mm in length were grown in the Nikolaev Institute of Inorganic Chemistry (NIIC, Novosibirsk, Russia) by using the low-thermal-gradient Czochralski technique.

Detector prototype of the cryogenic scintillating bolometer with a ZnMoO$_4$ scintillating crystal was constructed and tested in the Centre de Spectrométrie Nucléaire et de Spectrométrie de Masse (CSNSM, Orsay, France). Detector consists of ZnMoO$_4$ crystal with a mass of 23.8 g, faced by the thin light-detecting ultrapure germanium slab. Bolometer was surrounded by a highly reflective polymeric multilayer foil (3M VM2000/VM2002). The thermal signals from the ZnMoO$_4$ crystal and the germanium slab were read out by two neutron transmutation doped (NTD) germanium thermistors, with a mass of ≈ 10 mg.

The signals from the ZnMoO$_4$ crystal (the heat channel) corresponded to voltage pulses with an amplitude of ≈ 100 μV for 1 MeV deposited energy at 19 mK. The time structure was characterized by ≈ 10 ms rise time and ≈ 40 ms decay time (respectively from 10% to 90% and from 90% to 30% of the signal maximum amplitude). Intrinsic energy resolution of the heat channel was at the level of 1.3 keV. The α/β rejection factor was achieved better than 99.9% in the energy region of interest for $^{100}$Mo (see Fig. 1).

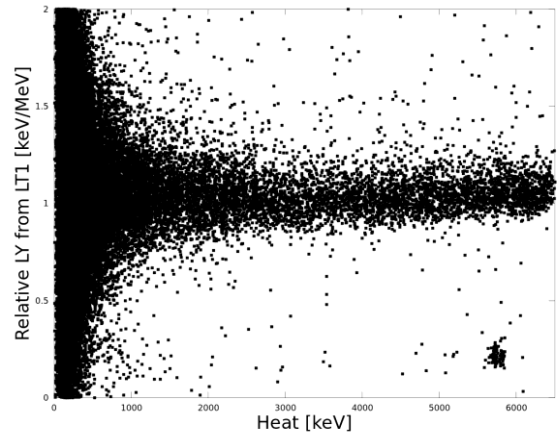

Fig. 1. The ratio of the light-to-heat energy as a function of the heat energy obtained with a 23.8 g ZnMoO$_4$ crystal working as a cryogenic bolometer. The upper band (caused by β/γ and cosmic muon events) and isolated lower band (populated by α decays of $^{210}$Po) are clearly separated.

## 3. RANDOM COINCIDENCES OF 2ν2β DECAY EVENTS AS A BACKGROUND SOURCE

A disadvantage of cryogenic bolometers is their slowness which causes poor time resolution. This can lead to a background component at the energy $Q_{2\beta}$ due to random coincidences of lower energy signals, in particular those due to the unavoidable two-neutrino double β decay events. The pile-up phenomenon was studied by generating light pulses with the pulse shapes and noise from a real light detector coupled to a ZnMoO$_4$ scintillating crystal.

We defined a 90% efficiency in accepting a pulse from the light detector as a potentially good 0ν2β pulse using three different pulse-shape indicators: (1) the risetime from 15% to 90% of the maximum amplitude; (2) the $\chi^2$ evaluated using an average pulse as a standard shape function; (3) the pulse shape parameter defined in [3], which also uses a standard pulse-shape function. In each pulse pair, the amplitude of the first pulse $A_1$ was extracted by sampling the 2ν2β distribution, while the amplitude of the second pulse $A_2$ was chosen as $Q_{2\beta}(^{100}\text{Mo}) - A_1 + \Delta E$, where $\Delta E$ is a random component in the interval [−5, +5] keV. Pairs of pulses were generated with random time distances with a flat distribution up to 10 ms. Signal-to-noise ratio was fixed at the level expected for a 0ν2β signal, i.e., of the order of 30.

The piled-up pulses generated in the simulation were analyzed with the mentioned pulse-shape indicators. Using the risetime method, rejection efficiency on the level of 80% − 90% was achieved. The other two indicators provide equivalent or even better results.

## 4. POTENIAL OF ZnMoO$_4$ SCINTILLATING BOLOMETERS TO SEARCH FOR 0ν2β DECAY OF $^{100}$Mo

Large ZnMoO$_4$ crystal scintillator (313 g of mass) was produced using the low-thermal-gradient Czochralski technique in the NIIC (Novosibirsk, Russia). Cryogenic scintillating bolometer with the ZnMoO$_4$ crystal was tested at the aboveground level in the CSNSM (Orsay, France). Preliminary measurements have shown possibility to use large ZnMoO$_4$ crystals as cryogenic bolometers. Measurements and further investigation are in progress.

According to the current study of a ZnMoO$_4$ scintillating bolometer prototype and the simulation of the background made in [4], we can evaluate the sensitivities of future 0ν2β searches (see Table 1). For the next tests an enriched 200 g ZnMoO$_4$ crystal is in preparation. In case of a positive outcome of the 200 g test, four large crystals of about 400–500 g each will be produced and measured in one of the existing underground low-activity dilution refrigerators in Modane or Gran Sasso laboratories (option (1) in Table 1). This pilot experiment would have sensitivities comparable to present searches and would constitute the general test bench for scaling up this technology. A further step could exploit existing enriched material, of the order of several kg. This search (option (2) in Table 1) would approach the inverted hierarchy (IH) region and could be housed by the former Cuoricino refrigerator or its equivalent. A major extension is required to substantially cover the IH mass pattern. The production of ≈ 350 kg of $^{100}$Mo fits the time and budget scale of a large next-generation 0ν2β search. Bolometric masses of the order of 1 ton could be housed by the CUORE (under construction) or EURECA (under design) dilution refrigerators, which could share dark matter and 0ν2β searches (options (3) and (4) in Table 1).

Table 1. Sensitivity (at 90% C.L.) of experiments based on ZnMoO$_4$ scintillating bolometers searching for 0ν2β of $^{100}$Mo. The live time is 5 yr. The enrichment level is 97% (except option (3) where natural molybdenum is considered) and the detection efficiency is 90%. The range in $m_{\beta\beta}$ takes into account three different approaches to the evaluation of the nuclear matrix elements (QRPA, ISM, IBM-2).

| Option | Number of 400 g crystals | Total isotope mass, kg | $T_{1/2}$ sensitivity, $10^{26}$ y | $m_{\beta\beta}$ sensitivity, eV |
|---|---|---|---|---|
| (1) | 4 | 0.68 | 0.05 | 0.1 – 0.5 |
| (2) | 40 | 6.8 | 0.5 | 0.05 – 0.2 |
| (3) | 2000 (nat.) | 33 | 1.5 | 0.03 – 0.09 |
| (4) | 2000 | 338 | 10 | 0.01 – 0.04 |

## 5. CONCLUSIONS

Large volume ZnMoO$_4$ crystals of improved quality were grown by the low-thermal gradient Czochralski technique. Development of the enriched Zn$^{100}$MoO$_4$ crystals is in progress.

Detector prototypes of the cryogenic scintillating bolometers with a 23.8 g and 313 g ZnMoO$_4$ scintillating crystals were constructed and tested. Both of them prove possibility of the large-scale next-generation experiment with ZnMoO$_4$ crystals.

The present bolometric detector technologies enable to control 2ν2β decay form of background at the required level, in particular the case of bolometers searching for the 0ν2β decay of $^{100}$Mo, which is characterized by a relatively short 2ν2β decay half-life.

ZnMoO$_4$ scintillating bolometers are extremely promising detectors for a next-generation 0ν2β experiment capable to explore the inverted hierarchy region of the neutrino mass pattern.

# Fine-tuning problem in five-dimensional brane world models

**S.H. Fakhr [1], M.V. Eingorn [1,2], A.I. Zhuk [1]**

[1] *Astronomical Observatory, Odessa National University, Dvoryanskaya st. 2, 65082 Odessa, Ukraine*
[2] *Department of Theoretical and Experimental Nuclear Physics, Odessa National Polytechnic University, Shevchenko av. 1, 65044 Odessa, Ukraine*

Fine tuning may be called a main disadvantage of the Randall-Sundrum model, being the most popular brane world model, constructed quite artificially in the five-dimensional space-time. It needs a bare multidimensional cosmological constant, which is related strictly to the four-dimensional tension (the brane physical property). We try to avoid this problem of naturalness, introducing a perfect fluid with arbitrary linear equations of state in both three-dimensional external and one-dimensional internal spaces. This model represents the direct generalization of the Randall-Sundrum one. We derive equations for background metric coefficients, determining a wide class of new exact solutions, and discuss uselessness of subsequent development of brane world models in view of their unjustified plurality.

## 1. INTRODUCTION

Extra spatial dimensions represent one of the approaches to solving such challenges of modern cosmology, astrophysics and elementary-particle physics as dark energy, dark matter and the hierarchy problem. In particular, they are introduced in extremely popular supersymmetric theories [1], pretending to being "theories of everything". Obviously, they should be examined experimentally for compatibility with the observations. In the recent paper [2] it was explicitly shown that Kaluza-Klein models with toroidal compactification of extra dimensions contradict such famous relativistic gravitational tests as the perihelion shift, the deflection of light and the time delay of radar echoes. Therefore, these models seem very doubtful, and one should look for their viable multidimensional alternatives. They are the brane world models (see the reviews [3, 4]), in which it is assumed that the Standard Model fields are localized on a three-dimensional subspace (a brane), embedded into the multidimensional space (the bulk), while gravity may propagate everywhere (in both external/our and internal/extra dimensions). The extra dimensions may be either compact or infinite, and a single brane or even several branes are admissible.

The most popular brane world model is the five-dimensional Randall-Sundrum model with one or two branes [3-5]. In this model the bulk is assumed to be empty (not counting dark energy, represented by the five-dimensional cosmological term), while the branes possess tensions (looking like the four-dimensional cosmological terms). Clearly, this scenario is very simple and unlikely well-grounded. Therefore, a natural question arises whether it can be generalized to the case of perfect fluids with arbitrary linear equations of state, filling the bulk as well as the branes. In this work we produce this important natural generalization and discuss it.

## 2. FIVE-DIMENSIONAL STATIC BACKGROUND SOLUTIONS

In the absence of the localized sources of the gravitational field (the gravitating masses) the five-dimensional static background metrics may be taken in the following general form:

$$ds^2 = Ac^2 dt^2 + Bdx^2 + Cdy^2 + Ddz^2 + Ed\xi^2,  \qquad (1)$$

where the functions $A(\xi)$, $B(\xi)$, $C(x,\xi) = x^2 B(\xi)$, $D(x,y,\xi) = C(x,\xi)\sin^2 y$, $E(\xi) = -1$ (the metric coefficients) are introduced (the coordinate $\xi$ corresponds to the only one extra spatial dimension, which may be either finite/compact, or infinite/noncompact). Our main aim lies in determining the metric coefficients in the most general acceptable case.

In order to write down the corresponding Einstein equations, let us first find the nonzero covariant Ricci tensor components:

$$R_{00} = -\frac{A_\xi}{4}\left(-\frac{2A_{\xi\xi}}{A_\xi} - \frac{3B_\xi}{B} + \frac{A_\xi}{A}\right), \quad R_{11} = -\frac{B_\xi}{4}\left(-\frac{2B_{\xi\xi}}{B_\xi} - \frac{A_\xi}{A} - \frac{B_\xi}{B}\right), \tag{2}$$

$$R_{22} = x^2 \cdot R_{11}, \quad R_{33} = \sin^2 y \cdot R_{22} = x^2 \sin^2 y \cdot R_{11}, \tag{3}$$

$$R_{44} = \frac{A_\xi}{4A}\left(-\frac{2A_{\xi\xi}}{A_\xi} + \frac{A_\xi}{A}\right) + \frac{3B_\xi}{4B}\left(-\frac{2B_{\xi\xi}}{B_\xi} + \frac{B_\xi}{B}\right). \tag{4}$$

Now let us assume that the bulk (the four-dimensional space) is filled with the perfect fluid, possessing the energy density $\varepsilon$, the external pressure $p_0$ and the internal pressure $p_1$. Generally speaking, the quantity $p_1$ may differ essentially from the quantity $p_0$, and these two pressures may even have different signs. The mixed energy-momentum tensor components read

$$T_0^0 = \varepsilon, \quad T_1^1 = T_2^2 = T_3^3 = -p_0, \quad T_4^4 = -p_1. \tag{5}$$

Consequently, for the covariant components and the trace we have respectively

$$T_{00} = \varepsilon A, \quad T_{11} = -p_0 B, \quad T_{22} = x^2 \cdot T_{11}, \quad T_{33} = x^2 \sin^2 y \cdot T_{11}, \quad T_{44} = p_1, \quad T = \varepsilon - 3p_0 - p_1. \tag{6}$$

Substituting (2), (3), (4) and (6) into the Einstein equations

$$R_{ik} - (R/2)Rg_{ik} = \kappa(T_{ik} + \Lambda_5 g_{ik}), \tag{7}$$

where $R = R_{ik}g^{ik}$ is the scalar curvature, $\kappa$ is some positive constant, being directly proportional to the multidimensional gravitational constant, while $\Lambda_5$ is the five-dimensional cosmological constant, we obtain the following system of three independent equations:

$$\kappa\varepsilon = -3B_{\xi\xi}/(2B) - \kappa\Lambda_5, \tag{8}$$

$$\kappa p_0 = \frac{B_{\xi\xi}}{B} + \frac{A_\xi B_\xi}{2AB} - \frac{B_\xi^2}{4B^2} + \frac{A_{\xi\xi}}{2A} - \frac{A_\xi^2}{4A^2} + \kappa\Lambda_5 = \omega_0 \cdot \kappa\varepsilon, \tag{9}$$

$$\kappa p_1 = \frac{3B_\xi^2}{4B^2} + \frac{3A_\xi B_\xi}{4AB} + \kappa\Lambda_5 = \omega_1 \cdot \kappa\varepsilon, \tag{10}$$

where the linear equations of state

$$p_0 = \omega_0\varepsilon, \quad p_1 = \omega_1\varepsilon \tag{11}$$

with the constant parameters $\omega_0$ and $\omega_1$ are assumed. Concerning the appropriate boundary conditions, for the brane, situated at $\xi = 0$, we take them in the form

$$[K_{mn} - g_{mn}K]|_{\xi=0} = (K_{mn} - g_{mn}K)|_{\xi=+0} - (K_{mn} - g_{mn}K)|_{\xi=-0} = \kappa S_{mn}|_{\xi=0}, \quad m,n = 0,1,2,3, \tag{12}$$

$$K_{00} = -\frac{A_\xi}{2}, \quad K_{11} = -\frac{B_\xi}{2}, \quad K_{22} = x^2 \cdot K_{11}, \quad K_{33} = x^2 \sin^2 y \cdot K_{11}, \quad K = -\frac{A_\xi}{2A} - \frac{3B_\xi}{2B}, \tag{13}$$

$$S_{00} = \epsilon_0 g_{00} = \epsilon_0 A, \quad S_{11} = -\pi_0 g_{11} = -\pi_0 B, \quad S_{22} = x^2 \cdot S_{11}, \quad S_{33} = x^2 \sin^2 y \cdot S_{11}. \tag{14}$$

In other words, on the considered brane the three-dimensional perfect fluid with the energy density $\epsilon_0$ and the pressure $\pi_0$ is situated. Substituting (13) and (14) into (12), we get

$$[B_\xi/B]|_{\xi=0} = 2\kappa\epsilon_0/3, \quad [A_\xi/A]|_{\xi=0} = -4\kappa\epsilon_0/3 - 2\kappa\pi_0. \tag{15}$$

If there is one more brane, situated at $\xi = a$ (where $a$ is some positive constant), then

$$[B_\xi/B]|_{\xi=a} = 2\kappa\epsilon_a/3, \quad [A_\xi/A]|_{\xi=a} = -4\kappa\epsilon_a/3 - 2\kappa\pi_a. \tag{16}$$

Let us start with the simplest case $\varepsilon = 0$, when the bulk is empty (excluding dark energy, represented by $\Lambda_5$). In this case from (8), (9) and (10) we obtain the following system of three independent equations:

$$B_{\xi\xi} + (2\kappa\Lambda_5/3)B = 0, \tag{17}$$

$$\frac{B_{\xi\xi}}{B} + \frac{A_\xi B_\xi}{2AB} - \frac{B_\xi^2}{4B^2} + \frac{A_{\xi\xi}}{2A} - \frac{A_\xi^2}{4A^2} + \kappa\Lambda_5 = 0, \tag{18}$$

$$\frac{3B_\xi^2}{4B^2} + \frac{3A_\xi B_\xi}{4AB} + \kappa\Lambda_5 = 0. \tag{19}$$

If $\Lambda_5 > 0$ and there is only one brane at $\xi = 0$, then from this system with the boundary conditions (15) we obtain the following metric coefficients:

$$A = \alpha \cos^2\left(\sqrt{\frac{2}{3}\kappa\Lambda_5}\,|\xi| + \gamma\right)\Big/\sin\left(\sqrt{\frac{2}{3}\kappa\Lambda_5}\,|\xi| + \gamma\right), \quad B = \beta \sin\left(\sqrt{\frac{2}{3}\kappa\Lambda_5}\,|\xi| + \gamma\right), \tag{20}$$

where $\alpha$ and $\beta$ are arbitrary constants, while the parameter $\gamma$ is defined by the condition

$$\frac{\sin\gamma}{\cos\gamma} = 3\sqrt{\frac{2}{3}\kappa\Lambda_5}\Big/(\kappa\in_0). \tag{21}$$

At the same time we have the following fine-tuning condition:

$$\Lambda_5 = \kappa\in_0^2/12 + \kappa\in_0 \pi_0/4 > 0. \tag{22}$$

If we have one more brane at $\xi = a$, then the distance $a$ between the branes is defined by

$$\sin\left(\sqrt{\frac{2}{3}\kappa\Lambda_5}\,a + \gamma\right)\Big/\cos\left(\sqrt{\frac{2}{3}\kappa\Lambda_5}\,a + \gamma\right) = -3\sqrt{\frac{2}{3}\kappa\Lambda_5}\Big/(\kappa\in_a), \tag{23}$$

and the following additional fine-tuning condition should hold true:

$$\Lambda_5 = \kappa\in_a^2/12 + \kappa\in_a \pi_a/4 > 0. \tag{24}$$

If $\Lambda_5 < 0$ and there is only one brane at $\xi = 0$, then from the system of equations (17), (18) and (19) with the boundary conditions (15) we obtain the following metric coefficients:

$$A = \alpha \cosh^2\left(\sqrt{-\frac{2}{3}\kappa\Lambda_5}\,|\xi| + \gamma\right)\Big/\sinh\left(\sqrt{-\frac{2}{3}\kappa\Lambda_5}\,|\xi| + \gamma\right), \quad B = \beta \sinh\left(\sqrt{-\frac{2}{3}\kappa\Lambda_5}\,|\xi| + \gamma\right), \tag{25}$$

where the parameter $\gamma$ is defined by the condition

$$\frac{\sinh\gamma}{\cosh\gamma} = 3\sqrt{-\frac{2}{3}\kappa\Lambda_5}\Big/(\kappa\in_0). \tag{26}$$

At the same time we have the following fine-tuning condition:

$$\Lambda_5 = \kappa\in_0^2/12 + \kappa\in_0 \pi_0/4 < 0. \tag{27}$$

If we have one more brane at $\xi = a$, then the distance $a$ between the branes is defined by

$$\sinh\left(\sqrt{-\frac{2}{3}\kappa\Lambda_5}\,a + \gamma\right)\Big/\cosh\left(\sqrt{-\frac{2}{3}\kappa\Lambda_5}\,a + \gamma\right) = -3\sqrt{-\frac{2}{3}\kappa\Lambda_5}\Big/(\kappa\in_a), \tag{28}$$

and the following additional fine-tuning condition should hold true:

$$\Lambda_5 = \kappa\in_a^2/12 + \kappa\in_a \pi_a/4 < 0. \tag{29}$$

Finally, if $\Lambda_5 = 0$ and there is only one brane at $\xi = 0$, from the system of equations (17), (18) and (19) with the boundary conditions (15) we obtain the following metric coefficients:

$$A = C_3/(C_1|\xi| + C_2), \quad B = C_1|\xi| + C_2, \quad C_1 \neq 0, \tag{30}$$

where $C_{1,2,3}$ are some constants,

$$C_1/C_2 = \kappa\in_0/3 = 2\kappa\in_0/3 + \kappa\pi_0. \tag{31}$$

At the same time we have the following fine-tuning condition:

$$\pi_0 = -\in_0/3. \tag{32}$$

If we have one more brane at $\xi = a$, then the distance $a$ between the branes is defined by

$$C_1/(aC_1 + C_2) = -\kappa\in_a/3 = -2\kappa\in_a/3 - \kappa\pi_a, \tag{33}$$

and the following additional fine-tuning condition should hold true:

$$\pi_a = -\in_a/3. \tag{34}$$

Now let us turn to the much more interesting and important case $\varepsilon \neq 0$ (when the bulk is nonempty). Then, excluding the quantity $\varepsilon$ from (8), (9) and (10), we come to the following system of two independent equations:

$$\frac{B_{\xi\xi}}{B} + \frac{A_\xi B_\xi}{2AB} - \frac{B_\xi^2}{4B^2} + \frac{A_{\xi\xi}}{2A} - \frac{A_\xi^2}{4A^2} + \kappa\Lambda_5 = \omega_0\left(-\frac{3B_{\xi\xi}}{2B} - \kappa\Lambda_5\right), \tag{35}$$

$$\frac{3B_\xi^2}{4B^2} + \frac{3A_\xi B_\xi}{4AB} + \kappa\Lambda_5 = \omega_1\left(-\frac{3B_{\xi\xi}}{2B} - \kappa\Lambda_5\right). \tag{36}$$

From the latter one we get

$$\frac{A_\xi}{A} = -2\omega_1\frac{B_{\xi\xi}}{B_\xi} - \frac{4}{3}\kappa\Lambda_5(1+\omega_1)\frac{B}{B_\xi} - \frac{B_\xi}{B}. \tag{37}$$

This equation defines the function $A(\xi)$, if the function $B(\xi)$ is known. Substituting (37) into (35), after some calculation we arrive at the following equation, defining the function $B(\xi)$:

$$-(1+3\omega_0)\frac{B_{\xi\xi}}{B} - \frac{2}{3}\kappa\Lambda_5(3\omega_0 - 2\omega_1 + 1) - 2\omega_1(\omega_1+1)\frac{B_{\xi\xi}^2}{B_\xi^2} -$$

$$-\frac{8}{9}[\kappa\Lambda_5(\omega_1+1)]^2\frac{B^2}{B_\xi^2} - \frac{4}{3}\kappa\Lambda_5(1+2\omega_1)(\omega_1+1)\frac{BB_{\xi\xi}}{B_\xi^2} = -2\omega_1\frac{B_{\xi\xi\xi}}{B_\xi}. \tag{38}$$

This equation is essentially nonlinear and very complicated. Consequently, we can not simply write down its general solution, then use it for determining $A(\xi)$ and, finally, satisfy the appropriate boundary conditions. However, one can easily solve this problem numerically.

### 3. CONCLUSION

In the case of the empty bulk we found six different static background solutions, generalizing the Randall-Sundrum metrics, and established fine-tuning conditions, imposed on the cosmological constant as well as energy densities and pressures of the three-dimensional perfect fluids, situated on the branes. Consequently, the fine-tuning problem, consisting in dubious reality of such conditions, remains here unsolved. At the same time in the case of the nonempty bulk we arrived at the situation, when there are a lot of static solutions, given by (37) and (38), and the unlikely solvable problem of the physically sensible choice of the background metrics arises. This unsatisfactory situation is a serious drawback of brane world models.

# Quintessence and phantom field inhomogeneities at late stages of Universe evolution

**A.Yu. Burgazli [1], M.V. Eingorn [2], A.I. Zhuk [3]**

[1] *Department of Theoretical Physics, Odessa National University, st. Dvoryanskaya 2, 65082 Odessa, Ukraine*
[2] *Department of Theoretical and Experimental Nuclear Physics, Odessa National Polytechnic University, Shevchenko av. 1, 65044 Odessa, Ukraine*
[3] *Astronomical Observatory, Odessa National University, st. Dvoryanskaya 2, 65082 Odessa, Ukraine*

As it directly follows from the theory of scalar cosmological perturbations, applied to the late stages of evolution of the Universe, filled with dark energy, nonrelativistic matter and, possibly, quintessence or phantom field, the last two components can not be homogeneous. We demonstrate that $\omega = -1/3$ is the only admissible negative parameter in the non-vacuum linear equation of state, then determine the corresponding gravitational potentials and discuss their main properties.

## 1. INTRODUCTION

At the present time the most popular cosmological model, describing evolution of the homogeneous and isotropic Universe, is the well known $\Lambda$CDM-model [1]. In its framework it is assumed that the Universe is spatially flat (in other words, the spatial curvature $\mathrm{K}$ is equal to zero) and filled (at late stages) with the nonrelativistic matter (~27% now) and the dark energy (~73% now), represented by the cosmological constant $\Lambda$. The dark energy in the $\Lambda$CDM-model can be presented in the form of a perfect fluid with the linear equation of state $p = \omega\varepsilon$ ($\varepsilon$ is the energy density and $p$ is the pressure), where the parameter $\omega$ is constant and equal to $-1$.

At the same time in a great number of modern papers their authors consider different perfect fluids with the general linear equation of state $p = \omega\varepsilon$, where $\omega$ is arbitrary. In particular, taking into consideration the accelerating character of the Universe expansion, $\omega$ may be negative (but not necessarily equal to $-1$). Such perfect fluids are usually called quintessence ($-1 < \omega < 0$) or the phantom field ($\omega < -1$) [2, 3]. In these papers $\omega$ is often assumed to be constant or, in other words, time-independent (see also [4, 5]). The acceleration is achieved for $\omega < -1/3$.

This work is devoted to the test of quintessence and phantom field with the constant negative parameter $\omega$ for compatibility with the theory of scalar cosmological perturbations at late stages of the Universe evolution. Inside the cell of uniformity (the spatial region with the scale ~ 150 Mpc) hydrodynamics is inapplicable, so we use the mechanical approach [6]. According to this paper, the investigated fluids can not be homogeneous. Therefore, we perturb the background values of their energy density and pressure. Then we try to derive formulas for gravitational potentials of usual point-like masses in closed ($\mathrm{K}=1$), flat ($\mathrm{K}=0$) and open ($\mathrm{K}=-1$) Universes. This procedure leads to severe constraints imposed on the parameter $\omega$. As a result, we single out unforbidden cases deserving further investigation.

## 2. SCALAR PERTURBATIONS AND GRAVITATIONAL POTENTIALS

Let us start with the background metrics

$$ds^2 = a^2\left(d\eta^2 - \gamma_{\alpha\beta}dx^\alpha dx^\beta\right), \quad \gamma_{\alpha\beta} = \frac{\delta_{\alpha\beta}}{\left[1+\frac{1}{4}\mathrm{K}(x^2+y^2+z^2)\right]^2}, \quad \alpha,\beta=1,2,3, \quad (1)$$

and the corresponding Friedmann equations

$$\frac{3(H^2+K)}{a^2} = \kappa \bar{T}_0^0 + \Lambda + \kappa\varepsilon, \qquad (2)$$

$$\frac{2H'+H^2+K}{a^2} = \Lambda - \kappa p = \Lambda - \kappa\omega\varepsilon, \qquad (3)$$

where $\varepsilon$ and $p = \omega\varepsilon$ are the energy density and the pressure of quintessence or the phantom field, $\bar{T}_0^0$ is the average energy density of the usual nonrelativistic dust-like matter (its pressure is equal to zero), K is the spatial curvature, $H \equiv a'/a \equiv (da/d\eta)/a$ and $\kappa \equiv 8\pi G_N/c^4$ ($c$ is the speed of light and $G_N$ is Newton's gravitational constant), $\Lambda$ is the cosmological constant and $a$ is the time-dependent scale factor. Let us remind that we consider the late stages and, hence, neglect the contribution of radiation and all other contributions of the same or close order. In the framework of the theory of scalar cosmological perturbations after some substitutions we get

$$\frac{1}{2}\left(-\kappa\bar{T}_0^0 - \kappa(1+\omega)\varepsilon\right)\varphi = \frac{1}{2}\kappa c^2 a\delta p, \quad \left(-\kappa\frac{\bar{\rho}c^2}{a^3} - \kappa(1+\omega)\varepsilon_0\frac{a_0^{3(1+\omega)}}{a^{3(1+\omega)}}\right)\varphi = \kappa c^2 a\omega\delta\varepsilon. \qquad (4)$$

It should be noted that we consider quintessence or phantom field without thermal coupling to any other type of matter. It means that evolution of its average characteristics as well as scalar perturbations occurs adiabatically or, in other words, without change of entropy. Therefore, we preserve the same linear equation of state $\delta p = \omega\delta\varepsilon$ with the same constant parameter $\omega$ for the scalar perturbations $\delta p$ and $\delta\varepsilon$, as for their background values $p$ and $\varepsilon$.

We can neglect the first term in parenthesis in the left hand side of (4), which is of the order $1/a^3$ (because even the order $o(1/a^2)$ here is beyond our accuracy). In the case $\omega > -1/3$ the second term can be also neglected, being of the order $o(1/a^2)$. Therefore, each type of additional matter with the linear equation of state $p = \omega\varepsilon$ with the constant nonzero parameter $\omega > -1/3$ is homogeneous ($\delta\varepsilon = o(1/a^3) \approx 0$) within our accuracy and does not influence anything, except for the dependence of $a$ on $\eta$. At the same time for $\omega = 0$ the scalar perturbation $\delta\varepsilon$ turns out to be arbitrary. There are two acceptable possibilities in this case: (a) $\delta\varepsilon = o(1/a^3) \approx 0$ (then it influences nothing, except for the dependence of $a$ on $\eta$) and (b) $\delta\varepsilon = O(1/a^3)$ (then it contributes to the gravitational potential). We shall not consider the latter possibility of the additional "non-particle" dust. Let us also note that if $\omega = -1$, then $\delta\varepsilon = o(1/a^3) \approx 0$.

However, if $\omega \neq -1$ and simultaneously $\omega \leq -1/3$, then the second term in parenthesis should be taken into account and from (4) we get

$$\delta\varepsilon = -\frac{1+\omega}{\omega}\varepsilon_0\frac{a_0^{3(1+\omega)}}{a^{3(1+\omega)}}\frac{\varphi}{c^2 a} = -\frac{1+\omega}{c^2\omega}\varepsilon_0\frac{a_0^{3+3\omega}}{a^{4+3\omega}}\varphi. \qquad (5)$$

After some substitutions, taking into account that $\delta T_0^0 = \delta\rho c^2/a^3$, where $\delta\rho = \rho - \bar{\rho}$ is the difference between the rest mass density and its average value, we obtain

$$\Delta\varphi + 3K\varphi = \frac{1}{2}\kappa c^2 a^3 \frac{\delta\rho c^2}{a^3} - \frac{1}{2}\kappa c^2 a^3 \frac{1+\omega}{c^2\omega}\varepsilon_0\frac{a_0^{3+3\omega}}{a^{4+3\omega}}\varphi, \quad \Delta\varphi + 3K\varphi = \frac{1}{2}\kappa c^4\delta\rho - \frac{1+\omega}{2\omega}\kappa\varepsilon_0\frac{a_0^{3+3\omega}}{a^{1+3\omega}}\varphi. \qquad (6)$$

Now the only acceptable possibility is $\omega = -1/3$. This is exactly the boundary case between the accelerating and decelerating regimes of evolution, investigated intensively in a lot of articles (see, e.g., [7-9]). At the same time it turns out that the phantom field with $\omega < -1$ is completely forbidden (together with its numerous popular particular cases). In the most interesting boundary case $\omega = -1/3$ we get

$$\delta\varepsilon = \frac{2\varepsilon_0 a_0^2}{c^2 a^3}\varphi, \quad \Delta\varphi + \left(3K - \frac{8\pi G_N}{c^4}\varepsilon_0 a_0^2\right)\varphi = 4\pi G_N(\rho - \bar{\rho}). \qquad (7)$$

Then we have focused attention on finding exact solutions of this Helmholtz equation. For the flat space $K = 0$, $dl^2 = dr^2 + r^2 d\Omega^2$ from (7) we obtain

$$\Delta\varphi - \frac{8\pi G_N}{c^4}\varepsilon_0 a_0^2 \varphi = 4\pi G_N(\rho - \bar{\rho}). \tag{8}$$

For the spherical space ($K = +1$, $dl^2 = d\chi^2 + \sin^2\chi d\Omega^2$)

$$\Delta\varphi + \left(3 - \frac{8\pi G_N}{c^4}\varepsilon_0 a_0^2\right)\varphi = 4\pi G_N(\rho - \bar{\rho}). \tag{9}$$

For the hyperbolic space ($K = -1$, $dl^2 = d\chi^2 + \sinh^2\chi d\Omega^2$)

$$\Delta\varphi - \left(3 + \frac{8\pi G_N}{c^4}\varepsilon_0 a_0^2\right)\varphi = 4\pi G_N(\rho - \bar{\rho}). \tag{10}$$

We considered all possible cases of the solutions of these equations and there were founded physically reasonable expressions for the gravitational potential of a resting point-like mass.

### 3. CONCLUSIONS

The produced investigation of inhomogeneous quintessence and phantom field with a constant parameter $\omega$ in the linear equation of state allows to draw the following three main conclusions:

1) the phantom field ($\omega < -1$) is completely forbidden;
2) the models containing quintessence may be viable only if $-1/3 < \omega < 0$;
3) in the boundary case $\omega = -1/3$ the gravitational potentials are defined for all generally accepted types of spatial topology, and some of them demonstrate satisfactory asymptotical behaviour and allow the averaging procedure, i.e. these gravitational potentials have clear physical meaning.

Now it is possible to investigate gravitational interaction between cosmic bodies and their motion in the Universe in the presence of quintessence, applying the derived formulas for the gravitational potentials. There are opportunities for numerical simulations. In the future the obtained results will be generalized to the case of a variable parameter when quintessence or phantom field represent some scalar field $\phi$ with a certain potential $V(\phi)$. The comparison with the experimental data can answer a question whether these material constituents of the Universe are viable and deserving serious scientific attention. Any answer to this urgent question will lead to extending of our astrophysical and cosmological knowledge.

# Motion of astrophysical objects in the open Universe

M.V. Eingorn [1], A.V. Kudinova [2], A.I Zhuk [3]

[1] *Department of Theoretical and Experimental Nuclear Physics, Odessa National Polytechnic University, Shevchenko av. 1, 65044 Odessa, Ukraine*
[2] *Department of Theoretical Physics, Odessa National University, st. Dvoryanskaya 2, 65082 Odessa, Ukraine*
[3] *Astronomical Observatory, Odessa National University, st. Dvoryanskaya 2, 65082 Odessa, Ukraine*

We consider dynamical behavior of astrophysical objects (galaxies) taking into account both the gravitational attraction between them and the cosmological expansion of the Universe. We obtain the general system of equations and apply them to some abstract systems of galaxies. We also define the region in the vicinity of our Local Group where the formation of the Hubble flows starts. For such processes, the zero-acceleration surface (where the gravitational attraction is balanced by the cosmological accelerated expansion) plays the crucial role. We show that such surface is absent for the Local Group. Instead, we find two points and one circle with zero acceleration. Nevertheless, there is a nearly closed area around the MW and M31 where the absolute value of the acceleration is approximately equal to zero. The Hubble flows are formed outside of this area.

## 1. INTRODUCTION

According to recent observations, there is no clear evidence of spatial homogeneity of the Universe up to sizes ~ 150 Mpc [1]. Deep inside of such scales and on late stages of evolution, the Universe consists of a set of discrete inhomogeneities (galaxies and their groups), which perturb the background Friedmann Universe. Hence, classical mechanics of discrete objects provides more adequate approach than hydrodynamics with its continuous flows. We have elaborated this approach for an arbitrary number of randomly distributed inhomogeneities on the cosmological background and found the gravitational potential of this system. We have shown that this potential has the most natural form in the case of the hyperbolic space [2]. For a system of gravitating masses $m_i$ it reads

$$\varphi = -G_N \sum_i m_i \frac{\exp(-2l_i)}{\sinh l_i} + \frac{4\pi G_N \bar{\rho}}{3}, \qquad (1)$$

where $G_N$ is the Newtonian gravitational constant, $\bar{\rho} = const$ is the comoving average rest mass density and $l_i$ denotes the comoving geodesic distance between the $i$-th mass $m_i$ and the point of observation in the open Universe, i.e. in the hyperbolic space.

We consider the potential (1) against the cosmological background deep inside the cell of uniformity, i.e. for physical distances $R \sim 150$ Mpc. As we have shown in [2], the comoving distances in the cell of uniformity are much less than 1: $l_i \ll 1$. For such small distances, we can use the Cartesian coordinates. The Lagrange equations for the $i$-th mass take the form

$$-G_N \sum_{j \neq i} \frac{m_j (X_i - X_j)}{[(X_i - X_j)^2 + (Y_i - Y_j)^2 + (Z_i - Z_j)^2]^{3/2}} = \frac{1}{a}(\ddot{X}_i a - \ddot{a} X_i), \qquad (2)$$

$$-G_N \sum_{j \neq i} \frac{m_j (Y_i - Y_j)}{[(X_i - X_j)^2 + (Y_i - Y_j)^2 + (Z_i - Z_j)^2]^{3/2}} = \frac{1}{a}(\ddot{Y}_i a - \ddot{a} Y_i), \qquad (3)$$

$$-G_N \sum_{j \neq i} \frac{m_j (Z_i - Z_j)}{[(X_i - X_j)^2 + (Y_i - Y_j)^2 + (Z_i - Z_j)^2]^{3/2}} = \frac{1}{a}(\ddot{Z}_i a - \ddot{a} Z_i). \qquad (4)$$

Now, we can apply these equations to real astrophysical systems such as a group of galaxies. To illustrate this, we consider first a number of abstract simplified examples.

## 2. ILLUSTRATIVE EXAMPLES

First, we study dynamics of three gravitating masses ($N=3$). The first diagram (Fig. 1) demonstrates the situation of zero initial velocities. On the left panel of Fig. 1 the cosmological expansion prevails the gravitational attraction between the test masses and all the points recede from each other. On the right panel of the Fig. 1 the gravitational attraction dominates in the interaction between two points, that is why these points get closer to each other eventually.

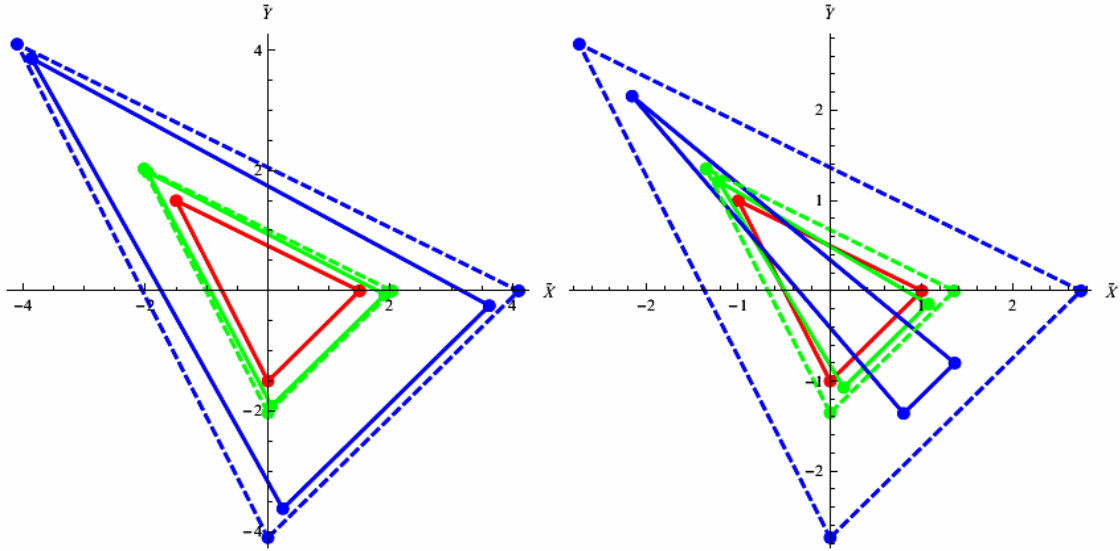

Fig. 1. The dynamics of three points.

The same analysis may be taken out for four gravitating masses ($N=4$). The second diagram (Fig. 2) illustrates the situation of non-zero initial velocities of test masses. Presence of these nonzero values influences the dynamics of the points: the whole system starts to turn around. The left panel of the figure illustrates the cosmological expansion domination (all the points recede from each other). The right panel illustrates the gravitational attraction domination, so that the system collapses.

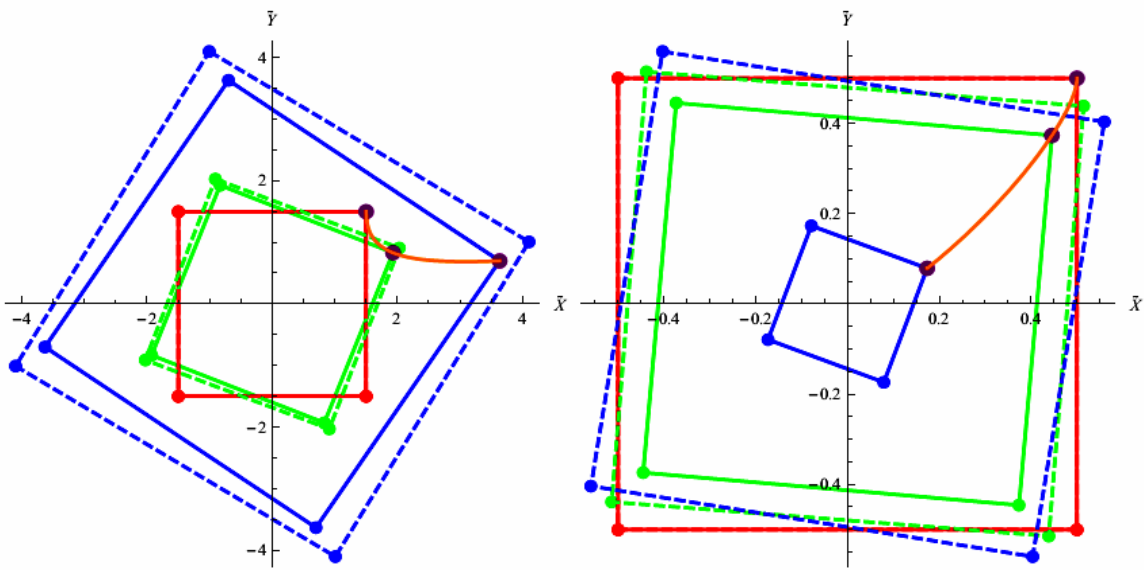

Fig. 2. The dynamics of four points.

## 3. FORMATION OF HUBBLE FLOWS IN THE VICINITY OF THE LOCAL GROUP

Now we consider a test particle/dwarf galaxy in the gravitational field of Andromeda and Milky Way. The picture is studied at present time when the separation distance between M31 and MW is $L = 0.78$ Mpc, and we do not take into account the relative motion of these galaxies. The Lagrange equation for a test particle is

$$\frac{d}{dt}\left(\vec{V} - \frac{\dot{a}}{a}\vec{R}\right) = -\frac{1}{a}\frac{\partial \varphi}{\partial \vec{R}} + \frac{\dot{a}^2}{a^2}\vec{R} - \frac{\dot{a}}{a}\vec{V} \quad , \tag{5}$$

or, equivalently,

$$\dot{\vec{V}} - \frac{\ddot{a}}{a}\vec{R} = -\frac{1}{a}\frac{\partial \varphi}{\partial \vec{R}} \quad . \tag{6}$$

In the case of our Local Group, the gravitational potential is:

$$\varphi = \varphi_A + \varphi_B \quad , \quad \varphi_A = -aG_N \frac{m_A}{\left|\vec{R}_A - \vec{R}\right|} \quad , \quad \varphi_B = -aG_N \frac{m_B}{\left|\vec{R}_B - \vec{R}\right|} \quad , \tag{7}$$

where we mark MW and M31 by letters A and B, respectively. The absolute value of the dimensionless acceleration is

$$\left|\vec{\tilde{W}}\right| = \left|\frac{d\vec{\tilde{V}}}{d\tilde{t}}\right| = \frac{1}{\left(H_0^4 G_N \overline{m}\right)^{1/3}}\left|\frac{d\vec{V}}{dt}\right| = \sqrt{\left(\tilde{W}_x\right)^2 + \left(\tilde{W}_y\right)^2 + \left(\tilde{W}_z\right)^2} \quad . \tag{8}$$

There is a characteristic region defined by the condition

$$\left|\frac{1}{\tilde{a}}\frac{d^2\tilde{a}}{d\tilde{t}^2}\vec{\tilde{R}}\right| \sim \left|\frac{1}{\tilde{a}}\frac{\partial \tilde{\varphi}}{\partial \vec{\tilde{R}}}\right| \quad , \tag{9}$$

where the gravitational attraction is balanced by the cosmological expansion. The gravitational attraction is stronger inside of this region (for smaller distances) and the cosmological expansion prevails outside of this area (for larger distances) where the Hubble flows are formed.

For our Local Group consisting of two giant galaxies MW and M31, we choose the origin of coordinates in the barycenter of these galaxies and X-axis along the line connecting MW and M31. Therefore, $X_A = L_0 m_B/(m_A + m_B)$, $X_B = -L_0 m_A/(m_A + m_B)$ and $Y_A = Z_A = 0$, $Y_B = Z_B = 0$. We investigate the distribution of the test body acceleration in the plane $Z = 0$. For the masses MW and M31, we take the following values: $m_A \approx 10^{12} M_\odot$ and $m_B \approx 1.6 \times 10^{12} M_\odot$.

In Fig. 3 we depict the absolute value and the vector field of the acceleration (8) of a test body in the plane $Z = 0$.

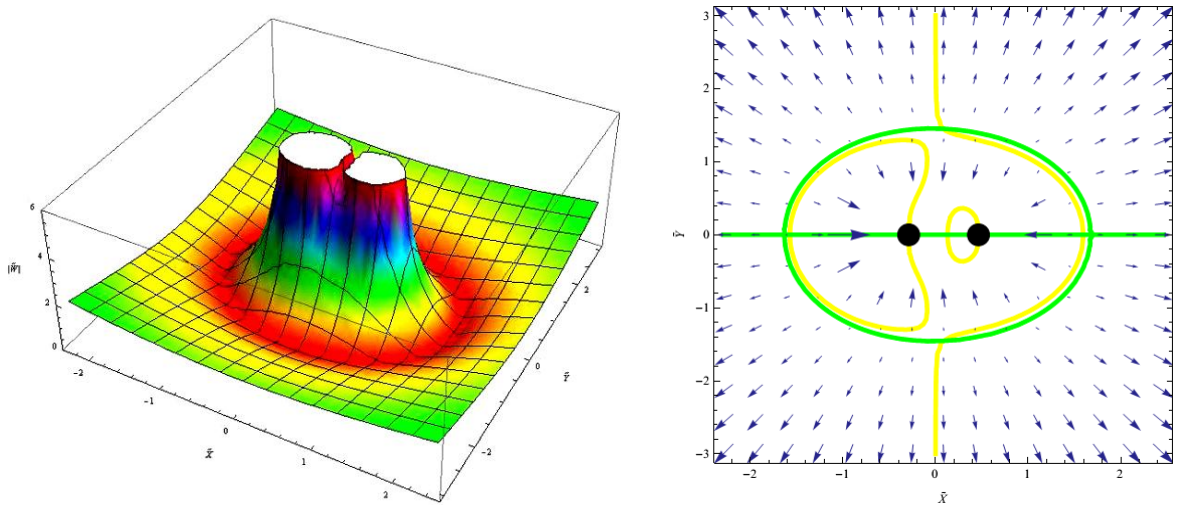

Fig. 3. The absolute value and the vector field of the acceleration.

To define more exactly the structure of the zero-acceleration surface, we draw Fig. 4. The yellow and green lines correspond to the conditions $W_x = 0$ and $W_y = 0$. Black points define the positions of the Milky Way (the right point) and Andromeda (the left point). Red points are defined by the condition $W_x = W_y = 0 \Rightarrow |\vec{W}| = 0$. The zero acceleration surface is absent. We have two additional red points on the $X$-axis and two vertical points. Clearly, due to the rotational symmetry, the latter two points are just the section of a zero acceleration circle by the plane $Z = 0$. The elliptic-like green and yellow lines are very close to each other and define the region where $|\vec{W}| \approx 0$. Also there are two regions where this approximate zero acceleration surface has discontinuity. It is clear from the rotational symmetry that these two regions belong to the round belt (they are the section of this belt by the plane $Z = 0$). Therefore, inside of this surface the gravitational attraction is stronger than the cosmological expansion while outside of this surface the cosmological expansion prevails over the gravitational attraction. Obviously, the Hubble flows are formed in the latter region. Additionally, we can see that there is an asymmetry in directions along $X$ and $Y$ axes. The characteristic distances from the barycenter to the zero acceleration surface are $|\tilde{X}| \approx 1.62 \to |X| \approx 1.68$ Mpc and $|\tilde{Y}| \approx 1.41 \to |Y| \approx 1.46$ Mpc.

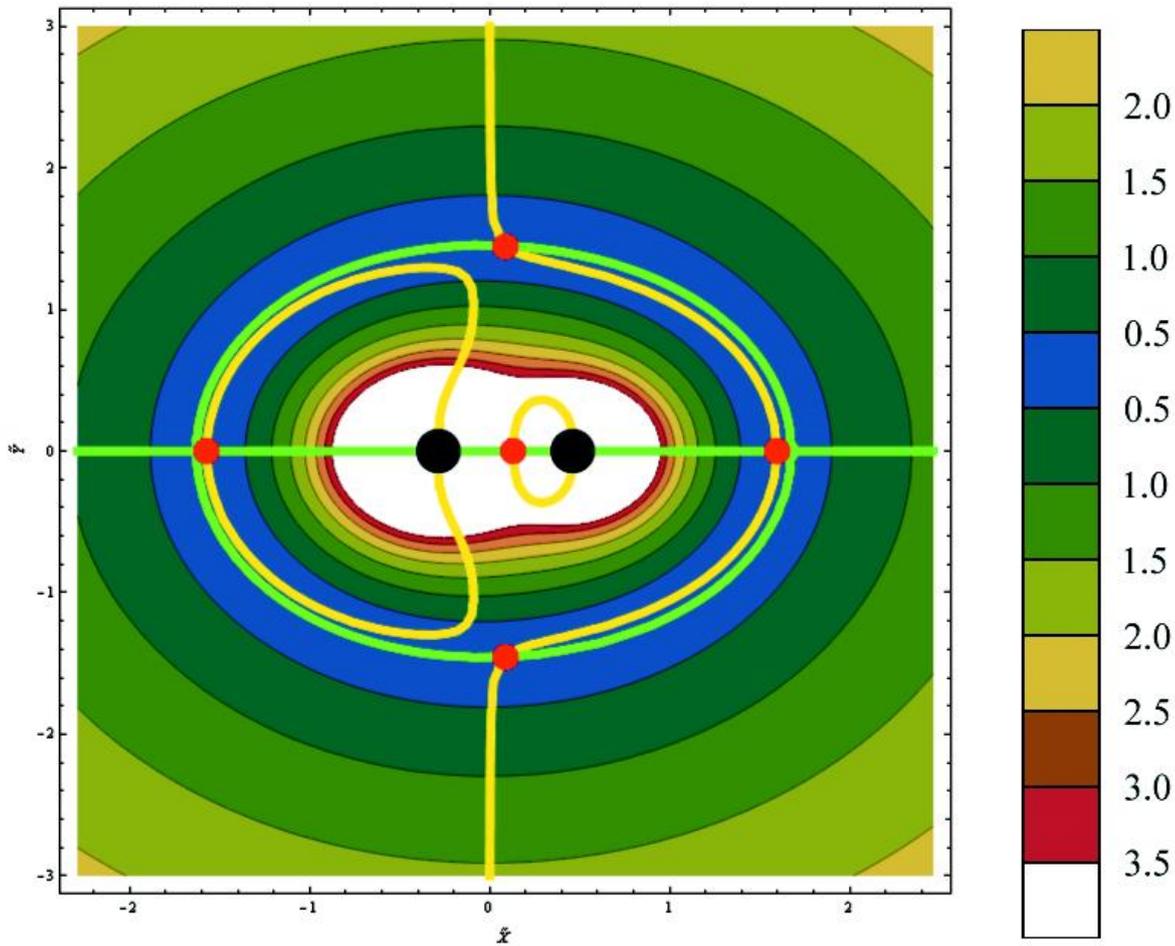

Fig. 4. Contour plot of the absolute value of the acceleration in our Local Group.

# Propagation of Ultra High Energy Cosmic Rays in Galactic and Extragalactic Magnetic Fields

**B.I. Hnatyk [1], O.O. Kobzar [2], V.V. Marchenko [2], O.B. Sushchov [2]**

[1] *Astronomical Observatory of Taras Shevchenko Kyiv National University, 04053 Kyiv, Ukraine*
[2] *T.G. Shevchenko Chernihiv National Pedagogical University, 14013 Chernihiv, Ukraine*

The propagation of ultra-high energy cosmic rays (UHECR) in Galactic and extragalactic magnetic fields is studied. The trajectories of charged particles of different chemical composition and energy are simulated with the use of different models of the Galactic magnetic field (GMF). The influence of both galactic and extragalactic turbulent fields was taken into account and thus the positions for the real sources of events registered by Auger Observatory were calculated. The possibility of their correlation with the nearest plausible sources such as radio galaxy Centaurus A is analyzed.

## 1. INTRODUCTION

Ultra-high energy (over $4 \cdot 10^{19}$ eV) cosmic rays are commonly believed to be of extragalactic origin due to absence of sources powerful enough to provide their acceleration to ultra-high energies within our galaxy. Their astrophysical nature is approved by registering the Greisen-Zatsepin-Kuzmin effect [1, 2] (the so called GZK cutoff of CR's energy spectrum) in the HiRes experiment [3], as well as in observations carried out by Auger Observatory [4].

The possible candidates to be UHECR sources include galaxies from the active galactic nuclei (AGN), gamma-ray bursts, young magnetars and shock waves in areas of large-scale structure's formation. Only a few classes of stable astrophysical objects (like Fanaroff-Riley radio galaxies of type I and II) are powerful enough to accelerate protons up to ultra-high energy values, but for heavier nuclei the list of possible accelerators is wider. Thus several closest radio galaxies could be regarded as contributors to the observed UHECR flux.

Analysis of UHECR chemical composition by Auger Observatory and Yakutsk EAS Array [5] data show a shift towards heavy nuclei (starting at $3 \times 10^{19}$ eV). On the other hand, HiRes and Telescope Array [6] report CR's predominantly proton composition. The reason of the discrepancy in the primary experiments' results is still to be discussed.

Updated Auger results [7] show that registered UHECR correlation with the AGN from the VCV catalogue [8] has decreased from 3σ to 2σ compared to the previous data. Thus only 30% of UHECR potentially correlate with AGN and the rest show the signs of isotropic distribution. The exception is the region near the closest to us active galaxy Centaurus A which contains more ultra-energy events than it could be statistically correct to expect.

We verify the possibility of UHECR observed in that area originate from Centaurus A.

## 2. MODELING

Magnetic field distorts the charged particle trajectory via Lorenz force. When it is static it doesn't affect the particle energy. As ultra-high energies far exceed the particles' rest energy, we assume that UHECR propagate with velocity close to the speed of light. In this case the equations of the motion of ultra-relativistic particles in magnetic field **B(r)** are:

$$\frac{dv}{dt} = \frac{qc^2}{E}[v \times B], \quad \frac{dr}{dt} = v$$

where $q$ is particle's charge, $E$ – its energy.

Due to the complex structure of the magnetic field, solving these equations analytically is impossible. Therefore trajectory was calculated numerically.

Modeling the UHECR propagation, we considered influence of Galactic as well as extragalactic magnetic fields. GMF consists of regular and turbulent components. The structure of the regular component is believed to generally follow the matter's distribution in the Galaxy. There are several GMF models [9-11] which differ in parameters' numerical values and presence of certain components. The structure of extragalactic magnetic field as well as of turbulent galactic field is not exactly clear, thus those fields were modeled as random ones [12].

### 3. RESULTS

Figure 1 shows the results of calculations by different regular GMF models. Circles with figures denote events registered near Centaurus A. Circles with the chemical elements symbols correspond to the calculated locations of UHECR's sources for the indicated particle types. Radii of all circles correspond to the Auger detectors' experiment error within 1σ. Results depicted in Fig. 1a were obtained via the use of model [9]; those in Fig. 1b were achieved as the result of using model [10]. The figures also demonstrate the outline of Centaurus A's radiation areas.

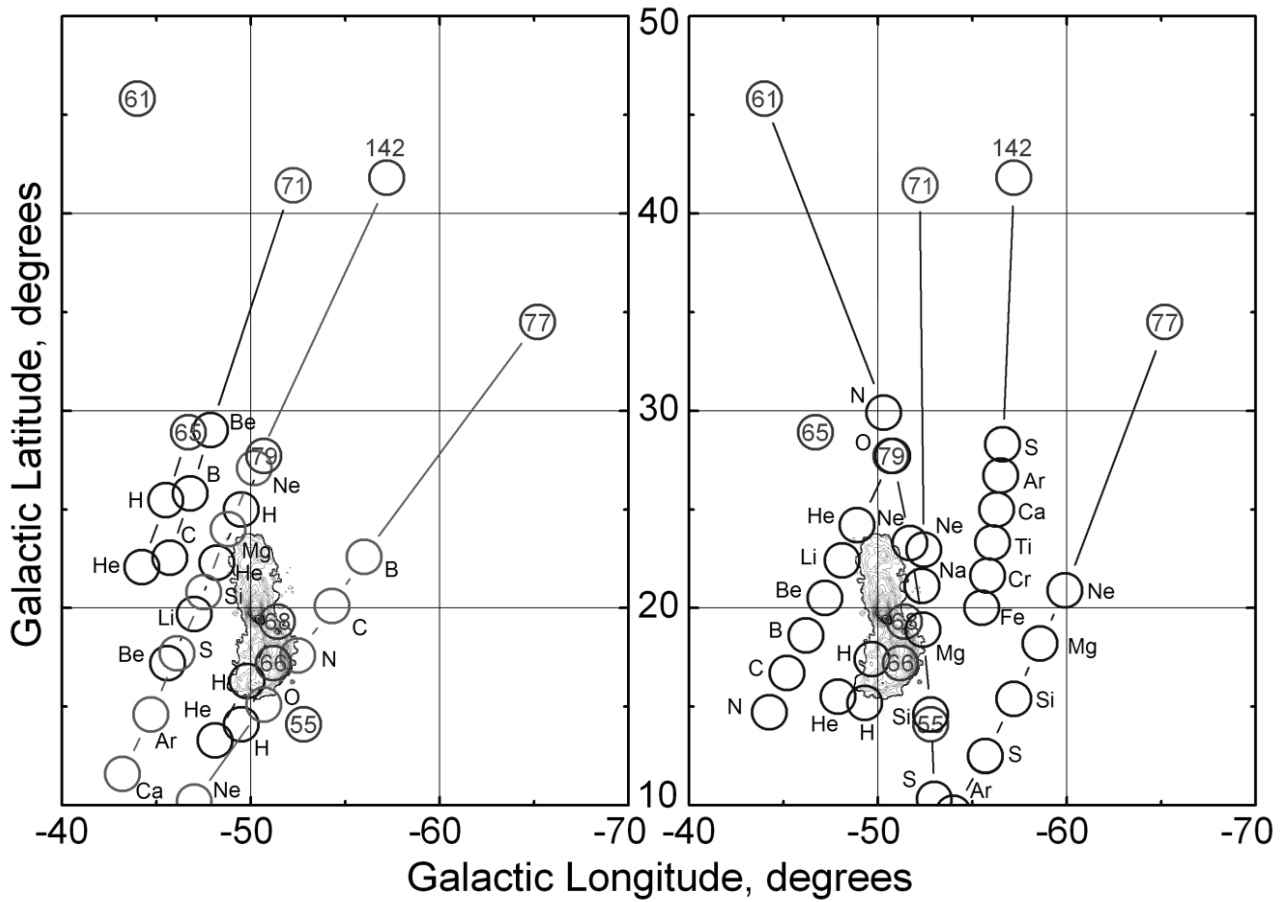

Fig. 1. UHECR sources' location calculated using the models: (a) model [9], (b) model [10].

We've found out that only six UHECR can in fact originate in Centaurus A – those with energies of 61, 66, 68, 77, 79 and 142 EeV. Table 1 shows chemical composition of the particles with the indicated energy. These particles correlate with Centaurus A following the two chosen GMF models and considering the impact of the magnetic field's various components (RegF – regular Galactic field; RmF – random Galactic field; EF - extragalactic field).

Table 1. UHECR correlation with Centaurus A

| CR's energy, EeV | CR's chemical composition | | | | | |
|---|---|---|---|---|---|---|
| | model [9] | | | model [10] | | |
| | RegF | RegF + RmF | RegF + RmF + EF | RegF | RegF + RmF | RegF + RmF + EF |
| 142 | Mg | Mg — Ar | Ne — Ca | — | Ca — Fe | S — Fe |
| 79 | He | He — Li | p — Be | He — Li | He — C | He — N |
| 77 | N — O | C — Ne | B — Ne | — | Mg — Ar | Ne — Ar |
| 68 | p | p | p — He | p | p — He | p — He |
| 66 | — | p | p | p | p | p |
| 61 | — | — | — | Ne — Mg | O — S | N — S |
| CR's energy, EeV | CR's chemical composition | | | | | |
| | model [11] —ASS symmetry | | | model [11] —BSS symmetry | | |
| | RegF | RegF + RmF | RegF + RmF + EF | RegF | RegF + RmF | RegF + RmF + EF |
| 68 | p | p | p | — | — | p |
| 66 | p | p | p | — | — | p |

# The Structure of X-ray Jet of Quasar 3C 273

A. Bogdan[1], V. Marchenko[2], B. Hnatyk[1]

[1]*Astronomical Observatory of Taras Shevchenko National University of Kyiv, 3 Observatorna Str., 04053 Kyiv, Ukraine*
[2]*Astronomical Scientific Research Center, Taras Shevchenko National Pedagogical University of Chernihiv, 53, Getmana Polubotka Str., 14013 Chernihiv, Ukraine*

The X-ray internal structure of extragalactic jets is analyzed. The Lucy-Richardson deconvolution algorithm is used to restore the image of X-ray sources. The analysis was done for Chandra observations of core-dominated quasar 3C 273. The transverse profiles are built for the jet knots.

## 1. INTRODUCTION

The active galactic nuclei (AGN) remain one of the most interesting and remarkable issues in modern astrophysics [1]. One of the manifestations of AGN activity is the existence of extragalactic jets that constitute the longest collimated structures in the Universe [2]. The detailed study of extragalactic jet structure in all wavelengths is important task for modeling the different astrophysical processes that take place in jet, for example the acceleration of cosmic rays up to the ultra high energies due to Fermi acceleration mechanisms [3, 4] and acceleration on the jet boundary [5].

## 2. OBSERVATIONAL DATA

We have used four Chandra observations of core-dominated quasar 3C 273 and its jet (ObsIDs: 4876, 4877, 4878, 4879) with total exposure time of 160 ks [6]. The X-ray data analysis was processed with CIAO 4.4 – a software package for Chandra interactive analysis of observations [7]. We have merged four observations for further analysis (ObsIDs: 4876, 4877, 4878, 4879) and have binned them with binning factor of 0.125 (it corresponds to bin size 0.0615"). The merged and binned image of jet is presented in Figure 1.

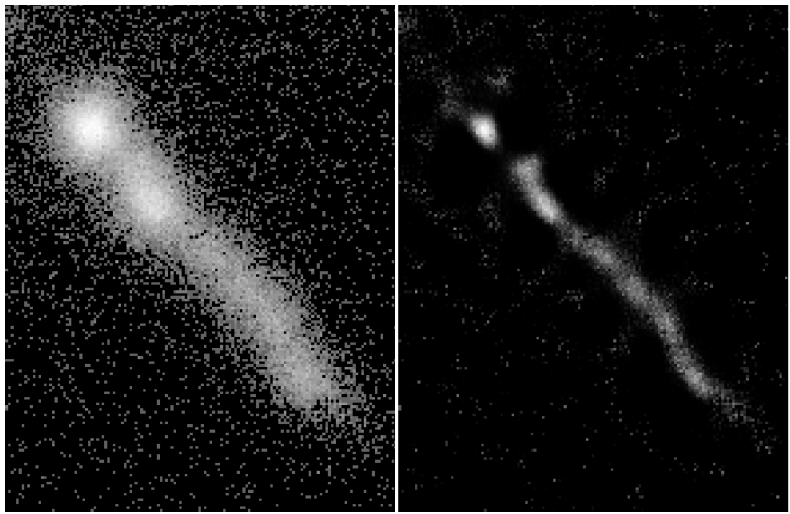

Fig. 1. The merged Chandra image of 3C 273 jet binned in 0.0615" bins (left) and restored image of 3C 273 jet using the Lucy-Richardson deconvolution algorithm with number of iteration $n = 120$ (right).

## 3. LUCY-RICHARDSON DECONVOLUTION ALGORITHM

Image of astronomical source observed from space can be mathematically described as a convolution of the intrinsic brightness distribution of the source with a blurring function, also known as a point spread function (PSF). Therefore one of the possibilities to restore the intrinsic source distribution from the observed data is to use some of deconvolution techniques.

In the present research we have used the Lucy-Richardson deconvolution algorithm that is implemented in CIAO 4.4 in "arestore" tool [8]. This algorithm requires the PSF image and we have used the modeled PSF generated by ChaRT and MARX - the programs for detailed ray-trace simulation [9, 10].

## 4. RESULTS AND CONCLUSIONS

Taking into account the fact, that the Lucy-Richardson deconvolution method does not produce reliable convergence and uncertainty information, users should be very cautious in interpreting and evaluating the results of the deconvolution, especially when applied to faint and/or extended sources. Also it's noticeable, that the characteristics of restored image (width of knots, amplitude, etc.) slightly vary for different number of iterations to perform, that is one of input parameters for Lucy-Richardson deconvolution algorithm.

In order to ensure that the result of deconvolution is in good correspondence with observed data and to choose reasonable number of iterations we have decided to inspect it by comparing the restored image with observed data. For this we have constructed the model of data as a result of convolution of the restored image with PSF:

$$M_i = S_i \circ P_i, \qquad (1)$$

where $M_i$, $S_i$ and $P_i$ is the numbers of counts per bin $i$ in model, restored image and PSF, respectively.

In order to estimate the optimal number of iterations, we need to explore what number of iterations produces the physical model that describes the data in the best way. To figure it out we have used $\chi^2$ statistics for different number of iterations

$$\chi^2 = \sum_i \frac{(N_i - M_i)^2}{\sigma_i^2}, \qquad (2)$$

where $N_i$ is the total number of observed counts in bin $i$ and $\sigma_i$ is the error in bin $i$.

At the present research we take into account only statistical noise as an error. We assume that counts are sampled from the Poisson distribution with a mean value equal to total number of observed counts $N_i$ in a bin. So the count standard deviation for a bin $i$ can be taken as $\sigma_i = \sqrt{N_i}$. But when the average number of counts in the bin is small ($N_i < 5$), then we cannot assume that the Poisson distribution − from which the counts are sampled − has a nearly Gaussian shape. The standard deviation for such low-count case has been derived as $\sigma_i = 1 + \sqrt{N_i + 0.75}$ [11].

We have calculated $\chi^2$ for different number of iterations and have shown that $\chi^2$ decreases with the number of iterations. We have decided to choose the reasonable number of iteration reaching the $\chi^2 = 1.1$. It corresponds to number of iterations $n \approx 120$ (Fig. 2).

Also to inspect the distribution of residuals between data $N_i$ and model $M_i$ across the image and to find the regions with significant deviations, we have analyzed the residuals $N_i - M_i$. To analyze the residual in bin $i$ one have to take into account that the significance of this residual depends on the error $\sigma$ in this bin. Therefore we have normalized the residuals by standard deviation $\sigma_i$, namely $R_i = (N_i - M_i)/\sigma_i$ (Fig. 2).

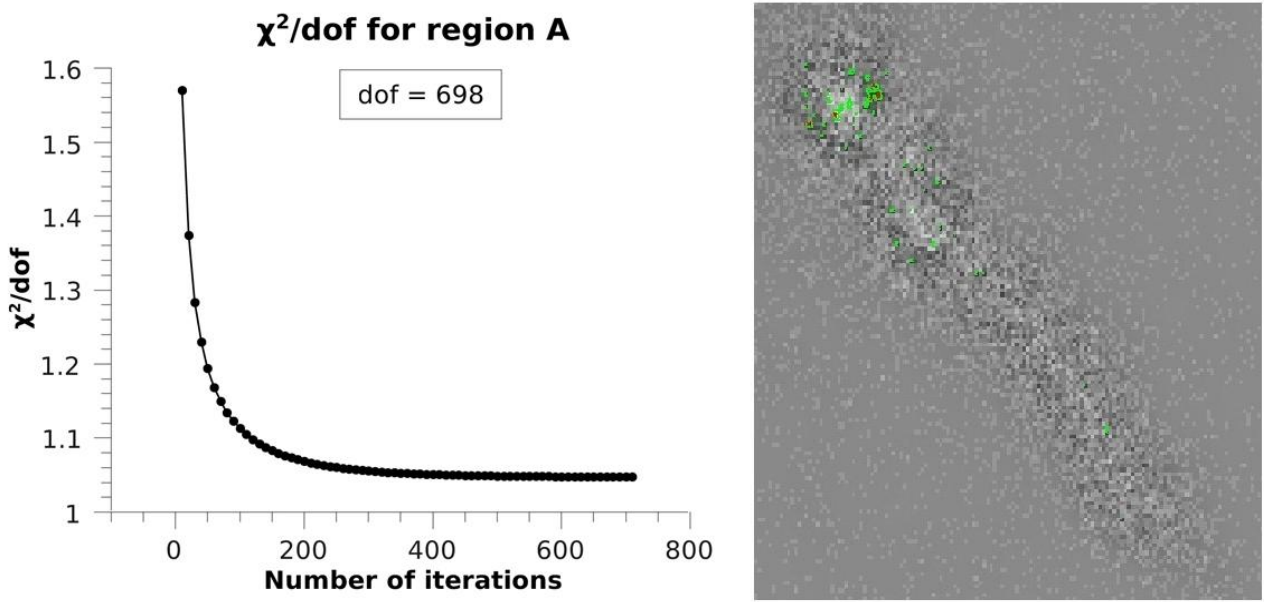

Fig. 2. $\chi^2$ for different number of iterations (left) and residual map normalized by standard deviation (right). The green contours correspond to $2\sigma$ deviation, red -- to $3\sigma$.

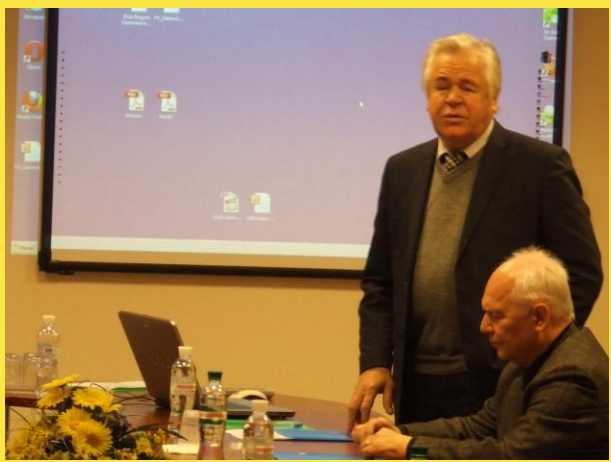 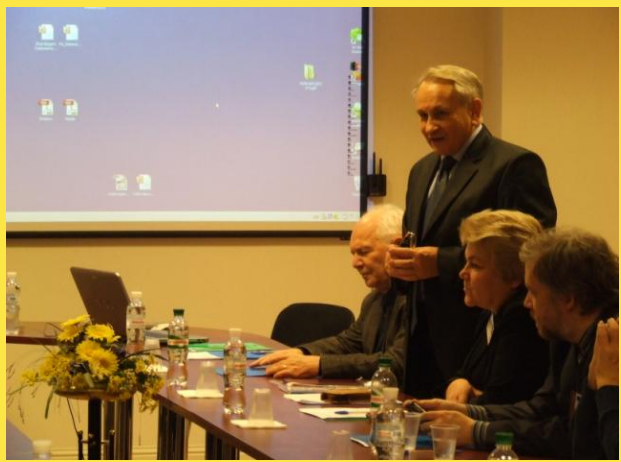
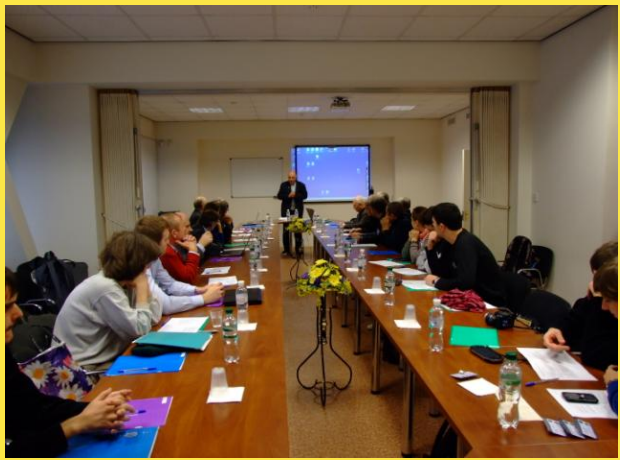 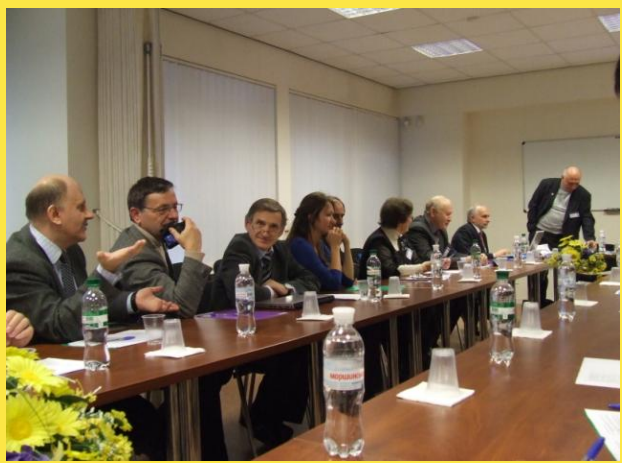
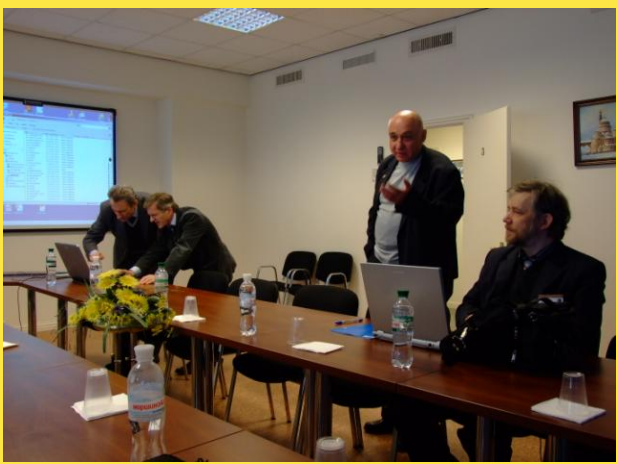 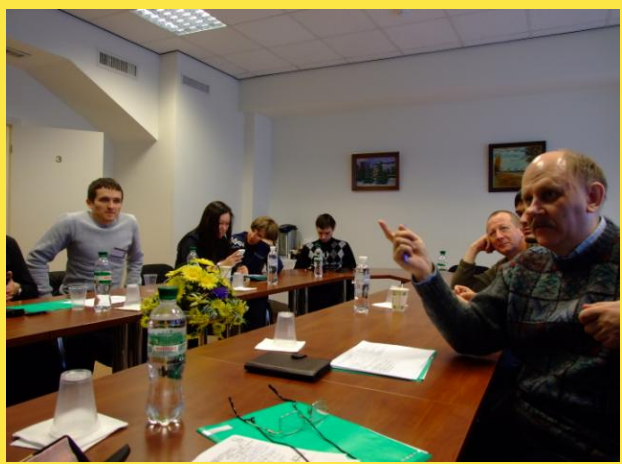
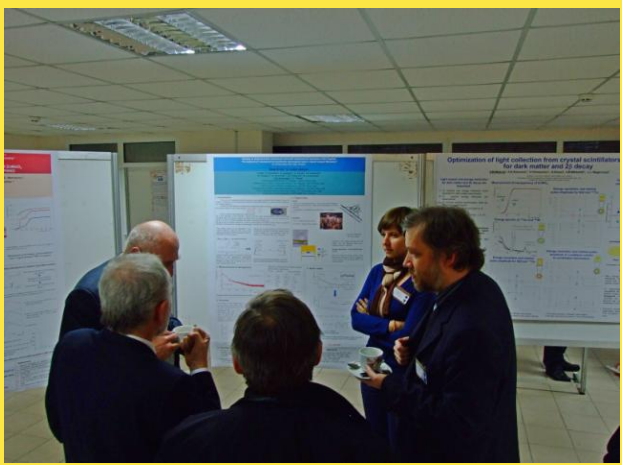 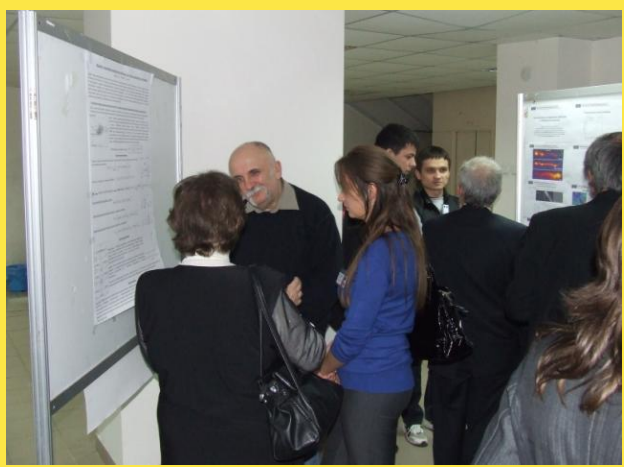